\documentclass[twocolumn]{aastex62}
\usepackage{etoolbox}

\hypersetup{linkcolor=blue,citecolor=blue,urlcolor=blue}

\accepted{to ApJS on June 17, 2018}

\shorttitle{MASSES 1.3 mm Subcompact Data Release}
\shortauthors{Stephens et al.}

\begin{document}
\newcommand{\ntdp}{\mbox{N$_2$D$^+$(3--2)}}
\newcommand{\ceo}{\mbox{C$^{18}$O(2--1)}}
\newcommand{\ceooz}{\mbox{C$^{18}$O(1--0)}}
\newcommand{\ttco}{\mbox{$^{13}$CO(2--1)}}
\newcommand{\coto}{\mbox{CO(2--1)}}

\newcommand{\ntdpnt}{\mbox{N$_2$D$^+$}}
\newcommand{\ceont}{\mbox{C$^{18}$O}}
\newcommand{\ttcont}{\mbox{$^{13}$CO}}
\newcommand{\cotont}{\mbox{CO}}

\newcommand{\httcop}{\mbox{H$^{13}$CO$^+$(4--3)}}
\newcommand{\hcop}{\mbox{HCO$^+$(4--3)}}
\newcommand{\cott}{\mbox{CO(3--2)}}

\newcommand{\httcopnt}{\mbox{H$^{13}$CO$^+$}}
\newcommand{\hcopnt}{\mbox{HCO$^+$}}
\newcommand{\cottnt}{\mbox{CO}}

\newcommand{\kms}{km\,s$^{-1}$}

\newcommand{\od}{$\tau_{\rm{353\,GHz}}$}
\newcommand{\pao}{$\rm{PA_{Out}}$}
\newcommand{\paf}{$\rm{PA_{Fil}}$}
\newcommand{\pafilf}{$\rm{PA_{Fil,F}}$}

\title{Mass Assembly of Stellar Systems and their Evolution with the SMA -- 1.3\,mm Subcompact Data Release}

\author{Ian W. Stephens}
\affiliation{Harvard-Smithsonian Center for Astrophysics, 60 Garden Street, Cambridge, MA, USA}

\author{Michael M. Dunham}
\affiliation{Department of Physics, State University of New York at Fredonia, 280 Central Ave, Fredonia, NY 14063, USA}
\affiliation{Harvard-Smithsonian Center for Astrophysics, 60 Garden Street, Cambridge, MA, USA}

\author{Philip C. Myers}
\affiliation{Harvard-Smithsonian Center for Astrophysics, 60 Garden Street, Cambridge, MA, USA}

\author{Riwaj Pokhrel}
\affiliation{Harvard-Smithsonian Center for Astrophysics, 60 Garden Street, Cambridge, MA, USA}
\affiliation{Department of Astronomy, University of Massachusetts, Amherst, MA 01003, USA}

\author{Tyler L. Bourke}
\affiliation{SKA Organization, Jodrell Bank Observatory, Lower Withington, Macclesfield, Cheshire SK11 9DL, UK}

\author{Eduard I. Vorobyov}
\affiliation{Research Institute of Physics, Southern Federal University, Stachki Ave. 194, Rostov-on-Don, 344090, Russia}
\affiliation{University of Vienna, Department of Astrophysics, Vienna, 1180, Austria}

\author{John J. Tobin}
\affiliation{Homer L. Dodge Department of Physics and Astronomy, University of Oklahoma, 440 W. Brooks Street, Norman, OK 73019, USA}
\affiliation{Leiden Observatory, Leiden University, P.O. Box 9513, 2300-RA Leiden, The Netherlands}

\author{Sarah I. Sadavoy}
\affiliation{Harvard-Smithsonian Center for Astrophysics, 60 Garden Street, Cambridge, MA, USA}

\author{Jaime E. Pineda}
\affiliation{Max-Planck-Institut f\"ur extraterrestrische Physik, D-85748 Garching, Germany}

\author{Stella S. R. Offner}
\affiliation{Department of Astronomy, The University of Texas at Austin, Austin, TX 78712, USA}

\author{Katherine I. Lee}
\affiliation{Harvard-Smithsonian Center for Astrophysics, 60 Garden Street, Cambridge, MA, USA}

\author{Lars E. Kristensen}
\affiliation{Centre for Star and Planet Formation, Niels Bohr Institute and Natural History Museum of Denmark, University of Copenhagen, \O ster Voldgade 5-7, DK-1350 Copenhagen K, Denmark}

\author{Jes K. J\o rgensen}
\affiliation{Niels Bohr Institute and Center for Star and Planet Formation, Copenhagen University, DK-1350 Copenhagen K., Denmark}

\author{Alyssa A. Goodman}
\affiliation{Harvard-Smithsonian Center for Astrophysics, 60 Garden Street, Cambridge, MA, USA}

\author{H\'{e}ctor G. Arce}
\affiliation{Department of Astronomy, Yale University, New Haven, CT 06520, USA}

\author{Mark Gurwell}
\affiliation{Harvard-Smithsonian Center for Astrophysics, 60 Garden Street, Cambridge, MA, USA}



\begin{abstract}

We present the Mass Assembly of Stellar Systems and their Evolution with the SMA (MASSES) survey, which uses the Submillimeter Array (SMA) interferometer to map the continuum and molecular lines for all 74 known Class~0/I protostellar systems in the Perseus molecular cloud. The primary goal of the survey is to observe an unbiased sample of young protostars in a single molecular cloud so that we can characterize the evolution of protostars. This paper releases the MASSES 1.3\,mm data from the subcompact configuration ($\sim$4$\arcsec$ or $\sim$1000\,au resolution), which is the SMA's most compact array configuration. We release both $uv$ visibility data and imaged data for the spectral lines \coto, \ttco, \ceo, and \ntdp, as well as for the 1.3\,mm continuum. We identify the tracers that are detected toward each source. We also show example images of continuum and \coto\ outflows, analyze \ceo\ spectra, and present data from the SVS~13 star-forming region. The calculated envelope masses from the continuum show a decreasing trend with bolometric temperature (a proxy for age). Typical \ceo\ linewidths are 1.45\,\kms, which is higher than the \mbox{C$^{18}$O} linewidths detected toward Perseus filaments and cores. We find that \ntdp\ is significantly more likely to be detected toward younger protostars. We show that the protostars in SVS~13 are contained within filamentary structures as traced by \ceo\ and \ntdp. We also present the locations of SVS~13A's high velocity (absolute line-of-sight velocities $>$150\,\kms) red and blue outflow components. Data can be downloaded from \url{https://dataverse.harvard.edu/dataverse/MASSES}.

\end{abstract}

\keywords{editorials, notices --- 
miscellaneous --- catalogs --- surveys}


\section{Introduction} Ê\label{sec:intro}
Stars are assembled in molecular clouds through the gravitational collapse of dense cores of gas and dust \citep[e.g.,][]{Shu1987}. The masses of stars are set during the protostellar stage by the complex interaction of many interrelated physical processes, including mass infall, core and disk fragmentation, ejection from multiple systems, the formation and evolution of protostellar disks, and mass loss through jets and outflows \citep[e.g.,][]{Offner2014}. While some progress has been made toward understanding these processes, studies have generally focused on small pieces of the puzzle using heterogeneous, small, and biased samples of well-studied protostars. A complete understanding of the interplay between these processes and their roles in assembling stars remains lacking.

Understanding core fragmentation, protostellar accretion, and outflows typically requires high spatial resolution ($\sim$1000\,au) line and continuum observations at (sub)millimeter wavelengths, and such observations can be accomplished with interferometers. Therefore, interferometric protostellar surveys can piece together the evolutionary sequence of protostars (defined here to be compact sources younger than the T Tauri/Class~II stage). Several spectral line and continuum interferometric surveys with sample sizes of about one to two dozen targets have already found important results. \citet{Arce2006} found evidence of erosion of protostellar envelopes by winds and that outflow cavities may widen as a protostar evolves. The PROSAC survey \citep{Jorgensen2007,Jorgensen2009,Jorgensen2015} also constrained protostellar evolution, with results that included finding evidence that disk masses are $\sim$0.05\,$M_\odot$ (with large scatter) during the Class~0/I stage and that accretion may be episodic. \citet{Yen2015} analyzed rotation kinematics at $\sim$1000\,au scales and suggested that magnetic braking may not be effective at stopping disk formation for most Class~0/I protostars.
Recent continuum-only interferometric surveys have also focused on protostellar evolution. For example, \citet{ChenX2013} found that in nearby clouds ($<$500~pc), Class~0 protostars exhibit a higher multiplicity fraction than Class~I protostars. The VLA Nascent Disk and Multiplicity (VANDAM) Perseus survey used the Karl G. Jansky Very Large Array (VLA) to observe continuum toward all protostars in the Perseus molecular cloud, and the survey showed that the protostellar companion separations follow a bimodal distribution \citep{Tobin2016}.


The spectral line interferometric surveys targeted a wide variety of sources in many different clouds. In particular, they focused on some of the brightest sources since they are easier to map with shorter integration times. However, considerable biases and problems may exist in these protostellar samples because these protostars 1)~are in widely varying star-forming environments, 2)~were mapped at different spatial resolutions, and 3)~were only the brightest sources. Such factors may greatly affect the statistical conclusions drawn from these observations.

One way to mitigate these problems is to survey all protostars within a single molecular cloud. Therefore, we used the Submillimeter Array \citep[SMA;][]{Ho2004} to map all the protostars in the Perseus molecular cloud  \citep[235 pc away;][]{Hirota2008} in a survey called the Mass Assembly of Stellar Systems and their Evolution with the SMA (MASSES). The MASSES survey observed both spectral lines and continuum toward more than 70 young stellar objects. Some early results from the survey have already been published. \citet{Lee2015} used survey data to characterize the well-known L1448N star-forming region and found consistency with thermal Jeans fragmentation. \citet{Lee2016} analyzed wide binaries (i.e., protostars separated by 1000 -- 10000 au) in the MASSES survey and found that their angular momentum axes (as probed by outflows) were either randomly aligned or perpendicularly aligned with each other. Models by \citet{Offner2016} found that such alignment is consistent with the predictions of turbulent fragmentation. \citet{Frimann2017} found evidence that accretion is episodic based on \ceo\ observations. \citet{Stephens2017b} investigated the alignment between filaments and outflows within Perseus, and found they may be randomly aligned rather than always parallel or perpendicular with each other. \citet{Pokhrel2018} found that, from the cloud scale down to the protostellar object/disk scale, sources with higher thermal Jeans numbers fragment into more sources than those with lower Jeans numbers; nevertheless, the number of detected fragments was lower than the expected Jeans number at every scale, suggesting the possibility of inefficient thermal Jeans fragmentation.

The studies above all focused on data using the SMA's subcompact (i.e., the most compact) array configuration and only used a subsample of all the data. In this paper we release all the MASSES subcompact 1.3\,mm data. The typical resolution of an observation is about 4$\arcsec$, or $\sim$1000\,au. For all protostellar objects in the sample, we release calibrated subcompact $uv$ visibility data and imaged data for the 1.3\,mm continuum and the spectral lines \coto, \ttco, \ceo, and \ntdp.

We describe the survey and data release in detail in this paper. In Section~\ref{observations} we describe the target selection and observations of the MASSES survey. In Section~\ref{data}, we describe the data calibration and imaging techniques. In Section~\ref{deliverables}, we describe the data that is delivered to the user. Section~\ref{example} presents some example observations with brief discussion, and in Section~\ref{summary} we summarize the paper. The data are publicly available at \url{https://dataverse.harvard.edu/dataverse/MASSES}.

\startlongtable
\begin{deluxetable*}{cccccccccccccccccccc}
\tabletypesize{\scriptsize}
\tablecaption{Source and Observing Information \label{tab:sources}}
\tablehead{\colhead{Source} & \colhead{$T_{\text{bol}}$\tablenotemark{a}} & \colhead{Other Names\tablenotemark{b}}  & \colhead{RA\tablenotemark{c}} & \colhead{DEC\tablenotemark{c}} & \colhead{Track(s)} & \colhead{Missing} & \colhead{Correlator} \vspace{-8pt} \\
\colhead{Name} & (K) & & \colhead{(J2000)} & \colhead{(J2000)} & & \colhead{Antennas} &  \colhead{for Track}}
\startdata
Per-emb-1 & 27 $\pm$ 1 & HH211-MMS & 03:43:56.53 & 32:00:52.90 & 141207\_05:11:03 & 6 & ASIC \\
Per-emb-2 & 27 $\pm$ 1 & IRAS 03292+3039 & 03:32:17.95 & 30:49:47.60 & 141122\_03:05:36 & 6 & ASIC \\
Per-emb-3 & 32 $\pm$ 2 & ... & 03:29:00.52 & 31:12:00.70 & 151022\_10:48:26 & 5,7 & ASIC \\
Per-emb-4 & 31 $\pm$ 3 & ... & 03:28:39.10 & 31:06:01.80 & 151102\_04:48:11 & 7 & ASIC \\
Per-emb-5 & 32 $\pm$ 2 & IRAS 03282+3035 & 03:31:20.96 & 30:45:30.205 & 141122\_03:05:36 & 6 & ASIC \\
Per-emb-6 & 52 $\pm$ 3 & ... & 03:33:14.40 & 31:07:10.90 & 151203\_05:02:22 & none & ASIC \\
Per-emb-7 & 37 $\pm$ 4 & ... & 03:30:32.68 & 30:26:26.50 & 160925\_08:16:53 & 2 & SWARM \\
Per-emb-8 & 43 $\pm$ 6 & ... & 03:44:43.62 & 32:01:33.70 & 151123\_03:56:56 & none & ASIC \\
-- & -- & -- & -- & -- & 151130\_04:08:59 & none & ASIC \\
Per-emb-9 & 36 $\pm$ 2 & IRAS 03267+3128, Perseus5 & 03:29:51.82 & 31:39:06.10 & 151023\_11:04:02 & 5,7 & ASIC \\
-- & -- & -- & -- & -- & 151023\_14:42:17 & 5,7 & ASIC \\
-- & -- & -- & -- & -- & 151024\_11:25:32 & 7,8 & ASIC \\
Per-emb-10 & 30 $\pm$ 2 & ... & 03:33:16.45 & 31:06:52.50 & 151203\_05:02:22 & none & ASIC \\
Per-emb-11 & 30 $\pm$ 2 & IC348MMS & 03:43:56.85 & 32:03:04.60 & 141207\_05:11:03 & 6 & ASIC \\
Per-emb-12 & 29 $\pm$ 2 & NGC 1333 IRAS4A & 03:29:10.50 & 31:13:31.00 & 141123\_04:09:39 & 6,7,8 & ASIC \\
-- & -- & -- & -- & -- & 141123\_07:49:31 & 6,7,8 & ASIC \\
-- & -- & -- & -- & -- & 141213\_03:41:25 & 6 & ASIC \\
Per-emb-13 & 28 $\pm$ 1 & NGC 1333 IRAS4B & 03:29:12.04 & 31:13:01.50 & 141120\_03:58:22 & 6 & ASIC \\
Per-emb-14 & 31 $\pm$ 2 & NGC 1333 IRAS4C & 03:29:13.52 & 31:13:58.00 & 141123\_04:09:39 & 6,7,8 & ASIC \\
-- & -- & -- & -- & -- & 141123\_07:49:31 & 6,7,8 & ASIC \\
-- & -- & -- & -- & -- & 141213\_03:41:25 & 6 & ASIC \\
Per-emb-15 & 36 $\pm$ 4 & RNO15-FIR & 03:29:04.05 & 31:14:46.60 & 151023\_11:04:02 & 5,7 & ASIC \\
-- & -- & -- & -- & -- & 151023\_14:42:17 & 5,7 & ASIC \\
-- & -- & -- & -- & -- & 151024\_11:25:32 & 7,8 & ASIC \\
-- & -- & -- & -- & -- & 160925\_08:16:53 & 2 & SWARM \\
Per-emb-16 & 39 $\pm$ 2 & ... & 03:43:50.96 & 32:03:16.70 & 141207\_05:11:03 & 6 & ASIC \\
Per-emb-17 & 59 $\pm$ 11 & ... & 03:27:39.09 & 30:13:03.00& 151102\_04:48:11 & 7 & ASIC \\
Per-emb-18 & 59 $\pm$ 12 & NGC 1333 IRAS7 & 03:29:10.99 & 31:18:25.50 & 141127\_02:21:26 & 6 & ASIC \\
Per-emb-19 & 60 $\pm$ 3 & ... & 03:29:23.49 & 31:33:29.50 & 141214\_03:50:32 & 6 & ASIC \\
Per-emb-20 & 65 $\pm$ 3 & L1455-IRS4 & 03:27:43.23 & 30:12:28.80& 151108\_04:20:52 & none & ASIC \\
Per-emb-21 & 45 $\pm$ 12 & ... & \multicolumn{3}{c}{Imaged in the same field as Per-emb-18} \\
Per-emb-22 & 43 $\pm$ 2 & L1448-IRS2 & 03:25:22.33 & 30:45:14.00 & 141129\_03:04:09 & 6 & ASIC \\
Per-emb-23 & 42 $\pm$ 2 & ASR 30 & 03:29:17.16 & 31:27:46.40 & 151206\_04:31:17 & none & ASIC \\
Per-emb-24 & 67 $\pm$ 10 & ... & 03:28:45.30 & 31:05:42.00 & 151122\_11:23:42 & none & ASIC \\
-- & -- & -- & -- & -- & 151122\_12:21:59 & none & ASIC \\
-- & -- & -- & -- & -- & 151127\_04:06:10 & none & ASIC \\
Per-emb-25 & 61 $\pm$ 12 & ... & 03:26:37.46 & 30:15:28.00 & 151026\_05:33:00 & 7,8 & ASIC \\
Per-emb-26 & 47 $\pm$ 7 & L1448C, L1448-mm & 03:25:38.95 & 30:44:02.00 & 141118\_02:15:14 & 6 & ASIC \\
Per-emb-27 & 69 $\pm$ 1 & NGC 1333 IRAS2A & 03:28:55.56 & 31:14:36.60 & 141120\_03:58:22 & 6 & ASIC \\
Per-emb-28 & 45 $\pm$ 2 & ... & \multicolumn{3}{c}{Imaged in the same field as Per-emb-16} \\
Per-emb-29 & 48 $\pm$ 1 & B1-c & 03:33:17.85 & 31:09:32.00 & 141128\_03:49:43 & 6 & ASIC \\
Per-emb-30 & 78 $\pm$ 6 & ... & 03:33:27.28 & 31:07:10.20 & 160917\_08:50:40 & 2 & SWARM \\
-- & -- & -- & -- & -- & 160927\_08:02:56 & 2,3,6 & SWARM \\
-- & -- & -- & -- & -- & 170122\_03:03:39 & 3 & SWARM \\
-- & -- & -- & -- & -- & 170122\_14:18:47 & 3 & SWARM \\
Per-emb-31 & 80 $\pm$ 13  & ... & 03:28:32.55 & 31:11:05.20& 151108\_04:20:52 & none & ASIC \\
Per-emb-32 & 57 $\pm$ 10 & ... & 03:44:02.40 & 32:02:04.90 & 151123\_03:56:56 & none & ASIC \\
-- & -- & -- & -- & -- & 151130\_04:08:59 & none & ASIC \\
Per-emb-33 & 57 $\pm$ 3 & L1448IRS3B, L1448N & 03:25:36.48 & 30:45:22.30 & 141118\_02:15:14 & 6 & ASIC \\
Per-emb-34 & 99 $\pm$ 13 & IRAS 03271+3013 & 03:30:15.12 & 30:23:49.20& 160917\_08:50:40 & 2 & SWARM \\
-- & -- & -- & -- & -- & 160927\_08:02:56 & 2,3,6 & SWARM \\
-- & -- & -- & -- & -- & 170122\_03:03:39 & 3 & SWARM \\
-- & -- & -- & -- & -- & 170122\_14:18:47 & 3 & SWARM \\
Per-emb-35 & 103 $\pm$ 26 & NGC 1333 IRAS1 & 03:28:37.09 & 31:13:30.70 & 141213\_03:41:25 & 6 & ASIC \\
Per-emb-36 & 106 $\pm$ 12 & NGC 1333 IRAS2B & 03:28:57.36 & 31:14:15.70 & 151124\_03:10:17 & none & ASIC \\
-- & -- & -- & -- & -- & 151129\_04:06:02 & none & ASIC \\
Per-emb-37 & 22 $\pm$ 1 & ... & 03:29:18.27 & 31:23:20.00 & 151203\_05:02:22 & none & ASIC  \\
Per-emb-38 & 115 $\pm$ 21 & ... & 03:32:29.18 & 31:02:40.90 & 170121\_04:28:59 & 3 & SWARM \\
Per-emb-39 & 125 $\pm$ 47 & ... & 03:33:13.78 & 31:20:05.20 & 160917\_08:50:40 & 2 & SWARM \\
-- & -- & -- & -- & -- & 160927\_08:02:56 & 2,3,6 & SWARM \\
-- & -- & -- & -- & -- & 170122\_03:03:39 & 3 & SWARM \\
-- & -- & -- & -- & -- & 170122\_14:18:47 & 3 & SWARM \\
Per-emb-40 & 132 $\pm$ 25 & B1-a & 03:33:16.66 & 31:07:55.20 & 151205\_04:33:28 & none & ASIC \\
Per-emb-41 & 157 $\pm$ 72 & B1-b & 03:33:20.96 & 31:07:23.80 & 141128\_03:49:43 & 6 & ASIC \\
Per-emb-42 & 163 $\pm$ 51 & L1448C-S & \multicolumn{3}{c}{Imaged in the same field as Per-emb-26} \\
Per-emb-43 & 176 $\pm$ 42 & ... & 03:42:02.16 & 31:48:02.10 & 160925\_08:16:53 & 2 & SWARM \\
Per-emb-44 & 188 $\pm$ 9 & SVS~13A & 03:29:03.42 & 31:15:57.72 & 151019\_06:11:24\tablenotemark{d} & 7 & ASIC \\
-- & -- & -- & -- & -- & 170127\_03:29:33 & 3 & SWARM \\
Per-emb-45 & 197 $\pm$ 93 & ... & 03:33:09.57 & 31:05:31.20 & 151205\_04:33:28 & none & ASIC \\
Per-emb-46 & 221 $\pm$ 7 & ... & 03:28:00.40 & 30:08:01.30& 151108\_04:20:52 & none & ASIC \\
Per-emb-47 & 230 $\pm$ 17 & IRAS 03254+3050 & 03:28:34.50 & 31:00:51.10 & 151019\_06:11:24\tablenotemark{d} & 7 &  ASIC \\
-- & -- & -- & -- & -- & 170127\_03:29:33 & 3 & SWARM \\
Per-emb-48 & 238 $\pm$ 14 & L1455-FIR2 & 03:27:38.23 & 30:13:58.80 & 151026\_05:33:00 & 7,8 & ASIC \\
Per-emb-49 & 239 $\pm$ 68 & ... & 03:29:12.94 & 31:18:14.40& 141127\_02:21:26 & 6 & ASIC \\
Per-emb-50 & 128 $\pm$ 23 & ... & 03:29:07.76 & 31:21:57.20 & 141127\_02:21:26 & 6 & ASIC \\
Per-emb-51 & 263 $\pm$ 115 & ... & 03:28:34.53 & 31:07:05.50 & 151026\_05:33:00 & 7,8 & ASIC \\
Per-emb-52 & 278 $\pm$ 119 & ... & 03:28:39.72 & 31:17:31.90 & 151122\_11:23:42 & none & ASIC \\
-- & -- & -- & -- & -- & 151122\_12:21:59 & none & ASIC \\
-- & -- & -- & -- & -- & 151127\_04:06:10 & none & ASIC \\
Per-emb-53 & 287 $\pm$ 8 & B5-IRS1 & 03:47:41.56 & 32:51:43.90 & 141130\_04:04:23 & 6 & ASIC \\
Per-emb-54 & 131 $\pm$ 63 & NGC 1333 IRAS6 & 03:29:01.57 & 31:20:20.70 & 151022\_10:48:26 & 5,7 & ASIC \\
Per-emb-55 & 309 $\pm$ 64 & IRAS 03415+3152 & \multicolumn{3}{c}{Imaged in the same field as Per-emb-8} \\
Per-emb-56 & 312 $\pm$ 1 & IRAS 03439+3233 & 03:47:05.42 & 32:43:08.40 & 141130\_04:04:23 & 6 & ASIC \\
Per-emb-57 & 313 $\pm$ 200 & ... & 03:29:03.33 & 31:23:14.60 & 151206\_04:31:17 & none & ASIC \\
Per-emb-58 & 322 $\pm$ 88 & ... & 03:28:58.44 & 31:22:17.40 & 151124\_03:10:17 & none & ASIC \\
-- & -- & -- & -- & -- & 151129\_04:06:02 & none & ASIC \\
Per-emb-59 & 341 $\pm$ 179 & ... & 03:28:35.04 & 30:20:09.90 & 151102\_04:48:11 & 7 & ASIC \\
Per-emb-60 & 363 $\pm$ 240 & ... & 03:29:20.07 & 31:24:07.50 & 151206\_04:31:17 & none & ASIC \\
Per-emb-61 & 371 $\pm$ 107 & ... & 03:44:21.33 & 31:59:32.60 & 141130\_04:04:23 & 6 & ASIC \\
Per-emb-62 & 378 $\pm$ 29 & ... & 03:44:12.98 & 32:01:35.40 & 151123\_03:56:56 & none & ASIC \\
-- & -- & -- & -- & -- & 151130\_04:08:59 & none & ASIC \\
Per-emb-63 & 436 $\pm$ 9 & ... & 03:28:43.28 & 31:17:33.00 & 151122\_11:23:42 & none & ASIC \\
-- & -- & -- & -- & -- & 151122\_12:21:59 & none & ASIC \\
-- & -- & -- & -- & -- & 151127\_04:06:10 & none & ASIC \\
Per-emb-64 & 438 $\pm$ 8 & ... & 03:33:12.85 & 31:21:24.10 & 151205\_04:33:28 & none & ASIC \\
Per-emb-65 & 440 $\pm$ 191 & ... & 03:28:56.31 & 31:22:27.80 & 151124\_03:10:17 & none & ASIC \\
-- & -- & -- & -- & -- & 151129\_04:06:02 & none & ASIC \\
Per-emb-66 & 542 $\pm$ 110 & ... & 03:43:45.15 & 32:03:58.60 & 170121\_04:28:59 & 3 & SWARM  \\
B1bN & 14.7 $\pm$ 1.0 & ... & 03:33:21.19 & 31:07:40.60 & 141128\_03:49:43 & 6 & ASIC \\
B1bS & 17.7 $\pm$ 1.0 & ... &  \multicolumn{3}{c}{Imaged in the same field as Per-emb-41} \\ 
L1448IRS2E & 15 & ... & 03:25:25.66 & 30:44:56.70 & 141129\_03:04:09 & 6 & ASIC \\
L1451-MMS & 15 & ... & 03:25:10.21 & 30:23:55.30 & 141129\_03:04:09 & 6 & ASIC \\
Per-bolo-45 & 15 & ... & 03:29:07.70 & 31:17:16.80 & 141125\_04:39:14 & 6,7,8 & SWARM \\
-- & -- & -- & -- & -- & 170121\_04:28:59 & 3 & SWARM \\
Per-bolo-58 & 15 & ... & 03:29:25.46 & 31:28:15.00 & 141125\_04:39:14 & 6,7,8 & ASIC \\
-- & -- & -- & -- & -- & 141214\_03:50:32 & 6 & ASIC \\
SVS~13B & 20 $\pm$ 20 & ... & \multicolumn{3}{c}{Imaged in the same field as Per-emb-44} \\  
SVS~13C & 21 $\pm$ 1 & ... & 03:29:01.97 & 31:15:38.05 & 151019\_06:11:24\tablenotemark{d} & 7 & SWARM \\
-- & -- & -- & -- & -- & 170127\_03:29:33 & 3 & SWARM 
\enddata
\tablenotetext{a}{The $T_{\text{bol}}$ values were taken from \citet{Tobin2016}. Sources with no errors were not detected by \emph{Herschel}, and \citet{Tobin2016} gave these sources approximate temperatures of 15~K.}
\tablenotetext{b}{Other names were taken directly from \citet{Tobin2016} and are not a complete list of other names for the target.}
\tablenotetext{c}{RA and DEC are given for the phase center of the observations.}
\tablenotetext{d}{This track was missing the ASIC chunks for \coto, \ttco, and the upper sideband s13.}
\end{deluxetable*}

\section{Observations}\label{observations}
\subsection{Target Selection}
We wanted the targeted cloud to be nearby and have a large protostellar population so that one can statistically constrain protostellar evolution, but not so large of a sample that a survey is impractical for the SMA (e.g., Orion). The Perseus molecular cloud has over 70 protostellar objects, ranging from candidate first hydrostatic cores that have just formed central, hydrostatic objects, all the way to evolved Class~I systems near the end of the protostellar stage. For star-forming clouds within $\sim$350\,pc, Perseus (and possibly Aquila; distance to cloud is uncertain) is the only star-forming cloud with more than 40 protostellar objects \citep{Dunham2015}. At DEC~=~+31$^\circ$, Perseus is ideally located in the sky for maximum SMA visibility and can be targeted by most telescopes in the world. Aquila, on the other hand, has a declination near 0$^\circ$, which causes difficulty in attaining sufficient SMA $uv$ coverage to produce high fidelity maps with the SMA. As one of the best-studied sites of nearby star formation, copious complementary data is available for Perseus to aid with analysis, including single-dish imaging at mid-IR ($Spitzer$), far-IR ($Herschel$), and (sub)mm (James Clerk Maxwell Telescope, Caltech Submillimeter Observatory; JCMT, CSO) wavelengths \citep[e.g,][]{Hatchell2005,Jorgensen2006,Kirk2006,Enoch2006,Evans2009,Sadavoy2014,Dunham2015,MChen2016,Zari2016}. Finally, the VANDAM Perseus survey had already observed the same targets to reveal multiplicity down to a projected separation of 15\,au \citep{Tobin2016}. The synergy between connecting the physical and kinematic properties of the dense gas and dust revealed by the SMA and the multiplicity revealed by the VLA is one of the key strengths of this survey. 


From 2014 to 2017, we used the SMA to observe all known protostars in the Perseus molecular cloud. We targeted 74 protostellar systems (some `systems' are multiples not resolved by $Spitzer$). $Spitzer$ was used to identify 66 of these targets, and they were identified as Per-emb-1 through Per-emb-66 \citep{Enoch2009}. Eight additional systems that were not identified in the \citet{Enoch2009} $Spitzer$ survey were observed as well. These systems are B1-bN and B1bS \citep[e.g.,][]{Pezzuto2012}, L1448-IRS2e \citep[e.g.,][]{ChenX2010}, L1451-mm \citep[e.g.,][]{Pineda2011}, Per-Bolo-45 \citep[e.g.,][]{Schnee2012}, Per-Bolo-58 \citep[e.g.,][]{Dunham2011}, and SVS~13B and 13C \citep[e.g.,][]{ChenX2009}. Except SVS~13B and 13C, these systems are candidate first hydrostatic cores (see \citealt{Dunham2014a} for a brief discussion on first hydrostatic cores), though some of the aforementioned studies mention they could be Class~0 protostars. The candidate first cores were not identified by \citet{Enoch2009} because they were deeply embedded and/or had low luminosities. The SVS 13B/13C sources were not identified since they lie near the SVS 13A diffraction spike and thus failed the 24\,$\mu$m signal to noise criteria set out in \citet{Enoch2009}. The vast majority of protostars are expected to be identified by \citet{Enoch2009}, unless a large population of protostars with luminosities substantially below 0.1\,$L_\odot$ exists \citep{Dunham2008}. A future $Herschel$ catalog of protostellar sources would better constrain the completeness of the MASSES protostellar sample.

The angular separation between some of these 74 protostellar systems was small enough so that a single pointing could observe both systems simultaneously. We needed a total of 68 pointings to survey every system. The phase centers of each target are given in Table~\ref{tab:sources}. Accurate positions of the protostars themselves (which are typically within the SMA envelopes), along with their multiplicity (resolved to a projected separation of 15\,au) are given in \citet{Tobin2016}.

\renewcommand{\tabcolsep}{0.1cm}
\begin{deluxetable*}{lcccccccc}
\tablecolumns{3}
\tabletypesize{\scriptsize}
\tablewidth{0pt}
\tablecaption{Spectral Lines Covered by the MASSES Survey \label{tab:lines}}
\tablehead{\colhead{Tracer} & \colhead{Transition} & \colhead{Frequency} & \colhead{ASIC} & \colhead{ASIC Channels} & \colhead{$\Delta v_{uv,\rm{ASIC}}$\tablenotemark{a}} & \colhead{$\Delta v_{uv,\rm{SWARM}}$\tablenotemark{a}} & \colhead{$\Delta v_{img}$\tablenotemark{a}} & \colhead{Number of Imaged} \\
& & \colhead{(GHz)} & \colhead{Chunk} & \colhead{Per Chunk} & \colhead{(\kms)} & \colhead{(\kms)} & \colhead{(\kms)} & \colhead{Channels}
}
\startdata
1.3\,mm cont & & 231.29\tablenotemark{b} & LSB s05 -- s12, s14 & 64\tablenotemark{c} & & & & 1 \\
& & & USB s05 -- s12 \\
\cotont & \emph{J} = 2 -- 1 & 230.53796 & USB s13, s14\tablenotemark{d} & 512 & 0.26 & 0.18 & 0.5 & ~\,220/430\tablenotemark{e}\\
\ttcont & \emph{J} = 2 -- 1 & 220.39868 & LSB s13 & 512 & 0.28 & 0.19 & 0.3 & 200\\
\ceont & \emph{J} = 2 -- 1 & 219.56036 & LSB s23 & 1024 & 0.14 & 0.19 & 0.2 & 200\\
\ntdpnt & \emph{J} = 3 -- 2 & 231.32183 & USB s23 & 1024 & 0.13 & 0.18 & 0.2 & ~125
\smallskip \\ 
\hline  
850\,$\mu$m cont & & 356.72/356.410\tablenotemark{f}  & LSB, USB s05 -- s12 & 64\tablenotemark{c} \\
\cottnt & \emph{J} = 3 -- 2 & 345.79599 & LSB s18 & 512 \\
\hcopnt & \emph{J} = 4 -- 3 & 356.73424 & USB s18 & 1024 & \multicolumn{4}{c}{ Future data release (Stephens et al. in prep)}\\
\httcopnt & \emph{J} = 4 -- 3 & 346.99835 & LSB s04 & 1024
\enddata
\tablenotetext{a}{Velocity resolution $\Delta v_{uv}$ and $\Delta v_{img}$ is for the $uv$ data and imaged data, respectively.}
\tablenotetext{b}{Tuning frequencies for the 1.3\,mm SMA observations. One track, 160927\_08:02:56, had a different tuning frequency of 230.538\,GHz. ASIC and SWARM tracks have a total continuum bandwidth of 1.394\,GHz and $\sim$16\,GHz, respectively.}
\tablenotetext{c}{The channel width for the LSB s13 is 512 channels. The delivered calibrated $uv$ continuum data for all chunks is delivered as 1 channel.}
\tablenotetext{d}{The central velocity and the majority of \coto\ line is in the s14 chunk. The s13 chunk contains higher, positive velocities.}
\tablenotetext{e}{The first value is for ASIC, and the second is for SWARM. Seven ASIC maps had slightly less than 220 channels due to noise spikes in higher velocity channels. The SWARM cubes for Per-emb-44/SVS~13B and SVS~13C were mapped with 695 channels due to a high velocity CO outflow.}
\tablenotetext{f}{Tuning frequencies for the SMA observations.  The first value is for ASIC, and the second is for SWARM.}
\end{deluxetable*}

\subsection{Observations and Correlator Setup}\label{setup}
Observations for the MASSES survey (project code 2014A-S093; Co-PIs M. Dunham and I. Stephens; searchable in SMA archive via Dunham) were conducted using the SMA \citep{Ho2004}, which is an eight-element array of 6.1\,m antennas located on Mauna Kea. While the SMA has eight antennas, only seven antennas were typically available for these observations. For the MASSES survey, we made observations in both the subcompact (SUB) and extended (EXT) SMA array configurations. The baselines covered by the SUB and EXT configurations at 230\,GHz were approximately \mbox{4 -- 55\,k$\lambda$} and \mbox{20 -- 165\,k$\lambda$}, respectively, although these ranges varied if certain antennas were missing from the array. The focus of this data release paper is on the SUB data, and the combined SUB plus EXT data will be presented in a forthcoming paper.

While the MASSES project was being observed, the SMA upgraded its correlator from the Application Specific Integrated Circuit (ASIC) correlator to the SMA Wideband Astronomical ROACH2 Machine (SWARM) correlator \citep{Primiani2016}. The SUB observations were predominately done with the ASIC correlator. Twenty-eight ASIC SUB tracks and six SWARM SUB tracks had usable data. More information on each correlator will be discussed below.

The SMA can observe simultaneously with two receivers that can be tuned to different frequencies. The spectral setup and line rest frequencies are indicated in Table~\ref{tab:lines}. In the SUB configuration, we used the dual receiver mode to tune the SMA's two receivers to different frequencies. For the ASIC data, the local oscillators of the receivers were tuned to 231.29\,GHz and 356.720\,GHz. For the SWARM data, the receivers were tuned to 231.29\,GHz and 356.410\,GHz. For some tracks, the higher frequency 356\,GHz tuning is missing due to technical difficulties with the SMA. In this paper, we focus solely on the 231.29\,GHz subcompact data; the $\sim$356\,GHz data will be presented in a future data release paper.

The ASIC correlator in dual receiver mode has a total bandwidth of 2\,GHz for each sideband, and the center of each sideband is separated by 10\,GHz. Each 2\,GHz sideband is divided into 24~chunks, each of which has a bandwidth of 104\,MHz. These 104\,MHz chunks slightly overlap in frequency, making the ``effective" bandwidth of each chunk 82\,MHz.

In dual receiver mode, the ASIC correlator is divided into 6 blocks, each with 4 chunks. Each block is allowed up to 1024 channels, which can be distributed to each chunk by a power of 2 between 64 and 1024. Chunks are also allowed 0 channels. If the user specifies 1024 channels in one block of one receiver (e.g., the low frequency receiver), the other receiver (e.g., the high frequency receiver) can have no channels assigned to its block. To maximize spectral resolution for the chosen spectral lines, Blocks 1 and 5 (chunks s01 to s04 and s16 to s19, respectively) had 1024 channels for the high frequency receiver and Blocks 4 and 6 (chunks s13 to s16 and s20 to s24, respectively) had 1024 channels for the low frequency receiver. Blocks 2 and 3 (chunks s05 to s12) were used for continuum. Chunk s14 in the lower sideband was also used for the continuum because it did not contain any lines. Table~\ref{tab:lines} shows the ASIC chunk number(s) assigned to the continuum and for each spectral line, the amount of channels per chunk, and the velocity resolution of the $uv$ data. The continuum has 8 chunks (total bandwidth of 656\,MHz) in the upper sideband, and 9 chunks in the lower sideband (738\,MHz). Combining the sidebands together, the total continuum bandwidth for the ASIC correlator is 1.394\,GHz.



\begin{deluxetable}{lcccccccc}
\tablecolumns{3}
\tabletypesize{\scriptsize}
\tablewidth{0pt}
\tablecaption{Other SWARM Lines Detected Toward Some Fields\label{tab:swarmlines}}
\tablehead{\colhead{Tracer} & \colhead{Transition} & \colhead{Frequency} & \colhead{$E_u$} \\
& & \colhead{(GHz)} & \colhead{(K)}
}
\startdata
SO & $J_N = 5_5 -4_4$ & 215.22065 & 44.1 \\
DCO$^+$ & $J = 3 - 2$ & 216.11258 & 20.7 \\
DCN	& $J = 3 -2$ & 217.23854 & 20.9 \\
c-C$_3$H$_2$ & $6_{0,6}$-5$_{1,5}$ & 217.82215 & 38.6 \\
c-C$_3$H$_2$ & $6_{1,6}$-5$_{0,5}$ & 217.82215 & 38.6 \\
H$_2$CO & $J_{K_a,K_b} = 3_{0,3} - 2_{0,2}$ & 218.22219 & 21.0\\
H$_2$CO & $J_{K_a,K_b} = 3_{2,2} - 2_{2,1}$ & 218.47563 & 68.1\\
H$_2$CO & $J_{K_a,K_b} = 3_{2,1} - 2_{2,0}$ & 218.76007 & 68.1\\
SO & $J_N = 6_5 -5_4$ & 219.94944 & 35.0
\enddata
\tablecomments{Frequencies and upper energy levels are from Splatalogue (\url{http://www.splatalogue.net/}). While all these lines are certainly detected toward some sources, other lines may exist in the data. Images for these lines are not provided in this data release. Only the full SWARM visibilities are delivered.}
\end{deluxetable}

The SWARM correlator allows for the SMA to observe 8\,GHz for each sideband simultaneously at a uniform spectral resolution of 140\,kHz (0.18\,\kms\ at 233\,GHz) across the entire bandwidth. The center of each sideband is separated by 16\,GHz. Each SWARM sideband is divided into 4 different chunks that slightly overlap in frequency, with each chunk containing 16384 channels. Combining the two sidebands together, the SWARM correlator provides a total bandwidth of 16\,GHz, which allows for tracks using SWARM to reach much better sensitivities in the 1.3\,mm continuum than those for ASIC. SWARM's high spectral resolution across its entire bandwidth along with its additional frequency coverage increases the likelihood that additional spectral lines are detected. These spectral lines were identified by looking at the uv-averaged spectrum, but are not mapped in this paper. These identified lines detected toward some targets are listed in Table~\ref{tab:swarmlines}. For MASSES targets using the SWARM correlator, these lines were the strongest for the fields targeting Per-emb-15, Per-emb-44/SVS~13B, and SVS~13C, and very weak or undetected toward other fields. It is certainly possible that additional lines that are not listed in this table were also detected toward some targets.

The full SWARM $uv$ data includes these additional lines. The SWARM frequency coverage for the lower sideband (lsb) is approximately 214.5 -- 222.5\,GHz and for the upper sideband (usb) is approximately 230.5 -- 238.5\,GHz.

For some observations, both correlators were used simultaneously and were tuned to similar frequencies. Given that using multiple correlators does not increase signal-to-noise (i.e., they use the same receivers), we discarded the ASIC correlator observations in these instances. These tracks are considered SWARM tracks in Table~\ref{tab:sources}.

The names of the tracks, as defined in the SMA Archive, are given in the ``Track(s)'' column of Table~\ref{tab:sources}. The format of the names is YYMMDD\_STARTTIME, where YY is the year, MM is the month, DD is the day, and STARTTIME denotes the start time of the track. Tracks that were taken on the same day (i.e., have the same YYMMDD prefix) were combined together during the data reduction process. We also indicate in this table which antenna number(s) are missing from the track, where the eight SMA antennas are assigned numbers 1 through 8.

\startlongtable
\renewcommand{\tabcolsep}{0.08cm}
\begin{deluxetable*}{@{}l@{}cccccccccccccccccccccccccccccccccccccccccccc}
\tablecaption{MASSES Subcompact Sensitivities and Beam Sizes of Images \label{tab:sens}}
\tablehead{\colhead{\tiny Source} & \multicolumn{4}{c}{\tiny \underline{~~~~~~~1.3\,mm continuum~~~~~~~}} & \multicolumn{4}{c}{\tiny \underline{\coto\,(0.5\,\kms)\tablenotemark{a}}} & \multicolumn{4}{c}{\tiny \underline{\tiny high\,$v$\,CO\,(0.5\,\kms)\tablenotemark{a}}} & \multicolumn{4}{c}{\tiny \underline{$^{\tiny 13}$CO(2--1)\,(0.3\,\kms)\tablenotemark{a}}} & \multicolumn{4}{c}{\tiny \underline{C$^{\tiny 18}$O(2--1)\,(0.2\,\kms)\tablenotemark{a}}} & \multicolumn{4}{c}{\tiny \underline{N$_2$D$^+$(3--2)\,(0.2\,\kms)\tablenotemark{a}}} \vspace{-8pt} \\%
\colhead{\tiny Name(s)} & \colhead{\tiny $\sigma_{\rm{1.3\,mm}}$\tablenotemark{b}} & \colhead{\tiny $\theta_{maj}$} & \colhead{\tiny $\theta_{min}$} & \colhead{\tiny PA} & \colhead{\tiny $\sigma$} & \colhead{\tiny $\theta_{maj}$} & \colhead{\tiny $\theta_{min}$} & \colhead{\tiny PA} & \colhead{\tiny $\sigma$} & \colhead{\tiny $\theta_{maj}$} & \colhead{\tiny $\theta_{min}$} & \colhead{\tiny PA} & \colhead{\tiny $\sigma$} & \colhead{\tiny $\theta_{maj}$} & \colhead{\tiny $\theta_{min}$} & \colhead{\tiny PA} & \colhead{\tiny $\sigma$} & \colhead{\tiny $\theta_{maj}$} & \colhead{\tiny $\theta_{min}$} & \colhead{\tiny PA}  & \colhead{\tiny $\sigma$} & \colhead{\tiny $\theta_{maj}$} & \colhead{\tiny $\theta_{min}$} & \colhead{\tiny PA} \vspace{-8pt}\\
\colhead{\tiny } & \colhead{\tiny (mJy\,bm$^{-1}$)} & \colhead{\tiny ($\arcsec$)} & \colhead{\tiny ($\arcsec$)} & \colhead{\tiny ($^\circ$)} & \colhead{\tiny (K)} & \colhead{\tiny ($\arcsec$)} & \colhead{\tiny ($\arcsec$)} & \colhead{\tiny ($^\circ$)} & \colhead{\tiny (K)} & \colhead{\tiny ($\arcsec$)} & \colhead{\tiny ($\arcsec$)} & \colhead{\tiny ($^\circ$)} & \colhead{\tiny (K)} & \colhead{\tiny ($\arcsec$)} & \colhead{\tiny ($\arcsec$)} & \colhead{\tiny ($^\circ$)} & \colhead{\tiny (K)} & \colhead{\tiny ($\arcsec$)} & \colhead{\tiny ($\arcsec$)} & \colhead{\tiny ($^\circ$)} & \colhead{\tiny (K)} & \colhead{\tiny ($\arcsec$)} & \colhead{\tiny ($\arcsec$)} & \colhead{\tiny ($^\circ$)}
}
\startdata
\tiny Per-emb-1 & \tiny 5.0 & \tiny 4.3 & \tiny 3.3 & \tiny -12 & \tiny 0.14 & \tiny 4.3 & \tiny 3.3 & \tiny -14 & \tiny 0.13 & \tiny 4.3 & \tiny 3.3 & \tiny -15 & \tiny 0.16 & \tiny 4.4 & \tiny 3.3 & \tiny -12 & \tiny 0.25 & \tiny 4.4 & \tiny 3.3 & \tiny -12 & \tiny 0.28 & \tiny 4.3 & \tiny 3.2 & \tiny -13 \\
\tiny Per-emb-2 & \tiny 8.8 & \tiny 4.3 & \tiny 3.3 & \tiny -16 & \tiny 0.12 & \tiny 4.3 & \tiny 3.4 & \tiny -19 & \tiny 0.13 & \tiny 4.3 & \tiny 3.4 & \tiny -19 & \tiny 0.14 & \tiny 4.3 & \tiny 3.4 & \tiny -15 & \tiny 0.24 & \tiny 4.1 & \tiny 3.4 & \tiny -5 & \tiny 0.24 & \tiny 4.3 & \tiny 3.3 & \tiny -19 \\
\tiny Per-emb-3 & \tiny 3.3 & \tiny 5.9 & \tiny 5.0 & \tiny 58 & \tiny 0.10 & \tiny 5.6 & \tiny 4.9 & \tiny 62 & \tiny 0.10 & \tiny 5.6 & \tiny 4.9 & \tiny 63 & \tiny 0.11 & \tiny 6.3 & \tiny 5.0 & \tiny 54 & \tiny 0.16 & \tiny 6.3 & \tiny 5.0 & \tiny 54 & \tiny 0.19 & \tiny 5.6 & \tiny 4.9 & \tiny 62 \\
\tiny Per-emb-4 & \tiny 2.0 & \tiny 5.1 & \tiny 2.9 & \tiny 41 & \tiny 0.17 & \tiny 5.0 & \tiny 3.0 & \tiny 41 & \tiny 0.16 & \tiny 5.0 & \tiny 3.0 & \tiny 41 & \tiny 0.18 & \tiny 5.3 & \tiny 3.0 & \tiny 41 & \tiny 0.26 & \tiny 5.3 & \tiny 3.0 & \tiny 41 & \tiny 0.30 & \tiny 5.0 & \tiny 3.0 & \tiny 41 \\
\tiny Per-emb-5 & \tiny 3.8 & \tiny 4.2 & \tiny 3.4 & \tiny -16 & \tiny 0.12 & \tiny 4.3 & \tiny 3.3 & \tiny -18 & \tiny 0.12 & \tiny 4.3 & \tiny 3.3 & \tiny -18 & \tiny 0.14 & \tiny 4.3 & \tiny 3.4 & \tiny -14 & \tiny 0.24 & \tiny 4.0 & \tiny 3.4 & \tiny -4 & \tiny 0.24 & \tiny 4.2 & \tiny 3.3 & \tiny -18 \\
\tiny Per-emb-6 & \tiny 1.5 & \tiny 3.9 & \tiny 3.6 & \tiny 75 & \tiny 0.11 & \tiny 3.9 & \tiny 3.7 & \tiny 73 & \tiny 0.12 & \tiny 3.9 & \tiny 3.7 & \tiny 73 & \tiny 0.13 & \tiny 4.0 & \tiny 3.8 & \tiny 75 & \tiny 0.23 & \tiny 4.0 & \tiny 3.8 & \tiny 75 & \tiny 0.25 & \tiny 3.9 & \tiny 3.6 & \tiny 71 \\
\tiny Per-emb-7 & \tiny 0.89 & \tiny 7.6 & \tiny 4.1 & \tiny 74 & \tiny 0.084 & \tiny 7.5 & \tiny 3.9 & \tiny 74 & \tiny ... & \tiny ... & \tiny ... & \tiny ... & \tiny 0.096 & \tiny 7.7 & \tiny 4.4 & \tiny 74 & \tiny 0.10 & \tiny 7.7 & \tiny 4.4 & \tiny 74 & \tiny 0.14 & \tiny 7.5 & \tiny 3.9 & \tiny 74 \\
\tiny Per-emb-8 & \tiny 2.3 & \tiny 4.0 & \tiny 3.8 & \tiny -76 & \tiny 0.14 & \tiny 4.0 & \tiny 3.8 & \tiny 84 & \tiny 0.14 & \tiny 4.0 & \tiny 3.8 & \tiny 83 & \tiny 0.16 & \tiny 4.0 & \tiny 3.9 & \tiny -86 & \tiny 0.23 & \tiny 4.1 & \tiny 3.9 & \tiny -76 & \tiny 0.27 & \tiny 4.0 & \tiny 3.8 & \tiny -77 \\
\tiny Per-emb-9 & \tiny 2.9 & \tiny 5.3 & \tiny 3.2 & \tiny -60 & \tiny 0.14 & \tiny 5.2 & \tiny 3.3 & \tiny -60 & \tiny 0.14 & \tiny 5.2 & \tiny 3.3 & \tiny -60 & \tiny 0.13 & \tiny 4.6 & \tiny 4.6 & \tiny -75 & \tiny 0.19 & \tiny 4.7 & \tiny 4.6 & \tiny 67 & \tiny 0.27 & \tiny 5.2 & \tiny 3.3 & \tiny -60 \\
\tiny Per-emb-10 & \tiny 1.7 & \tiny 3.9 & \tiny 3.7 & \tiny 74 & \tiny 0.11 & \tiny 3.9 & \tiny 3.7 & \tiny 73 & \tiny 0.11 & \tiny 3.9 & \tiny 3.7 & \tiny 73 & \tiny 0.13 & \tiny 4.0 & \tiny 3.8 & \tiny 74 & \tiny 0.23 & \tiny 4.0 & \tiny 3.8 & \tiny 73 & \tiny 0.25 & \tiny 3.9 & \tiny 3.7 & \tiny 73 \\
\tiny Per-emb-11 & \tiny 6.6 & \tiny 4.3 & \tiny 3.2 & \tiny -13 & \tiny 0.13 & \tiny 4.3 & \tiny 3.2 & \tiny -15 & \tiny 0.14 & \tiny 4.3 & \tiny 3.2 & \tiny -16 & \tiny 0.16 & \tiny 4.4 & \tiny 3.3 & \tiny -13 & \tiny 0.25 & \tiny 4.4 & \tiny 3.3 & \tiny -13 & \tiny 0.28 & \tiny 4.3 & \tiny 3.2 & \tiny -15 \\
\tiny Per-emb-12 & \tiny 58 & \tiny 5.0 & \tiny 3.2 & \tiny -31 & \tiny 0.095 & \tiny 5.0 & \tiny 3.2 & \tiny -31 & \tiny 0.097 & \tiny 5.1 & \tiny 3.2 & \tiny -32 & \tiny 0.11 & \tiny 5.1 & \tiny 3.3 & \tiny -31 & \tiny 0.17 & \tiny 5.1 & \tiny 3.3 & \tiny -31 & \tiny 0.19 & \tiny 5.0 & \tiny 3.2 & \tiny -31 \\
\tiny Per-emb-13 & \tiny 24 & \tiny 4.1 & \tiny 3.3 & \tiny -11 & \tiny 0.13 & \tiny 4.1 & \tiny 3.2 & \tiny -14 & \tiny 0.14 & \tiny 3.9 & \tiny 3.2 & \tiny -8 & \tiny 0.16 & \tiny 3.9 & \tiny 3.3 & \tiny -1 & \tiny 0.23 & \tiny 4.2 & \tiny 3.3 & \tiny -10 & \tiny 0.25 & \tiny 4.1 & \tiny 3.2 & \tiny -14 \\
\tiny Per-emb-14 & \tiny 11 & \tiny 5.1 & \tiny 3.2 & \tiny -32 & \tiny 0.098 & \tiny 5.1 & \tiny 3.2 & \tiny -32 & \tiny 0.098 & \tiny 5.1 & \tiny 3.2 & \tiny -33 & \tiny 0.11 & \tiny 5.1 & \tiny 3.3 & \tiny -32 & \tiny 0.17 & \tiny 5.1 & \tiny 3.3 & \tiny -32 & \tiny 0.19 & \tiny 5.1 & \tiny 3.2 & \tiny -32 \\
\tiny Per-emb-15 & \tiny 2.3 & \tiny 7.3 & \tiny 4.2 & \tiny 73 & \tiny 0.09 & \tiny 6.3 & \tiny 3.8 & \tiny 71 & \tiny 0.14 & \tiny 5.1 & \tiny 3.5 & \tiny -61 & \tiny 0.078 & \tiny 6.2 & \tiny 4.5 & \tiny 67 & \tiny 0.10 & \tiny 6.2 & \tiny 4.5 & \tiny 67 & \tiny 0.13 & \tiny 5.8 & \tiny 4.1 & \tiny 71 \\
\tiny Per-emb-16 & \tiny 2.2 & \tiny 4.3 & \tiny 3.3 & \tiny -13 & \tiny 0.14 & \tiny 4.3 & \tiny 3.3 & \tiny -14 & \tiny 0.14 & \tiny 4.3 & \tiny 3.3 & \tiny -15 & \tiny 0.16 & \tiny 4.4 & \tiny 3.3 & \tiny -13 & \tiny 0.25 & \tiny 4.4 & \tiny 3.3 & \tiny -12 & \tiny 0.28 & \tiny 4.3 & \tiny 3.2 & \tiny -14 \\
\tiny Per-emb-17 & \tiny 2.4 & \tiny 5.2 & \tiny 2.9 & \tiny 40 & \tiny 0.16 & \tiny 5.1 & \tiny 2.9 & \tiny 40 & \tiny 0.16 & \tiny 5.0 & \tiny 2.9 & \tiny 40 & \tiny 0.17 & \tiny 5.4 & \tiny 3.0 & \tiny 40 & \tiny 0.25 & \tiny 5.4 & \tiny 3.0 & \tiny 40 & \tiny 0.29 & \tiny 5.1 & \tiny 2.9 & \tiny 40 \\
\tiny Per-emb-18 & \tiny 4.7 & \tiny 4.7 & \tiny 3.3 & \tiny -26 & \tiny 0.19 & \tiny 4.8 & \tiny 3.3 & \tiny -28 & \tiny 0.19 & \tiny 4.8 & \tiny 3.3 & \tiny -28 & \tiny 0.23 & \tiny 4.7 & \tiny 3.4 & \tiny -25 & \tiny 0.34 & \tiny 4.7 & \tiny 3.4 & \tiny -25 & \tiny 0.36 & \tiny 4.8 & \tiny 3.3 & \tiny -28 \\
\tiny Per-emb-19 & \tiny 1.8 & \tiny 4.2 & \tiny 3.3 & \tiny -9 & \tiny 0.14 & \tiny 4.2 & \tiny 3.3 & \tiny -12 & \tiny 0.14 & \tiny 4.2 & \tiny 3.3 & \tiny -13 & \tiny 0.16 & \tiny 4.2 & \tiny 3.4 & \tiny -8 & \tiny 0.24 & \tiny 4.2 & \tiny 3.4 & \tiny -8 & \tiny 0.28 & \tiny 4.2 & \tiny 3.3 & \tiny -12 \\
\tiny Per-emb-20 & \tiny 1.9 & \tiny 4.0 & \tiny 3.5 & \tiny 56 & \tiny 0.15 & \tiny 4.4 & \tiny 3.2 & \tiny 44 & \tiny 0.15 & \tiny 4.4 & \tiny 3.2 & \tiny 43 & \tiny 0.17 & \tiny 4.4 & \tiny 3.3 & \tiny 44 & \tiny 0.23 & \tiny 4.1 & \tiny 3.6 & \tiny 54 & \tiny 0.27 & \tiny 4.0 & \tiny 3.5 & \tiny 54 \\
\tiny Per-emb-21\tablenotemark{c} & \tiny 4.7 & \tiny 4.7 & \tiny 3.3 & \tiny -26 & \tiny 0.19 & \tiny 4.8 & \tiny 3.3 & \tiny -28 & \tiny 0.19 & \tiny 4.8 & \tiny 3.3 & \tiny -28 & \tiny 0.23 & \tiny 4.7 & \tiny 3.4 & \tiny -25 & \tiny 0.34 & \tiny 4.7 & \tiny 3.4 & \tiny -25 & \tiny 0.36 & \tiny 4.8 & \tiny 3.3 & \tiny -28 \\
\tiny Per-emb-22 & \tiny 5.9 & \tiny 4.2 & \tiny 3.1 & \tiny -24 & \tiny 0.22 & \tiny 4.2 & \tiny 3.1 & \tiny -26 & \tiny 0.22 & \tiny 4.2 & \tiny 3.1 & \tiny -26 & \tiny 0.26 & \tiny 4.2 & \tiny 3.1 & \tiny -24 & \tiny 0.39 & \tiny 4.2 & \tiny 3.1 & \tiny -24 & \tiny 0.42 & \tiny 4.2 & \tiny 3.0 & \tiny -26 \\
\tiny Per-emb-23 & \tiny 1.7 & \tiny 4.1 & \tiny 3.8 & \tiny 76 & \tiny 0.11 & \tiny 4.1 & \tiny 3.8 & \tiny 80 & \tiny 0.11 & \tiny 4.1 & \tiny 3.8 & \tiny 77 & \tiny 0.12 & \tiny 4.2 & \tiny 3.9 & \tiny 72 & \tiny 0.19 & \tiny 4.2 & \tiny 3.9 & \tiny 70 & \tiny 0.23 & \tiny 4.1 & \tiny 3.8 & \tiny 80 \\
\tiny Per-emb-24 & \tiny 2.5 & \tiny 4.3 & \tiny 3.7 & \tiny 46 & \tiny 0.18 & \tiny 4.3 & \tiny 3.7 & \tiny 44 & \tiny 0.19 & \tiny 4.3 & \tiny 3.7 & \tiny 45 & \tiny 0.21 & \tiny 4.4 & \tiny 3.8 & \tiny 43 & \tiny 0.30 & \tiny 4.4 & \tiny 3.8 & \tiny 44 & \tiny 0.37 & \tiny 4.2 & \tiny 3.7 & \tiny 46 \\
\tiny Per-emb-25 & \tiny 3.2 & \tiny 3.3 & \tiny 3.0 & \tiny 64 & \tiny 0.35 & \tiny 3.3 & \tiny 3.0 & \tiny 60 & \tiny 0.36 & \tiny 3.4 & \tiny 3.0 & \tiny 52 & \tiny 0.40 & \tiny 3.4 & \tiny 3.0 & \tiny 52 & \tiny 0.57 & \tiny 3.4 & \tiny 3.1 & \tiny 52 & \tiny 0.67 & \tiny 3.3 & \tiny 3.0 & \tiny 59 \\
\tiny Per-emb-26 & \tiny 4.5 & \tiny 4.0 & \tiny 3.2 & \tiny -10 & \tiny 0.16 & \tiny 4.0 & \tiny 3.2 & \tiny -11 & \tiny 0.14 & \tiny 4.1 & \tiny 3.2 & \tiny -13 & \tiny 0.17 & \tiny 4.1 & \tiny 3.3 & \tiny -10 & \tiny 0.25 & \tiny 4.1 & \tiny 3.3 & \tiny -10 & \tiny 0.28 & \tiny 4.0 & \tiny 3.2 & \tiny -11 \\
\tiny Per-emb-27 & \tiny 7.6 & \tiny 4.1 & \tiny 3.3 & \tiny -11 & \tiny 0.13 & \tiny 4.1 & \tiny 3.3 & \tiny -14 & \tiny 0.13 & \tiny 4.1 & \tiny 3.2 & \tiny -15 & \tiny 0.15 & \tiny 4.1 & \tiny 3.3 & \tiny -10 & \tiny 0.23 & \tiny 4.1 & \tiny 3.4 & \tiny -10 & \tiny 0.25 & \tiny 4.1 & \tiny 3.2 & \tiny -14 \\
\tiny Per-emb-28\tablenotemark{c} & \tiny 2.2 & \tiny 4.3 & \tiny 3.3 & \tiny -13 & \tiny 0.14 & \tiny 4.3 & \tiny 3.3 & \tiny -14 & \tiny 0.14 & \tiny 4.3 & \tiny 3.3 & \tiny -15 & \tiny 0.16 & \tiny 4.4 & \tiny 3.3 & \tiny -13 & \tiny 0.25 & \tiny 4.4 & \tiny 3.3 & \tiny -12 & \tiny 0.28 & \tiny 4.3 & \tiny 3.2 & \tiny -14 \\
\tiny Per-emb-29 & \tiny 5.6 & \tiny 4.2 & \tiny 3.0 & \tiny -17 & \tiny 0.19 & \tiny 4.2 & \tiny 3.0 & \tiny -18 & \tiny 0.19 & \tiny 4.3 & \tiny 3.0 & \tiny -20 & \tiny 0.22 & \tiny 4.3 & \tiny 3.1 & \tiny -16 & \tiny 0.33 & \tiny 4.3 & \tiny 3.1 & \tiny -16 & \tiny 0.38 & \tiny 4.2 & \tiny 3.0 & \tiny -19 \\
\tiny Per-emb-30 & \tiny 1.2 & \tiny 7.0 & \tiny 4.0 & \tiny 83 & \tiny 0.096 & \tiny 6.5 & \tiny 5.1 & \tiny -13 & \tiny ... & \tiny ... & \tiny ... & \tiny ... & \tiny 0.098 & \tiny 7.0 & \tiny 4.2 & \tiny 82 & \tiny 0.12 & \tiny 7.0 & \tiny 4.2 & \tiny 82 & \tiny 0.15 & \tiny 7.2 & \tiny 3.8 & \tiny 83 \\
\tiny Per-emb-31 & \tiny 1.8 & \tiny 4.0 & \tiny 3.5 & \tiny 53 & \tiny 0.14 & \tiny 4.4 & \tiny 3.2 & \tiny 43 & \tiny 0.15 & \tiny 4.4 & \tiny 3.2 & \tiny 43 & \tiny 0.17 & \tiny 4.4 & \tiny 3.3 & \tiny 43 & \tiny 0.23 & \tiny 4.1 & \tiny 3.6 & \tiny 51 & \tiny 0.28 & \tiny 4.0 & \tiny 3.5 & \tiny 52 \\
\tiny Per-emb-32 & \tiny 2.2 & \tiny 4.0 & \tiny 3.8 & \tiny -80 & \tiny 0.14 & \tiny 4.0 & \tiny 3.8 & \tiny 78 & \tiny 0.14 & \tiny 4.0 & \tiny 3.8 & \tiny 76 & \tiny 0.19 & \tiny 4.1 & \tiny 3.9 & \tiny -80 & \tiny 0.23 & \tiny 4.1 & \tiny 3.9 & \tiny -79 & \tiny 0.27 & \tiny 4.0 & \tiny 3.8 & \tiny -81 \\
\tiny Per-emb-33 & \tiny 13 & \tiny 4.0 & \tiny 3.2 & \tiny -10 & \tiny 0.15 & \tiny 4.0 & \tiny 3.2 & \tiny -11 & \tiny 0.14 & \tiny 4.0 & \tiny 3.2 & \tiny -13 & \tiny 0.17 & \tiny 4.1 & \tiny 3.2 & \tiny -10 & \tiny 0.25 & \tiny 4.1 & \tiny 3.3 & \tiny -10 & \tiny 0.28 & \tiny 4.0 & \tiny 3.2 & \tiny -11 \\
\tiny Per-emb-34 & \tiny 0.88 & \tiny 7.0 & \tiny 4.0 & \tiny 81 & \tiny 0.095 & \tiny 6.4 & \tiny 5.1 & \tiny -14 & \tiny ... & \tiny ... & \tiny ... & \tiny ... & \tiny 0.095 & \tiny 7.0 & \tiny 4.2 & \tiny 81 & \tiny 0.11 & \tiny 7.0 & \tiny 4.2 & \tiny 81 & \tiny 0.15 & \tiny 7.1 & \tiny 3.8 & \tiny 81 \\
\tiny Per-emb-35 & \tiny 2.9 & \tiny 4.0 & \tiny 3.6 & \tiny 11 & \tiny 0.15 & \tiny 4.0 & \tiny 3.6 & \tiny 7 & \tiny 0.15 & \tiny 4.0 & \tiny 3.6 & \tiny 5 & \tiny 0.17 & \tiny 4.1 & \tiny 3.6 & \tiny 12 & \tiny 0.27 & \tiny 4.1 & \tiny 3.6 & \tiny 13 & \tiny 0.29 & \tiny 4.0 & \tiny 3.6 & \tiny 7 \\
\tiny Per-emb-36 & \tiny 4.1 & \tiny 4.2 & \tiny 3.7 & \tiny 53 & \tiny 0.11 & \tiny 4.1 & \tiny 3.6 & \tiny 53 & \tiny 0.11 & \tiny 4.1 & \tiny 3.6 & \tiny 53 & \tiny 0.12 & \tiny 4.3 & \tiny 3.7 & \tiny 49 & \tiny 0.19 & \tiny 4.2 & \tiny 3.8 & \tiny 52 & \tiny 0.22 & \tiny 4.1 & \tiny 3.7 & \tiny 57 \\
\tiny Per-emb-37 & \tiny 1.7 & \tiny 3.9 & \tiny 3.6 & \tiny 65 & \tiny 0.12 & \tiny 4.0 & \tiny 3.6 & \tiny 62 & \tiny 0.12 & \tiny 4.0 & \tiny 3.6 & \tiny 62 & \tiny 0.13 & \tiny 4.0 & \tiny 3.7 & \tiny 63 & \tiny 0.24 & \tiny 4.0 & \tiny 3.8 & \tiny 63 & \tiny 0.26 & \tiny 3.9 & \tiny 3.6 & \tiny 60 \\
\tiny Per-emb-38 & \tiny 1.2 & \tiny 6.7 & \tiny 4.5 & \tiny -14 & \tiny 0.09 & \tiny 6.6 & \tiny 4.8 & \tiny -16 & \tiny ... & \tiny ... & \tiny ... & \tiny ... & \tiny 0.13 & \tiny 6.9 & \tiny 4.4 & \tiny -12 & \tiny 0.13 & \tiny 6.9 & \tiny 4.4 & \tiny -12 & \tiny 0.19 & \tiny 6.6 & \tiny 4.8 & \tiny -17 \\
\tiny Per-emb-39 & \tiny 1.1 & \tiny 7.2 & \tiny 4.0 & \tiny 83 & \tiny 0.097 & \tiny 6.4 & \tiny 5.1 & \tiny -15 & \tiny ... & \tiny ... & \tiny ... & \tiny ... & \tiny 0.10 & \tiny 7.2 & \tiny 4.1 & \tiny 83 & \tiny 0.13 & \tiny 7.2 & \tiny 4.1 & \tiny 83 & \tiny 0.15 & \tiny 7.3 & \tiny 3.8 & \tiny 83 \\
\tiny Per-emb-40 & \tiny 2.1 & \tiny 4.2 & \tiny 3.8 & \tiny 54 & \tiny 0.15 & \tiny 4.2 & \tiny 3.9 & \tiny 57 & \tiny 0.15 & \tiny 4.2 & \tiny 3.9 & \tiny 57 & \tiny 0.16 & \tiny 4.4 & \tiny 3.9 & \tiny 51 & \tiny 0.25 & \tiny 4.4 & \tiny 4.0 & \tiny 51 & \tiny 0.31 & \tiny 4.2 & \tiny 3.8 & \tiny 57 \\
\tiny Per-emb-41 & \tiny 5.4 & \tiny 4.2 & \tiny 3.0 & \tiny -16 & \tiny 0.18 & \tiny 4.2 & \tiny 3.0 & \tiny -18 & \tiny 0.19 & \tiny 4.3 & \tiny 3.0 & \tiny -19 & \tiny 0.22 & \tiny 4.3 & \tiny 3.1 & \tiny -16 & \tiny 0.33 & \tiny 4.3 & \tiny 3.1 & \tiny -16 & \tiny 0.38 & \tiny 4.2 & \tiny 3.0 & \tiny -18 \\
\tiny Per-emb-42\tablenotemark{c} & \tiny 4.5 & \tiny 4.0 & \tiny 3.2 & \tiny -10 & \tiny 0.16 & \tiny 4.0 & \tiny 3.2 & \tiny -11 & \tiny 0.14 & \tiny 4.1 & \tiny 3.2 & \tiny -13 & \tiny 0.17 & \tiny 4.1 & \tiny 3.3 & \tiny -10 & \tiny 0.25 & \tiny 4.1 & \tiny 3.3 & \tiny -10 & \tiny 0.28 & \tiny 4.0 & \tiny 3.2 & \tiny -11 \\
\tiny Per-emb-43 & \tiny 0.75 & \tiny 7.6 & \tiny 4.1 & \tiny 74 & \tiny 0.083 & \tiny 7.5 & \tiny 3.9 & \tiny 74 & \tiny ... & \tiny ... & \tiny ... & \tiny ... & \tiny 0.098 & \tiny 7.7 & \tiny 4.4 & \tiny 74 & \tiny 0.10 & \tiny 7.7 & \tiny 4.4 & \tiny 74 & \tiny 0.14 & \tiny 7.5 & \tiny 3.9 & \tiny 74 \\
\tiny Per-emb-44 & \tiny 16 & \tiny 6.2 & \tiny 5.4 & \tiny -12 & \tiny 0.068 & \tiny 6.5 & \tiny 5.8 & \tiny -31 & \tiny ... & \tiny ... & \tiny ... & \tiny ... & \tiny 0.099 & \tiny 6.8 & \tiny 5.5 & \tiny -14 & \tiny 0.14 & \tiny 5.4 & \tiny 4.0 & \tiny 31 & \tiny 0.16 & \tiny 5.2 & \tiny 3.9 & \tiny 35 \\
\tiny Per-emb-45 & \tiny 2.0 & \tiny 4.3 & \tiny 3.8 & \tiny 51 & \tiny 0.14 & \tiny 4.2 & \tiny 3.8 & \tiny 53 & \tiny 0.15 & \tiny 4.2 & \tiny 3.8 & \tiny 53 & \tiny 0.17 & \tiny 4.4 & \tiny 3.9 & \tiny 49 & \tiny 0.26 & \tiny 4.4 & \tiny 3.9 & \tiny 49 & \tiny 0.32 & \tiny 4.2 & \tiny 3.8 & \tiny 54 \\
\tiny Per-emb-46 & \tiny 1.8 & \tiny 4.0 & \tiny 3.5 & \tiny 51 & \tiny 0.15 & \tiny 4.4 & \tiny 3.2 & \tiny 43 & \tiny 0.15 & \tiny 4.4 & \tiny 3.2 & \tiny 42 & \tiny 0.17 & \tiny 4.4 & \tiny 3.3 & \tiny 42 & \tiny 0.24 & \tiny 4.1 & \tiny 3.6 & \tiny 50 & \tiny 0.28 & \tiny 4.0 & \tiny 3.5 & \tiny 51 \\
\tiny Per-emb-47 & \tiny 0.79 & \tiny 6.2 & \tiny 5.5 & \tiny -15 & \tiny 0.067 & \tiny 6.5 & \tiny 5.8 & \tiny -34 & \tiny ... & \tiny ... & \tiny ... & \tiny ... & \tiny 0.10 & \tiny 6.7 & \tiny 5.6 & \tiny -16 & \tiny 0.14 & \tiny 5.5 & \tiny 4.0 & \tiny 31 & \tiny 0.17 & \tiny 5.3 & \tiny 3.9 & \tiny 35 \\
\tiny Per-emb-48 & \tiny 2.8 & \tiny 3.3 & \tiny 3.0 & \tiny 60 & \tiny 0.33 & \tiny 3.3 & \tiny 3.0 & \tiny 77 & \tiny 0.36 & \tiny 3.5 & \tiny 3.0 & \tiny 50 & \tiny 0.41 & \tiny 3.3 & \tiny 3.1 & \tiny 74 & \tiny 0.57 & \tiny 3.5 & \tiny 3.0 & \tiny 48 & \tiny 0.69 & \tiny 3.4 & \tiny 3.0 & \tiny 56 \\
\tiny Per-emb-49 & \tiny 4.4 & \tiny 4.6 & \tiny 3.3 & \tiny -27 & \tiny 0.19 & \tiny 4.7 & \tiny 3.3 & \tiny -28 & \tiny 0.20 & \tiny 4.8 & \tiny 3.3 & \tiny -29 & \tiny 0.22 & \tiny 4.6 & \tiny 3.4 & \tiny -26 & \tiny 0.34 & \tiny 4.6 & \tiny 3.4 & \tiny -25 & \tiny 0.37 & \tiny 4.7 & \tiny 3.3 & \tiny -28 \\
\tiny Per-emb-50 & \tiny 2.9 & \tiny 4.6 & \tiny 3.3 & \tiny -26 & \tiny 0.19 & \tiny 4.8 & \tiny 3.3 & \tiny -28 & \tiny 0.20 & \tiny 4.8 & \tiny 3.3 & \tiny -28 & \tiny 0.22 & \tiny 4.6 & \tiny 3.4 & \tiny -25 & \tiny 0.34 & \tiny 4.7 & \tiny 3.4 & \tiny -25 & \tiny 0.36 & \tiny 4.7 & \tiny 3.3 & \tiny -28 \\
\tiny Per-emb-51 & \tiny 3.4 & \tiny 3.3 & \tiny 3.0 & \tiny 62 & \tiny 0.36 & \tiny 3.3 & \tiny 3.0 & \tiny 77 & \tiny 0.36 & \tiny 3.4 & \tiny 3.0 & \tiny 52 & \tiny 0.41 & \tiny 3.3 & \tiny 3.0 & \tiny 75 & \tiny 0.60 & \tiny 3.4 & \tiny 3.0 & \tiny 51 & \tiny 0.69 & \tiny 3.3 & \tiny 3.0 & \tiny 57 \\
\tiny Per-emb-52 & \tiny 3.0 & \tiny 4.3 & \tiny 3.7 & \tiny 45 & \tiny 0.18 & \tiny 4.3 & \tiny 3.7 & \tiny 44 & \tiny 0.19 & \tiny 4.3 & \tiny 3.7 & \tiny 44 & \tiny 0.21 & \tiny 4.4 & \tiny 3.8 & \tiny 43 & \tiny 0.30 & \tiny 4.4 & \tiny 3.8 & \tiny 44 & \tiny 0.36 & \tiny 4.3 & \tiny 3.7 & \tiny 45 \\
\tiny Per-emb-53 & \tiny 2.4 & \tiny 4.3 & \tiny 3.5 & \tiny -8 & \tiny 0.18 & \tiny 4.3 & \tiny 3.4 & \tiny -10 & \tiny 0.18 & \tiny 4.3 & \tiny 3.4 & \tiny -15 & \tiny 0.22 & \tiny 4.4 & \tiny 3.5 & \tiny -7 & \tiny 0.31 & \tiny 4.4 & \tiny 3.5 & \tiny -7 & \tiny 0.34 & \tiny 4.3 & \tiny 3.4 & \tiny -10 \\
\tiny Per-emb-54 & \tiny 8.1 & \tiny 5.9 & \tiny 5.0 & \tiny 59 & \tiny 0.099 & \tiny 5.7 & \tiny 5.0 & \tiny 65 & \tiny 0.10 & \tiny 5.7 & \tiny 4.9 & \tiny 65 & \tiny 0.11 & \tiny 6.3 & \tiny 5.0 & \tiny 55 & \tiny 0.15 & \tiny 6.3 & \tiny 5.1 & \tiny 55 & \tiny 0.19 & \tiny 5.7 & \tiny 4.9 & \tiny 64 \\
\tiny Per-emb-55\tablenotemark{c} & \tiny 2.3 & \tiny 4.0 & \tiny 3.8 & \tiny -76 & \tiny 0.14 & \tiny 4.0 & \tiny 3.8 & \tiny 84 & \tiny 0.14 & \tiny 4.0 & \tiny 3.8 & \tiny 83 & \tiny 0.16 & \tiny 4.0 & \tiny 3.9 & \tiny -86 & \tiny 0.23 & \tiny 4.1 & \tiny 3.9 & \tiny -76 & \tiny 0.27 & \tiny 4.0 & \tiny 3.8 & \tiny -77 \\
\tiny Per-emb-56 & \tiny 2.4 & \tiny 4.3 & \tiny 3.4 & \tiny -9 & \tiny 0.19 & \tiny 4.3 & \tiny 3.4 & \tiny -10 & \tiny 0.19 & \tiny 4.3 & \tiny 3.4 & \tiny -16 & \tiny 0.22 & \tiny 4.4 & \tiny 3.5 & \tiny -8 & \tiny 0.32 & \tiny 4.4 & \tiny 3.5 & \tiny -8 & \tiny 0.34 & \tiny 4.3 & \tiny 3.4 & \tiny -10 \\
\tiny Per-emb-57 & \tiny 1.4 & \tiny 4.1 & \tiny 3.8 & \tiny 71 & \tiny 0.11 & \tiny 4.1 & \tiny 3.8 & \tiny 76 & \tiny 0.11 & \tiny 4.1 & \tiny 3.8 & \tiny 73 & \tiny 0.12 & \tiny 4.2 & \tiny 3.9 & \tiny 67 & \tiny 0.19 & \tiny 4.2 & \tiny 3.9 & \tiny 66 & \tiny 0.24 & \tiny 4.0 & \tiny 3.8 & \tiny 79 \\
\tiny Per-emb-58 & \tiny 1.6 & \tiny 4.2 & \tiny 3.6 & \tiny 50 & \tiny 0.12 & \tiny 4.2 & \tiny 3.6 & \tiny 49 & \tiny 0.12 & \tiny 4.2 & \tiny 3.6 & \tiny 49 & \tiny 0.13 & \tiny 4.3 & \tiny 3.7 & \tiny 47 & \tiny 0.19 & \tiny 4.3 & \tiny 3.7 & \tiny 49 & \tiny 0.22 & \tiny 4.1 & \tiny 3.6 & \tiny 52 \\
\tiny Per-emb-59 & \tiny 1.8 & \tiny 5.2 & \tiny 2.9 & \tiny 40 & \tiny 0.16 & \tiny 5.1 & \tiny 2.9 & \tiny 40 & \tiny 0.16 & \tiny 5.1 & \tiny 2.9 & \tiny 40 & \tiny 0.18 & \tiny 5.5 & \tiny 2.9 & \tiny 40 & \tiny 0.25 & \tiny 5.4 & \tiny 3.0 & \tiny 40 & \tiny 0.30 & \tiny 5.1 & \tiny 2.9 & \tiny 40 \\
\tiny Per-emb-60 & \tiny 1.5 & \tiny 4.1 & \tiny 3.8 & \tiny 67 & \tiny 0.11 & \tiny 4.0 & \tiny 3.8 & \tiny 71 & \tiny 0.11 & \tiny 4.0 & \tiny 3.8 & \tiny 68 & \tiny 0.12 & \tiny 4.2 & \tiny 3.9 & \tiny 61 & \tiny 0.19 & \tiny 4.2 & \tiny 3.9 & \tiny 61 & \tiny 0.24 & \tiny 4.0 & \tiny 3.8 & \tiny 71 \\
\tiny Per-emb-61 & \tiny 2.2 & \tiny 4.3 & \tiny 3.4 & \tiny -9 & \tiny 0.19 & \tiny 4.3 & \tiny 3.4 & \tiny -11 & \tiny 0.20 & \tiny 4.3 & \tiny 3.3 & \tiny -17 & \tiny 0.21 & \tiny 4.4 & \tiny 3.5 & \tiny -9 & \tiny 0.32 & \tiny 4.4 & \tiny 3.5 & \tiny -9 & \tiny 0.35 & \tiny 4.3 & \tiny 3.4 & \tiny -11 \\
\tiny Per-emb-62 & \tiny 1.9 & \tiny 4.0 & \tiny 3.8 & \tiny -87 & \tiny 0.14 & \tiny 4.0 & \tiny 3.8 & \tiny 71 & \tiny 0.14 & \tiny 4.0 & \tiny 3.8 & \tiny 70 & \tiny 0.16 & \tiny 4.0 & \tiny 3.9 & \tiny 71 & \tiny 0.23 & \tiny 4.1 & \tiny 3.9 & \tiny -85 & \tiny 0.27 & \tiny 4.0 & \tiny 3.8 & \tiny -83 \\
\tiny Per-emb-63 & \tiny 2.5 & \tiny 4.3 & \tiny 3.7 & \tiny 45 & \tiny 0.18 & \tiny 4.3 & \tiny 3.7 & \tiny 43 & \tiny 0.19 & \tiny 4.3 & \tiny 3.7 & \tiny 43 & \tiny 0.21 & \tiny 4.4 & \tiny 3.8 & \tiny 41 & \tiny 0.30 & \tiny 4.4 & \tiny 3.8 & \tiny 42 & \tiny 0.37 & \tiny 4.3 & \tiny 3.7 & \tiny 43 \\
\tiny Per-emb-64 & \tiny 2.0 & \tiny 4.3 & \tiny 3.8 & \tiny 50 & \tiny 0.15 & \tiny 4.2 & \tiny 3.8 & \tiny 52 & \tiny 0.15 & \tiny 4.2 & \tiny 3.8 & \tiny 52 & \tiny 0.17 & \tiny 4.4 & \tiny 3.9 & \tiny 48 & \tiny 0.25 & \tiny 4.4 & \tiny 3.9 & \tiny 47 & \tiny 0.32 & \tiny 4.2 & \tiny 3.8 & \tiny 52 \\
\tiny Per-emb-65 & \tiny 1.6 & \tiny 4.2 & \tiny 3.6 & \tiny 49 & \tiny 0.11 & \tiny 4.2 & \tiny 3.6 & \tiny 48 & \tiny 0.12 & \tiny 4.2 & \tiny 3.6 & \tiny 48 & \tiny 0.13 & \tiny 4.3 & \tiny 3.7 & \tiny 46 & \tiny 0.19 & \tiny 4.3 & \tiny 3.7 & \tiny 48 & \tiny 0.23 & \tiny 4.1 & \tiny 3.6 & \tiny 50 \\
\tiny Per-emb-66 & \tiny 0.90 & \tiny 6.8 & \tiny 4.5 & \tiny -13 & \tiny 0.091 & \tiny 6.6 & \tiny 4.7 & \tiny -16 & \tiny ... & \tiny ... & \tiny ... & \tiny ... & \tiny 0.14 & \tiny 6.9 & \tiny 4.3 & \tiny -11 & \tiny 0.13 & \tiny 6.9 & \tiny 4.4 & \tiny -11 & \tiny 0.19 & \tiny 6.6 & \tiny 4.7 & \tiny -16 \\
\tiny B1bN & \tiny 5.2 & \tiny 4.2 & \tiny 3.0 & \tiny -15 & \tiny 0.18 & \tiny 4.3 & \tiny 3.0 & \tiny -17 & \tiny 0.19 & \tiny 4.3 & \tiny 3.0 & \tiny -18 & \tiny 0.21 & \tiny 4.3 & \tiny 3.1 & \tiny -15 & \tiny 0.32 & \tiny 4.3 & \tiny 3.1 & \tiny -15 & \tiny 0.37 & \tiny 4.2 & \tiny 3.0 & \tiny -17 \\
\tiny B1bS\tablenotemark{c} & \tiny 5.4 & \tiny 4.2 & \tiny 3.0 & \tiny -16 & \tiny 0.18 & \tiny 4.2 & \tiny 3.0 & \tiny -18 & \tiny 0.19 & \tiny 4.3 & \tiny 3.0 & \tiny -19 & \tiny 0.22 & \tiny 4.3 & \tiny 3.1 & \tiny -16 & \tiny 0.33 & \tiny 4.3 & \tiny 3.1 & \tiny -16 & \tiny 0.38 & \tiny 4.2 & \tiny 3.0 & \tiny -18 \\
\tiny L1448IRS2E & \tiny 2.7 & \tiny 4.2 & \tiny 3.1 & \tiny -23 & \tiny 0.23 & \tiny 4.2 & \tiny 3.1 & \tiny -25 & \tiny 0.22 & \tiny 4.2 & \tiny 3.1 & \tiny -25 & \tiny 0.26 & \tiny 4.2 & \tiny 3.1 & \tiny -23 & \tiny 0.39 & \tiny 4.2 & \tiny 3.1 & \tiny -23 & \tiny 0.43 & \tiny 4.2 & \tiny 3.0 & \tiny -25 \\
\tiny L1451-MMS & \tiny 2.6 & \tiny 4.2 & \tiny 3.0 & \tiny -24 & \tiny 0.23 & \tiny 4.2 & \tiny 3.0 & \tiny -25 & \tiny 0.23 & \tiny 4.2 & \tiny 3.0 & \tiny -25 & \tiny 0.28 & \tiny 4.2 & \tiny 3.1 & \tiny -23 & \tiny 0.39 & \tiny 4.2 & \tiny 3.1 & \tiny -23 & \tiny 0.45 & \tiny 4.2 & \tiny 3.0 & \tiny -25 \\
\tiny Per-bolo-45 & \tiny 1.9 & \tiny 6.7 & \tiny 4.4 & \tiny -13 & \tiny 0.086 & \tiny 6.4 & \tiny 3.9 & \tiny -16 & \tiny 0.15 & \tiny 6.4 & \tiny 3.3 & \tiny -20 & \tiny 0.11 & \tiny 6.6 & \tiny 3.9 & \tiny -15 & \tiny 0.14 & \tiny 6.6 & \tiny 3.9 & \tiny -15 & \tiny 0.17 & \tiny 6.3 & \tiny 3.9 & \tiny -16 \\
\tiny Per-bolo-58 & \tiny 1.6 & \tiny 4.2 & \tiny 3.4 & \tiny -9 & \tiny 0.10 & \tiny 4.8 & \tiny 3.3 & \tiny -19 & \tiny 0.15 & \tiny 6.4 & \tiny 3.3 & \tiny -20 & \tiny 0.13 & \tiny 4.8 & \tiny 3.4 & \tiny -18 & \tiny 0.19 & \tiny 4.8 & \tiny 3.4 & \tiny -17 & \tiny 0.21 & \tiny 4.7 & \tiny 3.3 & \tiny -19 \\
\tiny SVS~13B & \tiny 16 & \tiny 6.2 & \tiny 5.4 & \tiny -12 & \tiny 0.068 & \tiny 6.5 & \tiny 5.8 & \tiny -31 & \tiny ... & \tiny ... & \tiny ... & \tiny ... & \tiny 0.099 & \tiny 6.8 & \tiny 5.5 & \tiny -14 & \tiny 0.14 & \tiny 5.4 & \tiny 4.0 & \tiny 31 & \tiny 0.16 & \tiny 5.2 & \tiny 3.9 & \tiny 35 \\
\tiny SVS~13C & \tiny 12 & \tiny 6.2 & \tiny 5.5 & \tiny -17 & \tiny 0.072 & \tiny 6.6 & \tiny 5.9 & \tiny -38 & \tiny ... & \tiny ... & \tiny ... & \tiny ... & \tiny 0.097 & \tiny 6.8 & \tiny 5.6 & \tiny -18 & \tiny 0.14 & \tiny 5.4 & \tiny 4.0 & \tiny 32 & \tiny 0.17 & \tiny 5.2 & \tiny 3.9 & \tiny 36 
\enddata
\tablenotetext{a}{The velocity in parentheses after the spectral line name indicates the velocity resolution of the imaged spectral line. The high $v$ CO label refers to the high velocity \coto\ emission probed by the upper sideband of ASIC chunk s13.} 
\tablenotetext{b}{Continuum sensitivities were frequently limited by dynamic range.} 
\tablenotetext{c}{This source was imaged simultaneously with another source, as indicated in Table~\ref{tab:sources}.} 
\end{deluxetable*}

\section{Data Processing}\label{data}
\subsection{Data Calibration}\label{calibration}
All MASSES data were calibrated using the MIR software package, partially following the general reduction process outlined in the MIR cookbook.\footnote{\url{https://www.cfa.harvard.edu/~cqi/mircook.html}}  We discuss our reduction process in detail here. 

With MIR, we first applied a baseline correction to tracks requiring such a correction, as noted in the Radio Telescope Data Center website.\footnote{\url{https://www.cfa.harvard.edu/rtdc/data/baseline/}} We then flagged irrelevant data, such as slew and pointing scans. We then applied a system temperature correction to the data. For bandpass calibration, we typically used 3C454.3 and/or 3C84. Sometimes the SMA operator observed additional bandpass calibrators in conjunction with these two, and if they were sufficiently bright and high quality, we often included them in the bandpass calibration. We first applied a phase only bandpass calibration, followed by an amplitude only bandpass calibration.

Next, we used flux calibrators to measure the fluxes of our gain calibrator, 3C84. We typically used Uranus as a flux calibrator, but when it was not available, we used Neptune, Callisto, or Ganymede. To maximize the accuracy of the flux calibration, we first tried to find where 3C84 elevation and integration time best matched that of the flux calibrator. We then used only these parts of the 3C84 data, and flux calibrated only on the shorter baselines ($\lesssim$10\,k$\lambda$). In doing so, we are only using the highest signal-to-noise data that is the most representative of the observing conditions of the flux calibrator. Using the Submillimeter Calibrator List,\footnote{\url{http://sma1.sma.hawaii.edu/callist/callist.html}} we checked the flux measured by the SMA for a nearby day, and found that our measured flux for 3C84 was similar (within $\sim$20\%) even though the source is quite variable.


After the flux was measured, we used the 3C84 flux to scale the fluxes for the MASSES targets during gain calibration. This was done by specifying the measured flux of 3C84 during MIR's \texttt{gain\_cal} task. We first performed phase-only gain calibration, followed by amplitude-only gain calibration.

\begin{figure*}[ht!]
\begin{center}
\includegraphics[width=1.9\columnwidth]{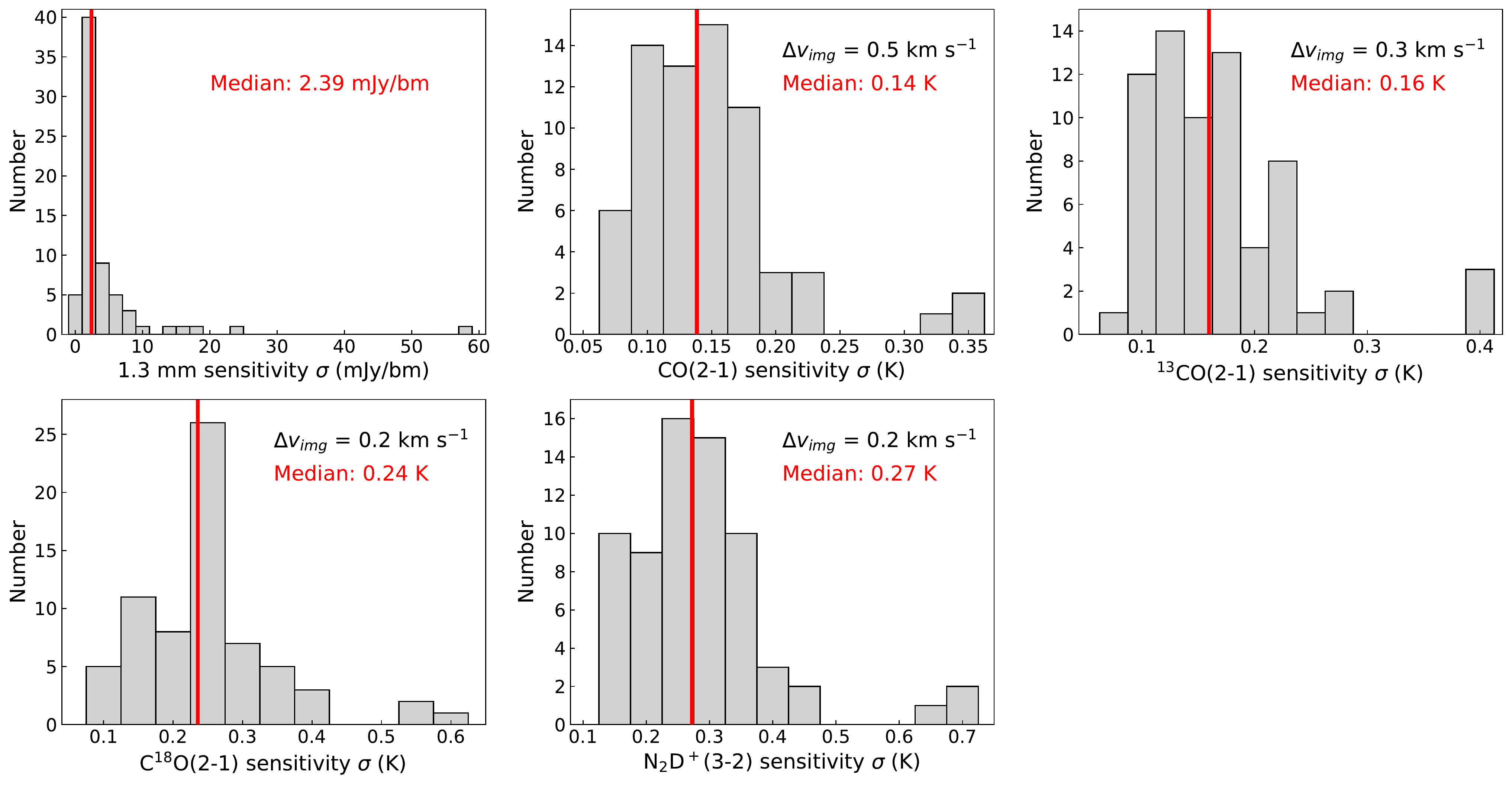}
\end{center}
\caption{Subcompact configuration sensitivities for the imaged continuum and spectral line data. The median of each distribution is marked with a vertical red line. The imaged velocity resolution ($\Delta v_{img}$) for each spectral line is indicated in the figure. Note that the 1.3\,mm continuum sensitivity is often limited by dynamic range, such as Per-emb-12 (NGC~1333 IRAS4A), which is the histogram bar at 58 mJy\,bm$^{-1}$.
}
\label{sensitivities} 
\end{figure*}

At this point, the calibration was complete, and the data was converted to MIRIAD format \citep{Sault1995}. For ASIC data, the high spectral resolution chunks containing \coto\ (including the upper sideband chunk s13), \ttco, \ceo, and \ntdp\ were exported with their full resolution, while the rest of the chunks were averaged together to generate the 1.3\,mm continuum. For SWARM, which has uniform spectral resolution across the entire spectrum, the data was exported in its entirety in MIRIAD. Lines were then split out of the bandwidth manually using MIRIAD, and the rest of the channels were averaged as continuum. In some cases, other high signal-to-noise spectral lines (besides the four mentioned above) were serendipitously detected with SWARM (Table~\ref{tab:swarmlines}), and thus were also removed from the continuum.

\startlongtable
\begin{deluxetable*}{l@{}c@{}c@{}c@{}c@{}c@{}c@{}c@{}c@{}c@{}c@{}c@{}c@{}c@{}c@{}c@{}c@{}c@{}c@{}c}
\tablecaption{Detection and Contour Information \label{tab:detection}}
\tablehead{ & & & & &  &  \colhead{\underline{1.3\,mm Contours\tablenotemark{b}}} & \multicolumn{2}{c}{\underline{\coto\ Blue Contours\tablenotemark{b}}} & \multicolumn{2}{c}{\underline{\coto\ Red Contours\tablenotemark{b}}} \vspace{-8pt} \\
& & & & & & \colhead{Contour Levels} & \colhead{Contour Levels} & \colhead{Velocity Range}  & \colhead{Contour Levels} & \colhead{Velocity Range} \vspace{-8pt} \\
\colhead{Source} & \multicolumn{5}{c}{\underline{~~Tracer Detected Toward Field?\tablenotemark{a}~~}} & \colhead{[start, step]} & \colhead{[start, step]} &\colhead{[$v_{\text{min}}$, $v_{\text{max}}$]}  & \colhead{[start, step]} & \colhead{[$v_{\text{min}}$, $v_{\text{max}}$]} \vspace{-8pt} \\
\colhead{Name} & \colhead{1.3\,mm} & \colhead{\cotont} & \colhead{\ttcont}  & \colhead{\ceont} & \colhead{\ntdpnt} & \colhead{(mJy\,bm$^{-1}$)} & \colhead{(Jy\,bm$^{-1}$\,\kms)} & \colhead{(\kms)} & \colhead{(Jy\,bm$^{-1}$\,\kms)}  & \colhead{(\kms)}    
} 
\startdata
Per-emb-1 & Y & Y & Y & Y & Y & [13, 20] & [3, 5] & [--13.5, 6.5] & [6.6, 10] & [10.5, 48.5]\\
Per-emb-2 & Y & Y & Y & Y & Y & [25, 50] & [1.4, 1.4] & [--1.6, 4.4] & [2, 2] & [8.9, 16.9]\\
Per-emb-3 & Y & Y & Y & Y & Y & [10, 10] & [3, 1] & [3.1, 3.6] & [2.5, 2.5] & [11.1, 19.1]\\
Per-emb-4 & N & Y & Y & Y & N & [5, 2] & [0.4, 0.2] & [2.5, 3.5] & [1, 1] & [11.5, 14.5]\\
Per-emb-5 & Y & Y & Y & Y & Y & [12, 24] & [2.6, 2.6] & [-11, 4] & [3.7, 3.7] & [10, 33]\\
Per-emb-6 & Y & Y & Y & Y & Y & [3.5, 2] & [3.1, 1.5] & [--3.35, 4.65] & [2.8, 2.5] & [9.65, 18.65]\\
Per-emb-7 & Y & Y & Y & Y & Y & [2.2, 1] & [0.5, 0.25] & [0, 2] & [0.65, 0.3] & [7, 9]\\
Per-emb-8 & Y & Y & Y & Y & Y & [6, 10] & [1.4, 0.7] & [4.65, 6.65] & [1.8, 1.8] & [12.65, 17.15]\\
Per-emb-9 & Y & Y & Y & Y & Y & [7.5, 5] & [0.9, 0.9] & [3.65, 6.15] & [1.3, 1.3] & [10.65, 14.65]\\
Per-emb-10 & Y & Y & Y & Y & Y & [4.5, 2] & [1.5, 1.5] & [--15.85, 5.65] & [2.2, 2] & [7.65, 37.15]\\
Per-emb-11 & Y & Y & Y & Y & Y & [18, 50] & [1.5, 2.2] & [--5.35, 5.65] & [5, 5] & [11.15, 25.15]\\
Per-emb-12 & Y & Y & Y & Y & Y & [200, 300] & [9, 20] & [--24.6, 3.9] & [12, 20] & [10.4, 46.4]\\
Per-emb-13 & Y & Y & Y & Y & Y & [80, 120] & [16, 10] & [--14.1, 4.9] & [5, 5] & [10.4, 25.4]\\
Per-emb-14 & Y & Y & Y & Y & Y & [25, 25] & [0.6, 0.2] & [1.5, 3] & [6.2, 2] & [13, 35.5]\\
Per-emb-15 & Y & Y & Y & Y & Y & [8, 4] & [8, 4] & [--18.35, 4.65] & [1.4, 1] & [11.15, 19.15]\\
Per-emb-16 & Y & Y & Y & Y & Y & [6, 4] & [1.3, 3.7] & [0.4, 6.4] & [1, 1.5] & [11.4, 14.9]\\ 
Per-emb-17 & Y & Y & Y & Y & Y & [7.5, 7.5] & [1.7, 1.7] & [--2.85, 0.65] & [2.3, 2] & [7.65, 9.15]\\
Per-emb-18 & Y & Y & Y & Y & Y & [15, 15] & [7, 7] & [--15.1, 1.4] & [14, 6] & [10.9, 24.4]\\
Per-emb-19 & Y & Y & Y & Y & Y & [5, 2] & [1.2, 1.2] & [3.65, 7.15] & [0.37, 0.37] & [10.15, 12.15]\\
Per-emb-20 & Y & Y & Y & Y & Y & [5, 2] & [3.1, 3.1] & [--5.35, 3.65] & [2.7, 2.7] & [7.65, 17.15]\\
Per-emb-21 & \multicolumn{10}{c}{Imaged in the same field as Per-emb-18} \\
Per-emb-22 & Y & Y & Y & Y & Y & [16, 20] & [8, 5] & [--32.9, --0.9] & [4.1, 5] & [8.1, 23.6]\\
Per-emb-23 & Y & Y & Y & Y & Y & [5, 2.5] & [2.1, 2.1] & [3.65, 6.65] & [0.6, 0.8] & [10.15, 12.65]\\
Per-emb-24 & M & Y & Y & Y & N & [5.5, 2] & [1.9, 1.9] & [0.65, 5.65] & [1.4, 1.4] & [11.65, 15.65]\\
Per-emb-25 & Y & Y & Y & Y & M & [10, 10] & [1.2, 1.2] & [0.65, 3.65] & [2, 1] & [6.15, 9.15]\\
Per-emb-26 & Y & Y & Y & Y & Y & [13, 30] & [13, 13] & [--59.6, 2.4] & [16, 16] & [7.4, 44.9]\\
Per-emb-27 & Y & Y & Y & Y & Y & [30, 50] & [5.4, 5.4] & [--19.5, 1.5] & [6, 6] & [11.5, 22.5]\\
Per-emb-28 & \multicolumn{10}{c}{Imaged in the same field as Per-emb-16} \\
Per-emb-29 & Y & Y & Y & Y & Y & [15, 20] & [7, 7] & [--12.2, 3.8] & [5, 5] & [10.3, 23.3]\\
Per-emb-30 & Y & Y & Y & Y & Y & [4, 6] & [3, 2] & [--2.5, 4] & [1.2, 1] & [11, 15.5]\\
Per-emb-31 & N & Y & Y & Y & N & [4, 2] & [1.7, 1.7] & [2.15, 6.15] & [0.5, 0.5] & [9.15, 17.15]\\
Per-emb-32 & M & Y & Y & Y & N & [6, 2] & [0.5, 0.5] & [3.65, 5.65] & [2.5, 2.5] & [10.15, 13.65]\\
Per-emb-33 & Y & Y & Y & Y & Y & [40, 100] & [14, 14] & [--45.1, 1.9] & [6.5, 6.5] & [7.4, 29.9]\\
Per-emb-34 & Y & Y & Y & Y & N & [2.2, 2] & [7.6, 7.6] & [--32.5, 5.5] & [4.8, 4.8] & [8, 38]\\
Per-emb-35 & Y & Y & Y & Y & N & [8, 8] & [2.2, 2.2] & [1.65, 6.15] & [4, 2] & [9.65, 15.65]\\
Per-emb-36 & Y & Y & Y & Y & N & [15, 15] & [6, 4] & [--10.35, 4.65] & [14, 8] & [10.65, 29.65]\\
Per-emb-37 & Y & Y & Y & Y & N & [4.6, 3] & [1.1, 0.5] & [2.65, 5.65] & [0.7, 0.5] & [11.15, 14.15]\\
Per-emb-38 & Y & Y & Y & Y & N & [3, 2] & [0.3, 0.2] & [1.5, 2.5] & [0.52, 0.52] & [8.5, 11.5]\\
Per-emb-39 & Y & Y & Y & M & Y & [3, 1] & [0.3, 0.1] & [2, 3] & [0.3, 0.1] & [12.5, 13.5]\\
Per-emb-40 & Y & Y & Y & Y & N & [6.5, 6.5] & [10, 20] & [--18.35, 4.65] & [1.3, 1.3] & [9.15, 14.15]\\
Per-emb-41 & Y & Y & Y & Y & Y & [20, 40] & [1.1, 1.1] & [--1.6, 4.4] & [3.5, 3.5] & [8.9, 12.9]\\
Per-emb-42 & \multicolumn{10}{c}{Imaged in the same field as Per-emb-26} \\
Per-emb-43 & N & Y & Y & Y & N & [2, 0.8] & [0.5, 0.3] & [3.5, 4.5] & [0.25, 0.1] & [11, 12]\\
Per-emb-44 & Y & Y & Y & Y & Y & [50, 70] & [30, 40] & [--153, 5.5] & [28, 28] & [11.5, 164]\\
Per-emb-45 & N & Y & Y & Y & N & [5, 2] & [0.3, 0.1] & [1.15, 2.15] & [0.67, 0.2] & [10.15, 11.15]\\
Per-emb-46 & Y & Y & Y & Y & N & [5.1, 1.3] & [1.2, 1.6] & [--0.35, 4.15] & [1, 0.5] & [6.15, 7.15]\\
Per-emb-47 & Y & Y & Y & Y & N & [2, 2] & [0.3, 0.08] & [--0.5, 0.5] & [1.3, 1.3] & [14.5, 17.5]\\
Per-emb-48 & M & Y & Y & Y & N & [6, 3] & [1, 0.5] & [--0.85, 2.65] & [0.4, 0.2] & [13.65, 14.65]\\
Per-emb-49 & Y & Y & Y & Y & Y & [15, 15] & [8, 8] & [--13.85, 5.15] & [5, 4] & [11.15, 20.15]\\
Per-emb-50 & Y & Y & Y & Y & N & [8, 16] & [5, 4] & [--0.7, 4.8] & [3, 2] & [11.3, 19.3]\\
Per-emb-51 & Y & Y & Y & M & Y & [7.5, 7.5] & [0.5, 0.1] & [1.65, 2.65] & [0.35, 0.1] & [11.65, 12.65]\\
Per-emb-52 & M & Y & Y & Y & Y & [6, 2] & [0.9, 0.9] & [6.15, 7.15] & [0.9, 0.9] & [10.15, 12.65]\\
Per-emb-53 & Y & Y & Y & Y & Y & [7, 5] & [4.5, 5] & [--18.8, 8.2] & [6, 8] & [11.7, 34.7]\\
Per-emb-54 & Y & Y & Y & Y & Y & [15, 15] & [5, 8] & [--12.85, 2.65] & [1.9, 1.9] & [14.15, 20.65]\\
Per-emb-55 & \multicolumn{10}{c}{Imaged in the same field as Per-emb-8} \\
Per-emb-56 & Y & Y & Y & Y & N & [7, 2] & [0.8, 0.6] & [2.65, 7.65] & [0.7, 0.6] & [13.15, 17.65]\\
Per-emb-57 & Y & Y & Y & Y & N & [4.5, 4.5] & [1.1, 1.1] & [2.65, 4.65] & [0.34, 0.15] & [13.15, 14.65]\\
Per-emb-58 & Y & Y & Y & Y & Y & [4.5, 2] & [0.23, 0.1] & [2.65, 3.65] & [1.3, 0.5] & [10.65, 11.15]\\
Per-emb-59 & N & Y & Y & M & N & [4, 1] & [0.2, 0.1] & [2.65, 3.65] & [0.2, 0.1] & [9.15, 10.15]\\
Per-emb-60 & M & Y & Y & Y & N & [3.3, 1.5] & [1, 1] & [2.15, 5.15] & [0.25, 0.2] & [12.15, 14.15]\\
Per-emb-61 & Y & Y & Y & Y & N & [6, 2] & [0.75, 0.5] & [5.65, 7.65] & [1.5, 1.5] & [11.15, 14.15]\\
Per-emb-62 & Y & Y & Y & Y & N & [5, 15] & [0.32, 0.32] & [5.15, 6.15] & [2, 6] & [10.65, 19.65]\\
Per-emb-63 & Y & Y & Y & Y & N & [7.5, 3] & [0.7, 0.5] & [0.65, 3.65] & [1, 1] & [11.15, 13.15]\\
Per-emb-64 & Y & Y & Y & Y & N & [5, 5] & [0.5, 0.3] & [3.65, 4.15] & [1.3, 0.5] & [11.15, 17.15]\\
Per-emb-65 & Y & Y & Y & Y & N & [3, 3] & [0.12, 0.12] & [3.15, 3.65] & [0.17, 0.17] & [11.65, 12.65]\\
Per-emb-66 & N & Y & Y & Y & N & [2, 1] & [0.7, 0.5] & [3, 5.5] & [1, 1] & [10, 12.5]\\
B1bN & Y & Y & Y & Y & Y  & [17, 40] & [1.7, 1.7] & [--1.6, 4.4] & [1.3, 0.5] & [8.9, 13.4]\\
B1bS & \multicolumn{10}{c}{Imaged in the same field as Per-emb-41} \\
L1448IRS2E & N & Y & Y & Y & M & [8, 4] & [0.4, 0.1] & [--3.9, --2.4] & [9, 9] & [6.6, 35.6]\\
L1451-MMS & Y & Y & Y & Y & Y & [8, 8] & [0.39, 0.39] & [3, 4] & [0.6, 0.3] & [5, 7]\\
Per-bolo-45 & Y & Y & Y & Y & Y & [5, 2] & [0.7, 0.2] & [2.65, 4.15] & [4.5, 4.5] & [11.15, 26.65]\\
Per-bolo-58 & Y & Y & Y & Y & Y & [4.5, 2] & [0.9, 0.2] & [3.8, 6.8] & [0.3, 0.2] & [10.8, 12.3]\\
SVS~13B & \multicolumn{10}{c}{Imaged in the same field as Per-emb-44} \\
SVS~13C & Y & Y & Y & Y & Y  & [35, 35] & [7.5, 7.5] & [108.5, 7] & [8, 8] & [11.5, 61]
\enddata
\tablenotetext{a}{Answers are: (Y)es, (N)o, or (M)arginal. Although \coto, \ttco, and \ceo\ are essentially detected toward every field, it does not mean that the line is associated with the protostar. Large-scale emission from the Perseus molecular cloud is frequently detected with the SMA even when emission is not associated with the protostar.}
\tablenotetext{b}{These contours are those shown in Figure~\ref{outflows}.}
\end{deluxetable*}

\subsection{Imaging}\label{imaging}
All targets of the MASSES survey were imaged with MIRIAD after exporting the calibrated data from MIR. For imaging of spectral lines, we first used the MIRIAD task \texttt{uvlin} to subtract a 0th order polynomial from the continuum. The delivered spectral line $uv$ data all have had their continuum subtracted. 

Dirty maps were then created via an inverse Fourier transform using the MIRIAD task \texttt{invert}. If a particular target was observed over multiple days (see Table~\ref{tab:sources}), the tracks were combined during this task. All targets were imaged using Briggs weighting with the robust parameter equal to~1. We also specified the pixel size to be 0$\farcs$8 with 100 pixels on each side of the map, resulting in 80$\arcsec$\,$\times$\,80$\arcsec$ maps. These maps image outside the full width at half maximum (FWHM) of the primary beam, which has a size of 48$\arcsec$ at 231\,GHz. For each line, we also specified the imaged channel width, which was slightly wider than the channel spectral resolution. The imaged spectral resolutions are specified in Table~\ref{tab:lines} under the $\Delta v_{img}$ column. The number of channels imaged are also listed in this table. For \coto\ data, which often have high velocity components, an additional ASIC chunk is available in the upper sideband (s13). This chunk provides for red-shifted velocities of \mbox{$\sim$26--160\,\kms}, but the user should note that for many targets, the noise in the chunk is extremely high at the edges. The s13 chunk has a slight overlap with the main \coto\ chunk, s14, which is imaged for velocities up to $\sim$48.5\,\kms. In the case that the \coto\ data was imaged with SWARM, we imaged channels for a larger velocity range (as indicated in Table~\ref{tab:lines}) than that which can done with a single ASIC chunk.

After creating dirty maps, we cleaned the maps using the MIRIAD task \texttt{clean}. For \coto\ and \ttco, which typically trace protostellar outflows, we used a three-step iterative cleaning algorithm. We first cleaned only the pixels and channels with emission to a level of 1.5 times the dirty map noise. This was completed by binning channels together with similar emission and making multiple regions with the task \texttt{cgcurs}. When selecting such pixels and channels, we typically excluded emission near the systemic velocity since the emission is often completely confused with the large scale emission of the Perseus molecular cloud. For the second step, we selected the area with \texttt{cgcurs} where emission is present over the entire cube, and cleaned this entire region for all channels to 2 times the dirty map noise. For the third and final step, we cleaned over all pixels and channels to 2.5 times the dirty map noise. We find that this three-step cleaning process recovers the extended emission well while minimizing the creation of interferometric imaging artifacts.

For the continuum, \ceo, and \ntdp, all of which typically trace protostellar envelopes, we used a two-step iterative cleaning algorithm. This algorithm is identical to that described above, except we skip the second step since the step had a negligible effect for imaging the compact emission that is usually found by these three tracers. For both the three-step and two-step cleaning processes, if no emission was obviously associated with a source, we only cleaned the channels to 3 times the dirty map noise. After deriving the clean components, we used the MIRIAD task \texttt{restor} to create clean maps.

The sensitivity (standard deviation) and synthesized beam parameters for the continuum and spectral lines for each map are listed in Table~\ref{tab:sens}. The sensitivity for the continuum was measured using the IDL task \texttt{sky}.\footnote{Available in the IDL Astronomy User's Library, \url{https://idlastro.gsfc.nasa.gov/contents.html}} This task measures the standard deviation over the entire map, clips all values deviating from 3 times this standard deviation, and iteratively repeats this for a total of 5 iterations. Sensitivities for the continuum were often limited by dynamic range rather than integration time. For some sources, self-calibration may help the continuum sensitivity, but we choose to not self-calibrate because we want to deliver consistent data products to the user. We note that sensitivity improvements using self-calibration with the SMA is typically minimal (i.e., less than $\sim$10\%), in part because the SMA is only an eight element array. Note that the $uv$ data is provided in case the user wants to apply the calibration. For the sensitivity of spectral line observations, we measured the standard deviation of the pixels over many emission-free channels and converted the standard deviation to a brightness temperature. The distribution of sensitivities are shown in Figure~\ref{sensitivities}. 

\begin{figure}[ht!]
\begin{center}
\includegraphics[width=1\columnwidth]{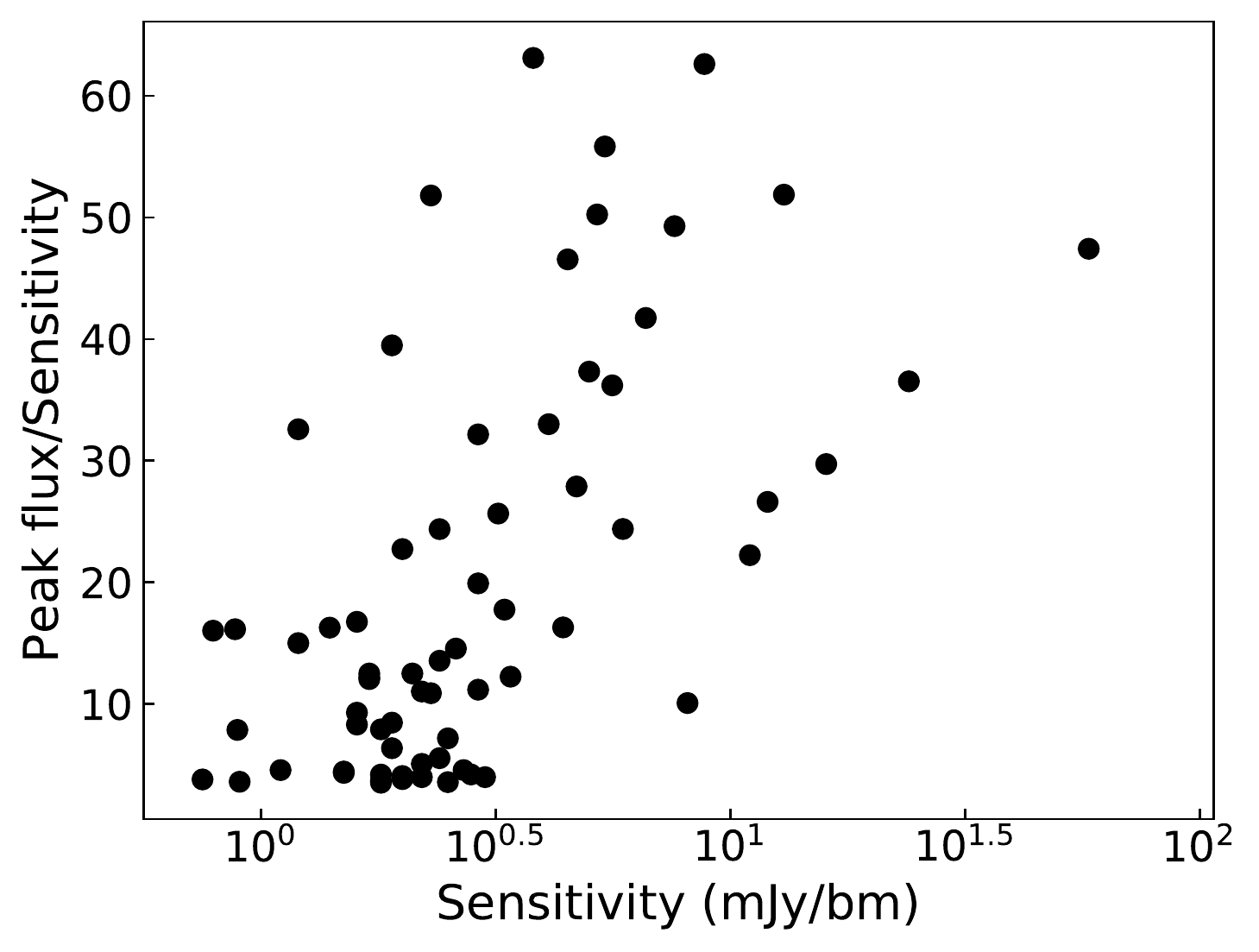}
\end{center}
\caption{Peak pixel flux of a 1.3\,mm continuum image divided by the sensitivity ($\sigma_{\rm{1.3\,mm}}$) of the image, versus the sensitivity of the image ($\sigma_{\rm{1.3\,mm}}$). The continuum sensitivity is frequently limited by dynamic range.
}
\label{dynamicrange} 
\end{figure}

The sensitivity of the continuum observations are frequently limited by dynamic range. Figure~\ref{dynamicrange} shows the peak 1.3\,mm pixel flux normalized by the sensitivity (a proxy for dynamic range) versus the sensitivity of the image. Targets that have higher dynamic range have worse sensitivities, even for good observing conditions, indicating that continuum sensitivities are often limited by dynamic range.

Table~\ref{tab:sens} also gives the FWHM of the synthesized beam's major and minor axes, $\theta_{maj}$, and $\theta_{min}$, and the position angle of the beam's major axis, PA, which is measured counterclockwise (east) from north.

\startlongtable
\begin{deluxetable*}{lccrrrrrrrrrr}
\tablecaption{Envelope masses, Dust Temperatures, and \ceo\ Fitting Information \label{tab:fits}}
\tablehead{ & \multicolumn{2}{c}{\underline{~~~~Envelope Position~~~~}} & \colhead{Envelope} & & \multicolumn{3}{c}{\underline{~~~~~~\ceo\ Spectra Fit\tablenotemark{c}~~~~~}} \vspace{-8pt} \\
\colhead{Source\tablenotemark{a}} & RA & DEC & \colhead{Mass\tablenotemark{b}} & \colhead{$T_{d,\,1000\,\text{au}}$} & Amplitude & $v_{\text{systemic}}$ & \colhead{$\Delta v$}  \vspace{-8pt} \\
\colhead{Name} & (J2000) & (J2000) & \colhead{($M_\odot$)} & (K) & (mJy\,bm$^{-1}$) & (\kms) & (\kms) 
} 
\startdata
Per-emb-1	&	03:43:56.77	&	+32:00:49.87	&	0.35	&	23	&	3.5	$\pm$	0.5	&	9.1	$\pm$	0.1	&	1.4	$\pm$	0.2	\\
Per-emb-2	&	03:32:17.92	&	+30:49:48.03	&	0.94	&	20	&	2.0	$\pm$	0.5	&	6.8	$\pm$	0.2	&	1.8	$\pm$	0.5	\\
Per-emb-3	&	03:29:00.55	&	+31:11:59.85	&	0.082	&	18	&	0.60	$\pm$	0.47	&	7.4	$\pm$	0.7	&	1.8	$\pm$	1.6	\\
Per-emb-4	&	...	&	...	&	$<$0.20		&	...	&	...	&	...	&	...	\\
Per-emb-5	&	03:31:20.93	&	+30:45:30.33	&	0.37	&	22	&	2.2	$\pm$	0.6	&	7.2	$\pm$	0.2	&	1.3	$\pm$	0.4	\\
Per-emb-6	&	...	&	...	&	$<$0.15		&	...	&	0.40	$\pm$	0.63	&	6.2	$\pm$	0.8	&	1.0	$\pm$	1.8	\\
Per-emb-7	&	03:30:32.54	&	+30:26:26.40	&	0.039	&	15	&	0.70	$\pm$	0.65	&	6.2	$\pm$	0.4	&	0.93	$\pm$	1.00	\\
Per-emb-8	&	03:44:43.98	&	+32:01:34.97	&	0.18	&	24	&	1.9	$\pm$	0.4	&	11.0	$\pm$	0.3	&	3.1	$\pm$	0.7	\\
Per-emb-9	&	03:29:51.88	&	+31:39:05.52	&	0.23	&	19	&	7.0	$\pm$	0.8	&	8.3	$\pm$	0.04	&	0.62	$\pm$	0.08	\\
Per-emb-10	&	03:33:16.41	&	+31:06:52.38	&	0.078	&	19	&	0.75	$\pm$	0.54	&	6.6	$\pm$	0.5	&	1.4	$\pm$	1.1	\\
-Per-emb-10-SMM	&	03:33:18.47	&	+31:06:33.63	&	0.025	&	19	&	...	&	...	&	...	\\						
Per-emb-11	&	03:43:57.06	&	+32:03:04.67	&	0.53	&	22	&	2.0	$\pm$	0.6	&	8.7	$\pm$	0.2	&	1.2	$\pm$	0.4	\\
 -IC 348 MMS2	&	03:43:57.74	&	+32:03:10.10	&	0.087	&	22	&	...	&	...	&	...	\\						
Per-emb-12	&	03:29:10.49	&	+31:13:31.37	&	3.3	&	29	&	7.4	$\pm$	0.6	&	6.7	$\pm$	0.05	&	1.2	$\pm$	0.1	\\
Per-emb-13	&	03:29:11.99	&	+31:13:08.14	&	1.1	&	26	&	3.3	$\pm$	0.6	&	6.8	$\pm$	0.1	&	1.1	$\pm$	0.2	\\
-IRAS 4B$^\prime$	&	03:29:12.83	&	+31:13:06.96	&	0.28	&	26	&	...	&	...	&	...	\\						
Per-emb-14	&	03:29:13.52	&	+31:13:57.75	&	0.16	&	19	&	1.4	$\pm$	0.5	&	7.7	$\pm$	0.3	&	1.8	$\pm$	0.7	\\
Per-emb-15	&	03:29:04.19	&	+31:14:48.43	&	0.12	&	18	&	1.2	$\pm$	0.5	&	6.5	$\pm$	0.3	&	1.6	$\pm$	0.8	\\
Per-emb-16	&	03:43:51.00	&	+32:03:23.86	&	0.14	&	18	&	1.7	$\pm$	0.6	&	8.5	$\pm$	0.2	&	1.1	$\pm$	0.5	\\
Per-emb-17	&	03:27:39.12	&	+30:13:02.53	&	0.10	&	26	&	1.6	$\pm$	0.5	&	5.8	$\pm$	0.3	&	1.8	$\pm$	0.6	\\
Per-emb-18	&	03:29:11.26	&	+31:18:31.33	&	0.21	&	25	&	3.7	$\pm$	0.5	&	8.1	$\pm$	0.1	&	1.6	$\pm$	0.2	\\
Per-emb-19	&	03:29:23.48	&	+31:33:28.94	&	0.022	&	17	&	1.7	$\pm$	0.6	&	7.5	$\pm$	0.2	&	1.0	$\pm$	0.4	\\
-Per-emb-19-SMM	&	03:29:24.33	&	+31:33:22.57	&	0.011	&	17	&	...	&	...	&	...	\\						
Per-emb-20	&	03:27:43.20	&	+30:12:28.96	&	0.061	&	22	&	2.2	$\pm$	0.6	&	5.1	$\pm$	0.1	&	0.98	$\pm$	0.33	\\
-Per-emb-20-SMM	&	03:27:42.78	&	+30:12:25.94	&	0.017	&	22	&	...	&	...	&	...	\\						
Per-emb-21	&	03:29:10.69	&	+31:18:20.15	&	0.19	&	25	&	1.7	$\pm$	0.5	&	8.7	$\pm$	0.2	&	1.7	$\pm$	0.6	\\
Per-emb-22	&	03:25:22.35	&	+30:45:13.21	&	0.37	&	26	&	3.6	$\pm$	0.6	&	3.9	$\pm$	0.1	&	1.3	$\pm$	0.2	\\
Per-emb-23	&	03:29:17.25	&	+31:27:46.34	&	0.098	&	20	&	3.5	$\pm$	0.6	&	7.6	$\pm$	0.09	&	1.1	$\pm$	0.2	\\
Per-emb-24	&	...	&	...	&	$<$0.25	&	...	&	0.99	$\pm$	0.61	&	7.6	$\pm$	0.3	&	1.1	$\pm$	0.8	\\
Per-emb-25	&	03:26:37.49	&	+30:15:27.90	&	0.10	&	21	&	1.5	$\pm$	0.7	&	5.5	$\pm$	0.2	&	0.75	$\pm$	0.43	\\
Per-emb-26	&	03:25:38.87	&	+30:44:05.30	&	0.37	&	30	&	3.1	$\pm$	0.5	&	5.1	$\pm$	0.1	&	1.4	$\pm$	0.3	\\
Per-emb-27	&	03:28:55.56	&	+31:14:37.17	&	0.47	&	34	&	3.1	$\pm$	0.5	&	7.8	$\pm$	0.1	&	1.9	$\pm$	0.3	\\
Per-emb-28	&	03:43:50.99	&	+32:03:07.97	&	0.086	&	18	&	0.65	$\pm$	0.59	&	8.4	$\pm$	0.5	&	1.1	$\pm$	1.2	\\
Per-emb-29	&	03:33:17.86	&	+31:09:32.31	&	0.43	&	26	&	3.5	$\pm$	0.6	&	6.1	$\pm$	0.09	&	1.1	$\pm$	0.2	\\
Per-emb-30	&	03:33:27.33	&	+31:07:10.29	&	0.074	&	23	&	2.2	$\pm$	0.5	&	7.1	$\pm$	0.2	&	1.6	$\pm$	0.4	\\
Per-emb-31	&	...	&	...	&	$<$0.18	&	...	&	0.43	$\pm$	0.42	&	7.2	$\pm$	1.1	&	2.3	$\pm$	2.6	\\
Per-emb-32	&	...	&	...	&	$<$0.22	&	...	&	1.2	$\pm$	0.6	&	9.4	$\pm$	0.3	&	1.0	$\pm$	0.6	\\
Per-emb-33	&	03:25:36.32	&	+30:45:14.77	&	0.82	&	29	&	2.0	$\pm$	0.4	&	4.9	$\pm$	0.2	&	2.1	$\pm$	0.5	\\
-L1448IRS3	&	03:25:35.68	&	+30:45:35.16	&	0.26	&	29	&	...	&	...	&	...	\\						
-L1448NW	&	03:25:36.46	&	+30:45:21.43	&	0.17	&	29	&	...	&	...	&	...	\\						
Per-emb-34	&	03:30:15.19	&	+30:23:49.11	&	0.024	&	23	&	1.4	$\pm$	0.5	&	6.1	$\pm$	0.3	&	1.4	$\pm$	0.6	\\
Per-emb-35	&	03:28:37.12	&	+31:13:31.24	&	0.097	&	30	&	3.4	$\pm$	0.6	&	7.2	$\pm$	0.1	&	1.3	$\pm$	0.2	\\
Per-emb-36	&	03:28:57.36	&	+31:14:15.61	&	0.19	&	27	&	2.2	$\pm$	0.4	&	7.0	$\pm$	0.2	&	2.2	$\pm$	0.5	\\
Per-emb-37	&	03:29:18.94	&	+31:23:13.11	&	0.082	&	18	&	...	&	...	&	...	\\						
Per-emb-38	&	03:32:29.22	&	+31:02:42.73	&	0.044	&	19	&	0.92	$\pm$	0.56	&	7.0	$\pm$	0.4	&	1.3	$\pm$	0.9	\\
Per-emb-39	&	...	&	...	&	?\tablenotemark{d}	&	...	&	...	&	...	&	...	\\
Per-emb-40	&	03:33:16.65	&	+31:07:54.81	&	0.028	&	22	&	2.0	$\pm$	0.4	&	7.1	$\pm$	0.2	&	2.0	$\pm$	0.5	\\
Per-emb-41	&	...	&	...	&	$<$0.54	&	...	&	...	&	...	&	...	\\
Per-emb-42	&	...	&	...	&	$<$0.45\tablenotemark{e}	&	...	&	1.4	$\pm$	0.5	&	5.5	$\pm$	0.3	&	1.4	$\pm$	0.6	\\
Per-emb-43	&	...	&	...	&	$<$0.075	&	...	&	...	&	...	&	...	\\
Per-emb-44	&	03:29:03.76	&	+31:16:03.43	&	0.38	&	38	&	4.4	$\pm$	0.4	&	8.4	$\pm$	0.1	&	2.0	$\pm$	0.2	\\
Per-emb-45	&	...	&	...	&	$<$0.20		&	...	&	...	&	...	&	...	\\
Per-emb-46	&	...	&	...	&	$<$0.18		&	...	&	0.38	$\pm$	0.58	&	5.1	$\pm$	0.9	&	1.2	$\pm$	2.1	\\
Per-emb-47	&	03:28:33.87	&	+31:00:52.49	&	0.016	&	22	&	1.1	$\pm$	0.6	&	7.4	$\pm$	0.3	&	1.0	$\pm$	0.7	\\
Per-emb-48	&	...	&	...	&	$<$0.28		&	...	&	...	&	...	&	...	\\
Per-emb-49	&	...	&	...	&	$<$0.44		&	...	&	...	&	...	&	...	\\
Per-emb-50	&	03:29:07.76	&	+31:21:57.16	&	0.062	&	35	&	1.4	$\pm$	0.6	&	7.4	$\pm$	0.3	&	1.3	$\pm$	0.6	\\
Per-emb-51	&	03:28:34.52	&	+31:07:05.47	&	0.25	&	13	&	0.52	$\pm$	0.52	&	6.7	$\pm$	0.7	&	1.5	$\pm$	1.7	\\
Per-emb-52	&	...	&	...	&	?\tablenotemark{d}		&	...	&	0.99	$\pm$	0.83	&	7.9	$\pm$	0.2	&	0.58	$\pm$	0.56	\\
Per-emb-53	&	03:47:41.58	&	+32:51:43.75	&	0.065	&	27	&	2.9	$\pm$	0.5	&	10.0	$\pm$	0.1	&	1.5	$\pm$	0.3	\\
Per-emb-54	&	03:29:02.83	&	+31:20:41.32	&	0.13	&	33	&	8.5	$\pm$	0.5	&	8.0	$\pm$	0.04	&	1.5	$\pm$	0.1	\\
Per-emb-55	&	...	&	...	&	$<$0.23\tablenotemark{e}		&	...	&	...	&	...	&	...	\\
Per-emb-56	&	03:47:05.42	&	+32:43:08.33	&	0.019	&	19	&	0.76	$\pm$	0.72	&	11.0	$\pm$	0.4	&	0.76	$\pm$	0.84	\\
Per-emb-57	&	03:29:03.32	&	+31:23:14.34	&	0.048	&	14	&	...	&	...	&	...	\\						
Per-emb-58	&	03:28:58.36	&	+31:22:16.81	&	0.010	&	19	&	2.8	$\pm$	0.9	&	8.0	$\pm$	0.08	&	0.46	$\pm$	0.18	\\
Per-emb-59	&	...	&	...	&	$<$0.18		&	...	&	...	&	...	&	...	\\
Per-emb-60	&	...	&	...	&	?\tablenotemark{d}	&	...	&	...	&	...	&	...	\\
Per-emb-61	&	03:44:21.30	&	+31:59:32.53	&	0.019	&	16	&	1.0	$\pm$	0.5	&	9.5	$\pm$	0.3	&	1.3	$\pm$	0.8	\\
Per-emb-62	&	03:44:12.97	&	+32:01:35.29	&	0.080	&	23	&	1.2	$\pm$	0.8	&	8.3	$\pm$	0.2	&	0.56	$\pm$	0.47	\\
Per-emb-63	&	03:28:43.28	&	+31:17:33.25	&	0.019	&	23	&	...	&	...	&	...	\\						
Per-emb-64	&	03:33:12.85	&	+31:21:23.95	&	0.043	&	25	&	...	&	...	&	...	\\						
Per-emb-65	&	03:28:56.30	&	+31:22:27.69	&	0.049	&	15	&	...	&	...	&	...	\\						
Per-emb-66	&	...	&	...	&	$<$0.090		&	...	&	...	&	...	&	...	\\
B1bN	&	03:33:21.20	&	+31:07:43.93	&	0.38	&	17	&	...	&	...	&	...	\\						
B1bS	&	03:33:21.34	&	+31:07:26.44	&	0.38	&	24	&	...	&	...	&	...	\\						
L1448IRS2E	&	...	&	...	&	$<$0.27	&	...	&	...	&	...	&	...	\\
L1451-MMS	&	03:25:10.24	&	+30:23:55.01	&	0.092	&	12	&	0.20	$\pm$	0.47	&	7.5	$\pm$	2.0	&	1.8	$\pm$	4.8	\\
Per-bolo-45	&	03:29:06.77	&	+31:17:29.96	&	0.16	&	13	&	...	&	...	&	...	\\						
Per-bolo-58	&	03:29:25.42	&	+31:28:14.21	&	0.22	&	12	&	0.48	$\pm$	0.61	&	8.1	$\pm$	0.7	&	1.1	$\pm$	1.6	\\
SVS~13B	&	03:29:03.08	&	+31:15:50.98	&	0.86	&	21	&	3.3	$\pm$	0.6	&	8.3	$\pm$	0.09	&	0.95	$\pm$	0.22	\\
SVS~13C	&	03:29:02.03	&	+31:15:37.75	&	0.20	&	23	&	3.4	$\pm$	0.5	&	8.7	$\pm$	0.1	& 1.9	$\pm$	0.3	
\enddata
\tablenotetext{a}{Answers are: (Y)es, (N)o, or (M)arginal. Although \coto, \ttco, and \ceo\ are essentially detected toward every field, it does not mean that the line is associated with the protostar. Large-scale emission from the Perseus molecular cloud is frequently detected with the SMA even when emission is not associated with the protostar.}
\tablenotetext{b}{These contours are those shown in Figure~\ref{outflows}.}
\end{deluxetable*}

Table~\ref{tab:detection} shows whether the line is detected toward each MASSES field, as judged by analyzing the continuum images and the spectral cubes by eye. With the exception of some marginal \ceo\ detections, the three spectral lines \coto, \ttco, and \ceo\ are detected toward every field. Nevertheless, these detections do not imply that the emission is always associated with the source since the large-scale emission of the Perseus molecular cloud is detected with these observations. Indeed, spectral lines do not seem to be associated with many of the targets, which will be analyzed in more detail in Section~\ref{fake}.

\subsection{Continuum Mass Detection Limit}\label{completeness}
We estimate the minimum mass of a compact source that we expect to detect for a given continuum observations based on the measured sensitivity $\sigma_{\rm{1.3\,mm}}$. Following \citet{Hildebrand1983}, the mass of a source for optically thin dust continuum flux is

\begin{equation} \label{m_disk_eq}
	M = R_{\text{gd}} \frac{F_{\nu} d^2}{\kappa_{\nu} B_{\nu}(T_{\text{dust}})},
\end{equation}
where $R_{\text{gd}}$ is the gas to dust mass ratio, $F_\nu$ is the source's flux, $d$ is the distance to the source, $\kappa_\nu$ is the dust opacity, and $B_{\nu}(T_{\text{dust}})$ is the Planck function at dust temperature $T_{\text{dust}}$. We assume typical values of $R_{\text{gd}}$~=~100, $d$~=~235\,pc \citep{Hirota2008}, and $\kappa_{\rm{1.3\,mm}}$~=~0.899\,cm$^2$\,g$^{-1}$ \citep[][assuming thin ice mantles and a gas density of 10$^6$\,cm$^{-3}$]{Ossenkopf1994}. To be conservative with our minimum detected compact mass estimates, we assume $T_{\text{dust}}$~=~10\,K and require a three-sigma detection with all the flux in a single beam, i.e., $F_{\nu}$~=~3($\sigma_{\rm{1.3\,mm}} \times \text{bm})$. Note that if the source is larger than the beam, this assumption for $F_{\nu}$ is not valid, i.e., we are only concerned with the detection limit of a source smaller than the beam. Given these assumptions, the mass detection limit for each field is

\begin{equation} \label{m_disk_eq}
	M_{\text{limit}} = \left(\frac{\sigma_{\rm{1.3\,mm}}}{\rm{mJy\,bm}^{-1}}\right) \times 0.010\,M_\sun ,
\end{equation}
where $\sigma_{\rm{1.3\,mm}}$ is given in Table~\ref{tab:sens}. The mass sensitivity for the continuum varies dramatically, often due to dynamic range, but also due to observing conditions and the correlator that is used (ASIC vs. SWARM). To demonstrate a pessimistic mass detection limit for the majority of the observations, we select the observation for Per-emb-4 as an illustrative example. In this field, no continuum source was detected (i.e., we were not limited by dynamic range), and our estimated thermal noise is a bit worse than most MASSES observations due to unfavorable observing conditions. For this field, $\sigma_{\rm{1.3\,mm}}$~=~2.0\,mJy\,bm$^{-1}$, so $M_{\text{limit}}$~=~0.02\,$M_\sun$. Therefore, for most of our fields, we expect to detect a compact source greater than 0.020\,$M_\sun$ with $>$3$\sigma$ significance, if it exists. 

Many sources in Table~\ref{tab:sens} have $\sigma_{\rm{1.3\,mm}}$ values that are much higher than 2.0\,mJy\,bm$^{-1}$ because these observations were limited by dynamic range. In other words, while our detection limit is generally $\sim$0.02\,$M_\sun$, if there is a much brighter source in the field, we would not be able to detect a 0.02\,$M_\sun$ source in the same field.
\section{Deliverables}\label{deliverables}
The deliverables for the MASSES survey are available on the Harvard Dataverse. The data can be downloaded directly from \url{https://dataverse.harvard.edu/dataverse/MASSES}. The MASSES Dataverse contains two separate datasets for each observed source. One dataset contains $uv$ data while the other dataset contains images/cubes for the continuum and line observations. For each dataset, we include a README file that briefly summarizes the contents and explains how to use the dataset.

As mentioned in Section\,\ref{setup} and Table\,\ref{tab:lines}, the \coto\ line uses two chunks in the upper sideband, with the majority of the spectral line in the s14 chunk. The deliverables refer to the s14 chunk as the ``12CO21" chunk and the s13 chunk as the ``highvelCO" chunk.

\begin{figure*}[ht!]
\begin{center}
\includegraphics[width=2\columnwidth]{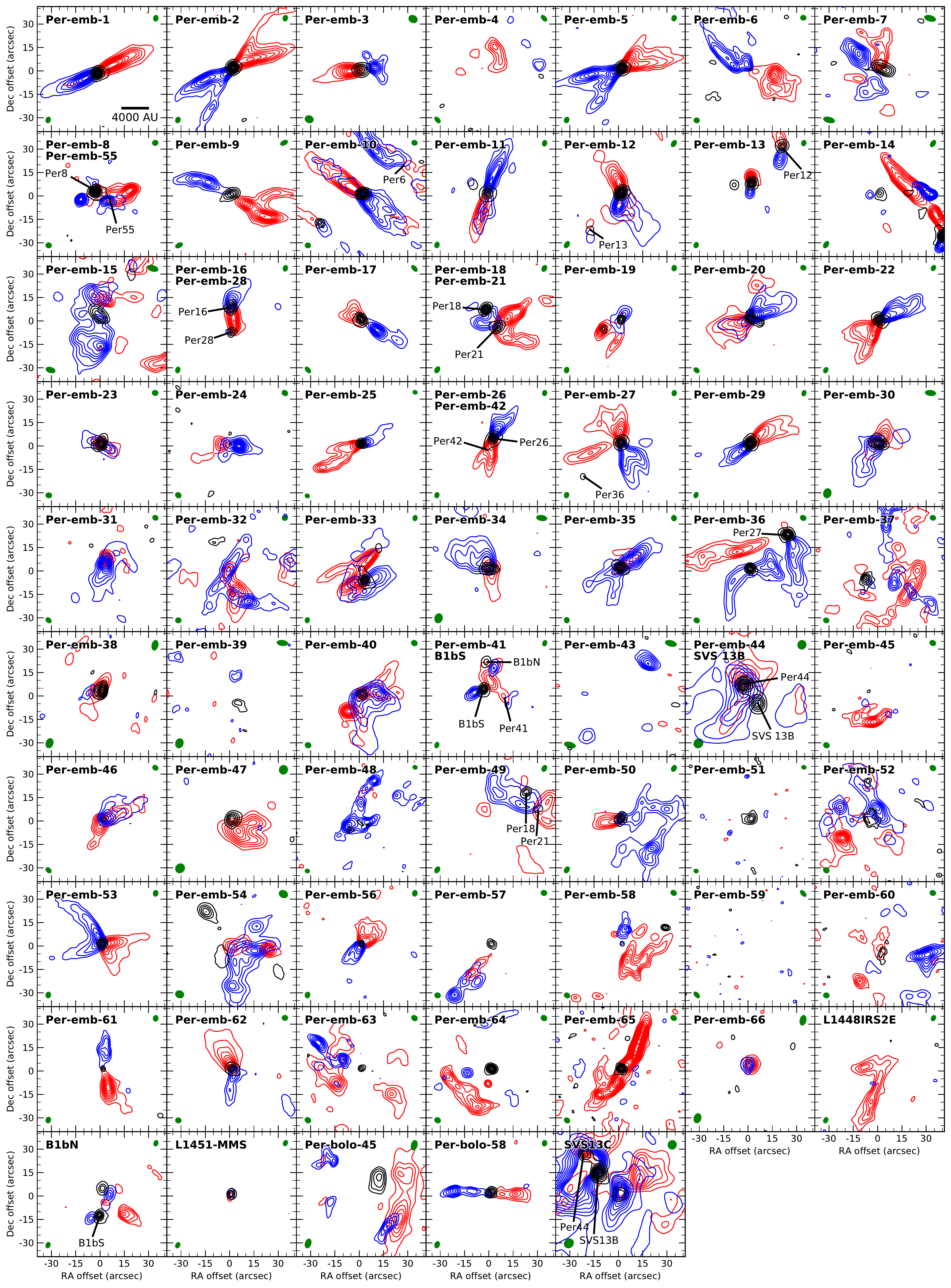}
\end{center}
\vspace{-12pt}
\caption{CO maps of all MASSES sources. Blue and red contours show integrated intensity (moment 0) maps of blue- and red-shifted emission, respectively. Black contours show the 1.3\,mm continuum. Contour levels are given in Table~\ref{tab:detection}. Bottom left and top right green ellipses are the synthesized beams for \coto\ and the 1.3\,mm continuum, respectively.
}
\label{outflows} 
\end{figure*}

\subsection{$uv$ data}
We provide the delivered $uv$ data for both the continuum observations and spectral line observations.  If tracks were taken on the same day (i.e., had the same YYMMDD prefix; see Section~\ref{setup}), they are combined into a single dataset during the MIR calibration. 
As explained in Section~\ref{imaging}, lines were subtracted when generating the continuum, and the continuum was subtracted for each spectral line observation. The line $uv$ data is delivered per spectral line for each track, while the continuum $uv$ data is delivered separately for the lower and upper sideband.

Finally, for the SWARM data, we also provide the full resolution $uv$ data, which is not continuum-subtracted. Many additional spectral lines (Table~\ref{tab:swarmlines}) may be found in these cubes, as discussed in Section~\ref{setup}. These lines were typically not detected in the ASIC cubes in part due to the smaller bandwidth coverage, although the coarse spectral resolution also makes such detections with ASIC less obvious. These SWARM $uv$ data are delivered separately for each of its 8 (4 per sideband) spectral chunks.

We also note that toward the edge of both SWARM and ASIC chunks, the noise in the spectra is extremely high, so the user should use these channels with caution. These channels were not used for the delivered imaged cubes.

The $uv$ data are delivered as uv-fits files. Examples of the delivered uv-fits filenames are given in Section~\ref{delivered}.


\subsection{Imaged data}
For the continuum and spectral line observations, we deliver both the images not corrected for the primary beam as well as the primary-beam corrected images. The images that are not corrected for the primary beam are typically used to better show structure throughout the entire map. The primary-beam corrected images allow the user to make accurate flux measurements.

Multiple tracks were combined during the MIRIAD \texttt{invert} task (Section~\ref{imaging}), allowing for two delivered products per continuum/spectral line (primary beam uncorrected and corrected). These data are delivered as fits files. 

The units for the continuum images are Jy\,bm$^{-1}$, and the units for the spectral line cubes are Jy\,bm$^{-1}$\,channel$^{-1}$.

\subsection{Examples of Delivered Data}\label{delivered}
An example of the delivered fits files for the Per-emb-24 dataset is shown below. 

The delivered $uv$ data for Per-emb-24 are:
\begin{itemize}\vspace{-6pt}
\item Per24.sub.cont1.3mm.lsb.151122.uvfits \vspace{-6pt}
\item Per24.sub.cont1.3mm.usb.151122.uvfits \vspace{-6pt}
\item Per24.sub.cont1.3mm.lsb.151127.uvfits \vspace{-6pt}
\item Per24.sub.cont1.3mm.usb.151127.uvfits \vspace{-6pt}
\item Per24.sub.12CO21.151122.uvfits \vspace{-6pt}
\item Per24.sub.12CO21.151127.uvfits \vspace{-6pt}
\item Per24.sub.highvelCO.151122.uvfits \vspace{-6pt}
\item Per24.sub.highvelCO.151127.uvfits \vspace{-6pt}
\item Per24.sub.13CO21.151122.uvfits \vspace{-6pt}
\item Per24.sub.13CO21.151127.uvfits \vspace{-6pt}
\item Per24.sub.C18O21.151122.uvfits \vspace{-6pt}
\item Per24.sub.C18O21.151127.uvfits \vspace{-6pt}
\item Per24.sub.N2DP.151122.uvfits \vspace{-6pt}
\item Per24.sub.N2DP.151127.uvfits 
\end{itemize}

The delivered images/data cubes for Per-emb-24 are:
\begin{itemize}\vspace{-6pt}
\item Per24.sub.cont1.3mm.fits \vspace{-6pt}
\item Per24.sub.cont1.3mm.pbcor.fits \vspace{-6pt}
\item Per24.sub.12CO21.cube.fits \vspace{-6pt}
\item Per24.sub.12CO21.cube.pbcor.fits \vspace{-6pt}
\item Per24.sub.highvelCO.cube.fits \vspace{-6pt}
\item Per24.sub.highvelCO.cube.pbcor.fits \vspace{-6pt}
\item Per24.sub.13CO21.cube.fits \vspace{-6pt}
\item Per24.sub.13CO21.cube.pbcor.fits \vspace{-6pt}
\item Per24.sub.C18O21.cube.fits \vspace{-6pt}
\item Per24.sub.C18O21.cube.pbcor.fits \vspace{-6pt}
\item Per24.sub.N2DP.cube.fits \vspace{-6pt}
\item Per24.sub.N2DP.cube.pbcor.fits 
\end{itemize}



Tracks with SWARM data will include include additional full spectral resolution $uv$ data, separated into 8 chunks (4 for each sideband). An example of the additional SWARM $uv$ data for Per-emb-7 is shown below.

\begin{itemize}\vspace{-6pt}
\item Per7.sub.SWARM.lsb.s1.160925.uvfits \vspace{-6pt}
\item Per7.sub.SWARM.lsb.s2.160925.uvfits \vspace{-6pt}
\item Per7.sub.SWARM.lsb.s3.160925.uvfits \vspace{-6pt}
\item Per7.sub.SWARM.lsb.s4.160925.uvfits \vspace{-6pt}
\item Per7.sub.SWARM.usb.s1.160925.uvfits \vspace{-6pt}
\item Per7.sub.SWARM.usb.s2.160925.uvfits \vspace{-6pt}
\item Per7.sub.SWARM.usb.s3.160925.uvfits \vspace{-6pt}
\item Per7.sub.SWARM.usb.s4.160925.uvfits
\end{itemize}

Note, the \emph{full} SWARM $uv$ data has not been continuum subtracted since the subtraction is best done by fitting the continuum near each spectral line rather than across the entire bandwidth. If a user analyzes the SWARM visibility data, the first and last $\sim$1200 channels for each chunk also should probably be discarded since fluxes toward the chunk edges of SWARM cannot be trusted. For sources with multiple protostellar systems in the same field, the names were combined in the filename, e.g., Per8Per55.

\begin{figure}[ht!]
\begin{center}
\includegraphics[width=1\columnwidth]{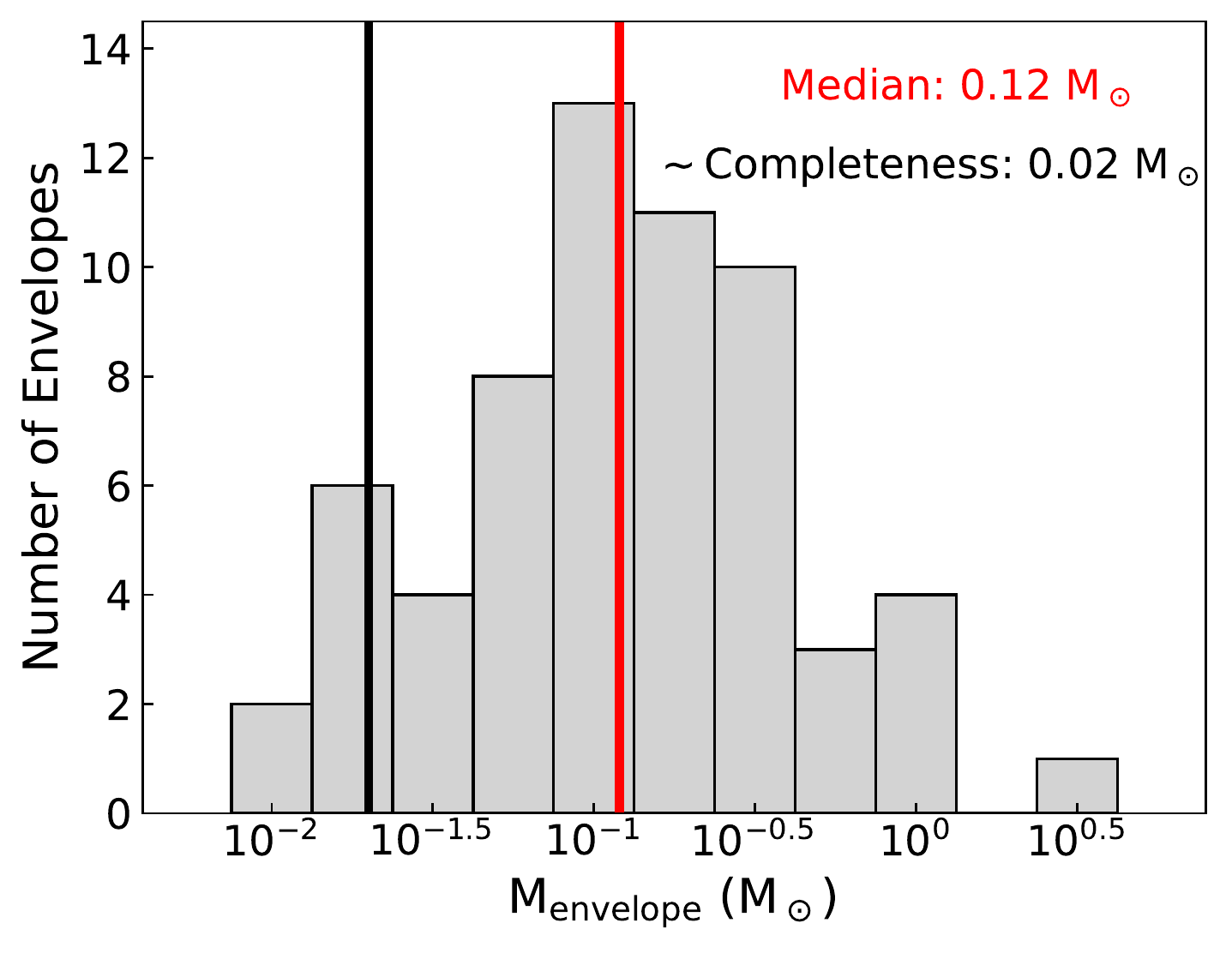}
\end{center}
\caption{Distribution of measured envelope masses (log scale) for protostars within the MASSES sample. The red line shows the median measured envelope mass, and the black line shows the approximate completeness level (see Section~\ref{completeness}).
}
\label{envelopemasses} 
\end{figure}

\begin{figure}[ht!]
\begin{center}
\includegraphics[width=1\columnwidth]{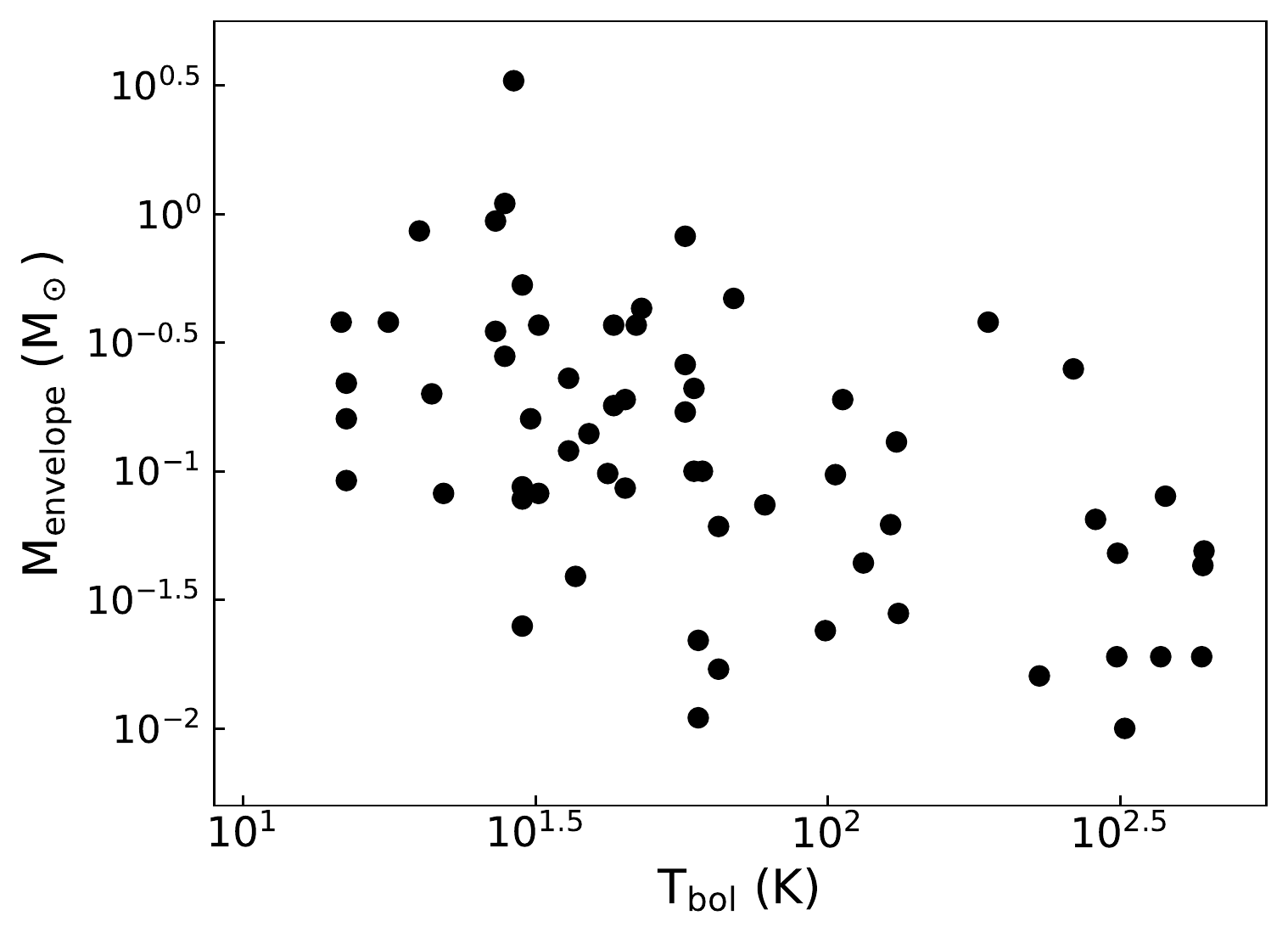}
\end{center}
\caption{Bolometric temperature ($T_{\text{bol}}$) versus envelope masses (log-log scale) for protostars within the MASSES sample. Additional envelopes detected near another protostar (indicated by a dash in Table~\ref{tab:fits}) are assumed to have the same $T_{\text{bol}}$ as the primary MASSES target.
}
\label{masstbol} 
\end{figure}

\begin{figure*}[ht!]
\begin{center}
\includegraphics[width=2\columnwidth]{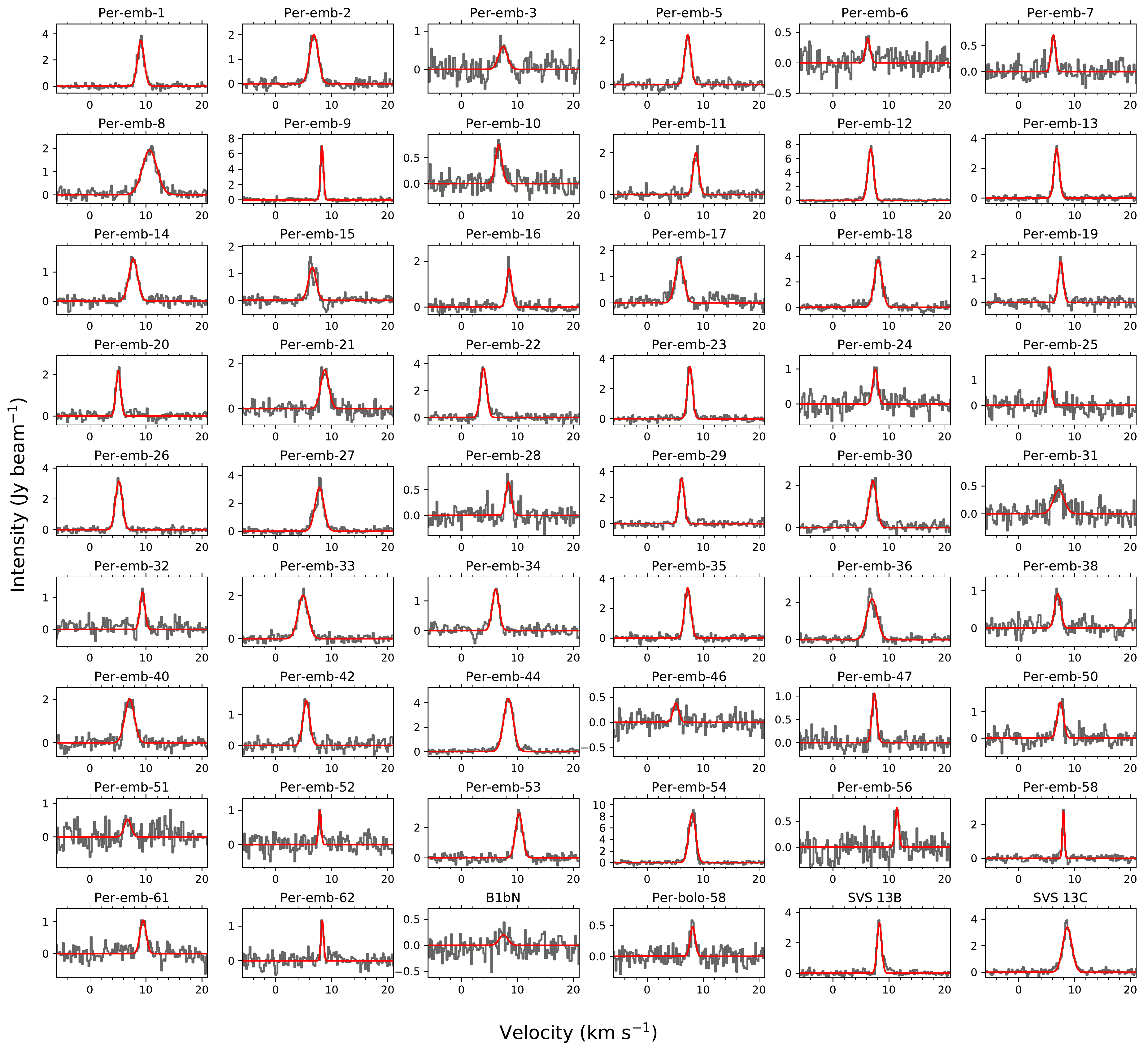}
\end{center}
\caption{\ceo\ spectra toward MASSES protostars in which we believe the emission is associated with the protostellar source. The red curve shows a single component Gaussian fit, with the fit parameters shown in Table~\ref{tab:fits}. 
}
\label{ceospectra} 
\end{figure*}

\begin{figure}[ht!]
\begin{center}
\includegraphics[width=1\columnwidth]{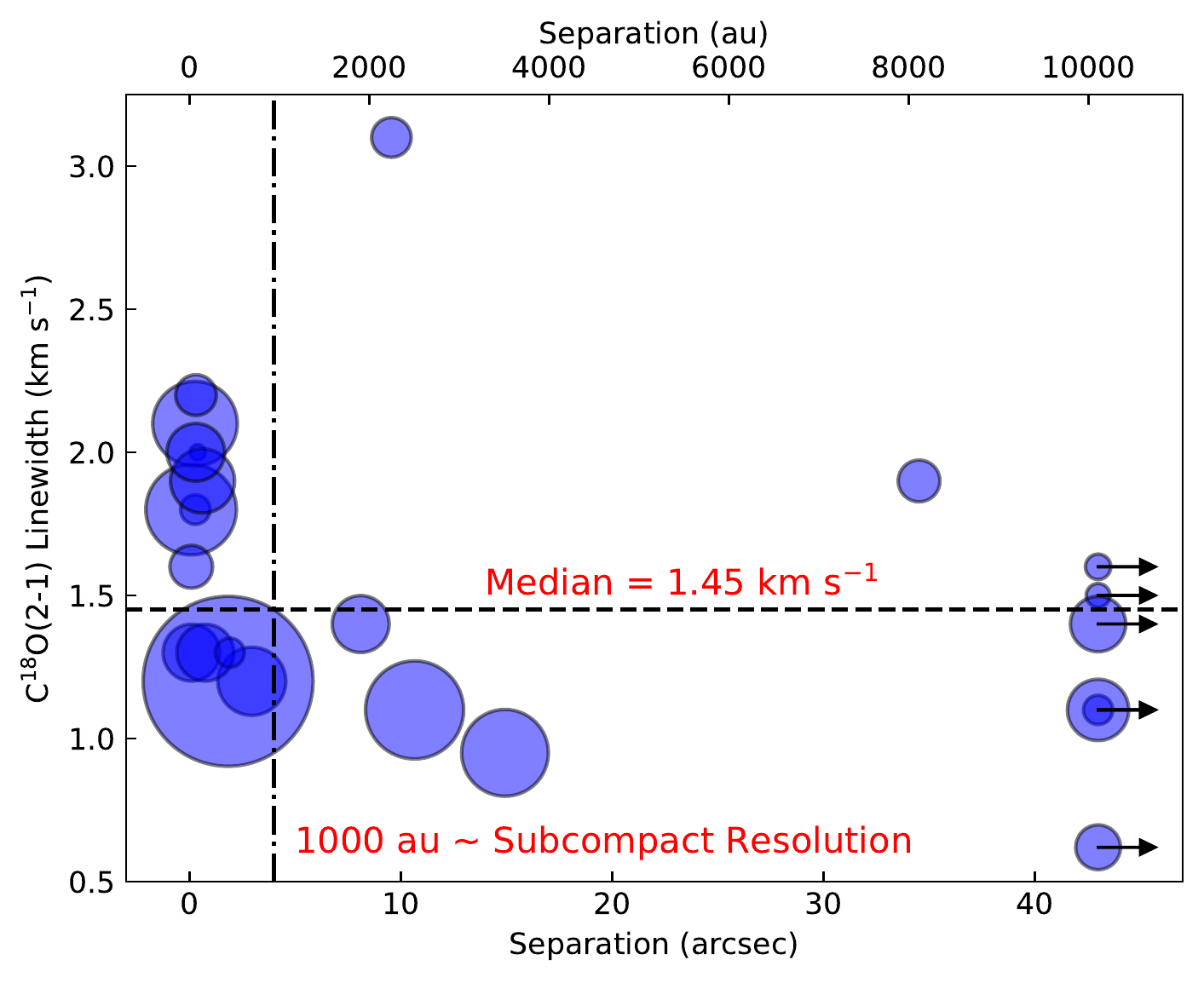}
\end{center}
\caption{Projected separation to the protostar's nearest companion \citep[from][]{Tobin2016} versus the measured \ceo\ linewidth (Table~\ref{tab:fits}). Only linewidths measured with 3$\sigma$ significance are shown. Protostars that do not have a known companion within 43$\arcsec$ (i.e., likely a single system) have an arrow indicating its separation lower limit. The size of the circle is directly proportional to the envelope mass (given in Table~\ref{tab:fits}).
}
\label{lw-sep} 
\end{figure}

\section{Example Survey Results}\label{example}

\subsection{\coto\ outflows}

Figure~\ref{outflows} shows CO(2--1) moment~0 maps of all protostars in the MASSES sample. For some maps, outflows are quite clear, while maps for other protostars, outflows can be confused with other emission (e.g., Per-emb-32), or they are completely absent (e.g., Per-emb-4). Note that we do not supply moment~0 maps since selecting integrated channel ranges for such maps depends on the features the user wants to extract from the map. Nevertheless, these maps can be reproduced by using the channel integration ranges shown in Table~\ref{tab:detection}, and these ranges were selected by eye in attempt to best show the outflow emission. Spectra for many of these outflows are presented in \citet{Stephens2017b}.

\subsection{1.3\,mm continuum Envelope Masses}\label{sec:envelopemasses}
Measurements of most of the envelope masses of the protostars were originally estimated in \citet{Pokhrel2018} by converting the integrated 1.3\,mm continuum flux to a gas mass. We briefly summarize the methodology below, and we refer the reader to \citet{Pokhrel2018} for additional details.

To measure the integrated fluxes of envelopes, we first analyzed the envelope's visibility plot of the amplitude versus $uv$-distance to determine which of three following models best describe the source. Some envelopes appeared more as point sources (amplitude versus $uv$-distance relation was flat), some were more Gaussian (relation is Gaussian), and others were a combination of the two (relation is a Gaussian with a flat tail). The total integrated flux of the source was then fit via miriad task \texttt{uvfit} by specifying one of these models.

The targets have a variety of luminosities and evolutionary stages, and therefore many targets are likely to have different temperatures. Temperatures at radius $r$ of each envelope is estimated via

\begin{equation}
T(r) = 60 \left(\frac{r}{2 \times 10^{15}\, \text{m}} \right )^{-q} \left(\frac{L_{\text{bol}}}{10^5 L_\odot} \right )^{q/2} \, \text{K} ,
\end{equation}
where $L_{\text{bol}}$ is the bolometric luminosity of the source \citep[taken from][]{Tobin2016} and $q$ is related to the dust emissivity index via $q = 2/(4+\beta)$. For our envelopes, we calculate $T_{d,\,1000\,\text{au}}$, which is the dust temperature at 1000\,au (approximately the resolution of the MASSES observations) assuming $q = 0.33$. The values for $T_{d,\,1000\,\text{au}}$ estimated for each envelope are listed in Table~\ref{tab:fits}. Targets with multiple envelopes in the same field (e.g., Per-emb-10 and Per-emb-10-SMM) are assigned with the same temperature $T_{d,\,1000\,\text{au}}$. The integrated intensity and $T_{d,\,1000\,\text{au}}$ are then used to estimate the gas mass via Equation~\ref{m_disk_eq}  (using the same values for $R_{\text{gd}}$, $d$, and, $\kappa_\nu$ discussed in Section~\ref{completeness}). These estimated envelope masses are given in Table~\ref{tab:fits}.

Three envelope masses (for Per-emb-7, Per-emb-34, and Per-emb-38) were not reported in \citet{Pokhrel2018} because those data were not yet calibrated at the time of that publication. In the same manner as \citet{Pokhrel2018}, Per-emb-7 was fit with a Gaussian, while Per-emb-34 and Per-emb-38 were each fit with a point source plus Gaussian. 

The envelope estimates for many sources were updated since newer, calibrated baselines were added to the visibility data (Per-emb-15, Per-emb-30, Per-emb-44, Per-emb-47, Per-bolo-45, SVS~13B, and SVS~13C). The calculated masses were similar to those calculated in \citet{Pokhrel2018}, except for Per-emb-44, whose mass was essentially halved. This change was due to a different dust temperature used for Per-emb-44, 38~K, rather than 21~K temperature (i.e., the temperature of the adjacent source SVS~13B) used in \citet{Pokhrel2018}. Moreover, we also update the \citet{Pokhrel2018} table with two other corrections; the measured \citet{Pokhrel2018} integrated flux for Per-emb-41 should have been for B1bS, and we also use update Per-emb-19-SMM's $T_{d,\,1000\,\text{au}}$ to be the same as Per-emb-19. Finally, we update all the masses to use 235\,pc as the distance rather than 230\,pc. These new masses do not change any of the conclusions found in \citet{Pokhrel2018}.




We show the distribution of envelope masses in Figure~\ref{envelopemasses}. The range of masses are from 0.010 to 3.2~$M_\odot$. The median value for the measured envelope masses is 0.13\,$M_\odot$. We do not provide the median value for all envelopes (i.e., including those undetected in the continuum) since we question the protostellar nature of some of these targets (see Section~\ref{fake}).

In Figure~\ref{masstbol}, we show the measured envelope masses versus the bolometric temperatures, $T_{\text{bol}}$. Envelope masses tend to slightly decrease with increasing bolometric temperature. Since protostars with higher bolometric temperatures tend to be older sources, this trend may suggest that envelopes typically become smaller as the protostar evolves. The decreasing envelope masses between the Class~0 and I stages agree with the results based on \ceooz\ observations in \citet{Arce2006} and submillimeter continuum observations in \citet{Jorgensen2009}. This trend, along with the large observed scatter, was also found via simulations in \citet{Frimann2016}.


\subsection{\ceo\ Spectra}
For the MASSES survey, \ceo\ is more frequently detected toward the peak of the continuum emission than \ntdp\ and typically has much higher signal to noise. Therefore, we fit \ceo\ spectra to estimate linewidths toward individual protostellar sources.

We show the \ceo\ spectra toward each protostar in which we believe \ceo\ emission is associated with the protostar. We deemed emission associated with the protostar if an integrated intensity map of \ceo\ overlapped significantly with the location of the protostar. In Figure~\ref{ceospectra}, we show the \ceo\ spectra, which is taken at the 1.3\,mm continuum peak of the protostar. If the 1.3\,mm continuum is not detected (Per-emb-31, Per-emb-32, Per-emb-42, Per-emb-52, and Per-emb-54) we instead take the spectra from the protostar's position given in \citet{Tobin2016}. Each spectrum is fit with a Gaussian, and the fit parameters (amplitude, systemic velocity $v_{\text{systemic}}$, and linewidth $\Delta v$) are presented in Table~\ref{tab:fits}. Sources with no fits in the table have 1.3\,mm continuum detected, but \ceo\ is not obviously associated with the source.


The fit parameters in Table~\ref{tab:fits} indicate that the amplitudes and linewidths do not always have great fits. However, the systemic velocities are fairly well constrained. The median linewidth of the \ceo\ spectra toward the protostars' positions is 1.45\,\kms, considering only fitted linewidths with 3-sigma significance. These \ceo\ linewidths are broader than the 0.6--1.0\,\kms\ linewidths measured for the Perseus filaments via \mbox{C$^{18}$O(1--0)} with the Five College Radio Astronomy Observatory 14\,m telescope at 1$\arcmin$ (0.07\,pc) resolution \citep{Hatchell2005}. Moreover, \citet{Kirk2007} observed \ceo\ at Perseus cores using the IRAM 30\,m telescope at 11$\arcsec$ ($\sim$2600\,au) resolution. They found that \ceo\ linewidths toward Perseus cores are also typically lower than 1\,\kms, with protostellar cores having slightly larger linewidths than starless cores. Other cores in nearby molecular clouds also typically have lower than~1\,\kms\ \mbox{C$^{18}$O(1--0)} linewidths \citep{Myers1983}.

We investigate whether multiplicity or the envelope mass (as measured by the 1.3\,mm continuum; Table~\ref{tab:fits}) affects the observed \ceo\ linewidths. \citet{Tobin2016} investigated multiplicity of all Perseus protostars with a resolution of $\sim$0$\farcs$065 (15\,au), and reported the projected separations of all multiples out to 43$\arcsec$ ($\sim$10,000\,au). In Figure~\ref{lw-sep}, we show the projected separation of the nearest known multiple of each protostar as a function of the \ceo\ linewidth. The size of each point in Figure~\ref{lw-sep} is directly proportional to the envelope mass. Protostellar systems with the largest linewidths tend to be multiple systems within the SMA beam ($\sim$1000\,au), but there are still many close multiples with smaller linewidths. Envelope mass does not seem to have a major effect on the observed linewidths. Other factors such as rotation, infall, and/or outflows could also broaden the \ceo\ line. At this time, it is unclear why the envelope linewidths are larger than the observed large-scale linewidths.



%

\begin{figure}[ht!]
\begin{center}
\includegraphics[width=1\columnwidth]{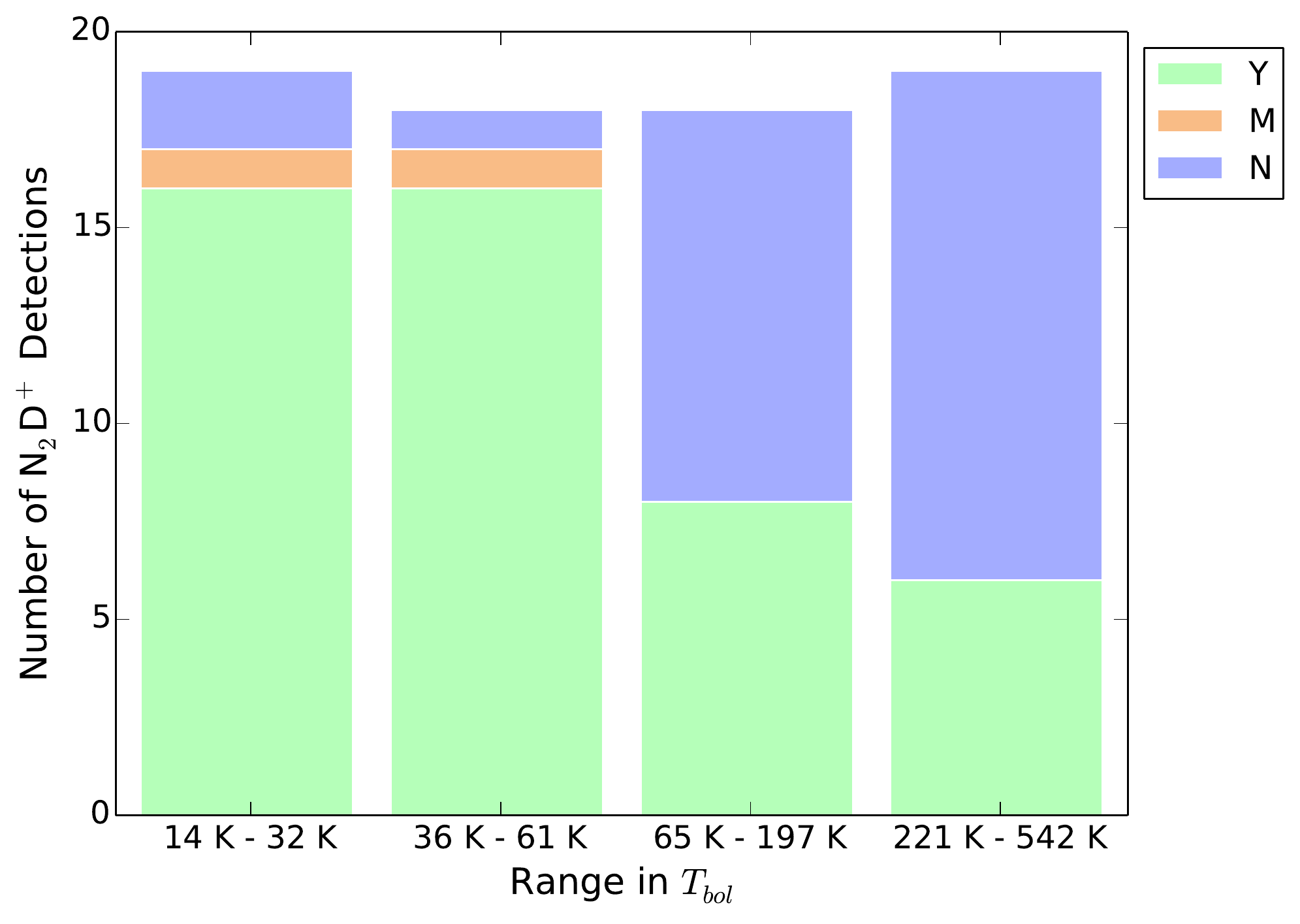}
\end{center}
\caption{A stacked histogram binned by $T_{\text{bol}}$, showing how many sources in each bin have N$_2$D$^+$ (Y)es detected, (M)arginally detected, or (N)ot detected (see Table~\ref{tab:detection}). For $T_{\text{bol}}$ ranges between these bins, e.g., between 32 and 36~K, there exists no MASSES sources with these temperatures. The bin ranges are chosen to put approximately the same amount of protostars in each bin. 
}
\label{n2dpdetect} 
\end{figure}

\begin{figure}[ht!]
\begin{center}
\includegraphics[width=0.95\columnwidth]{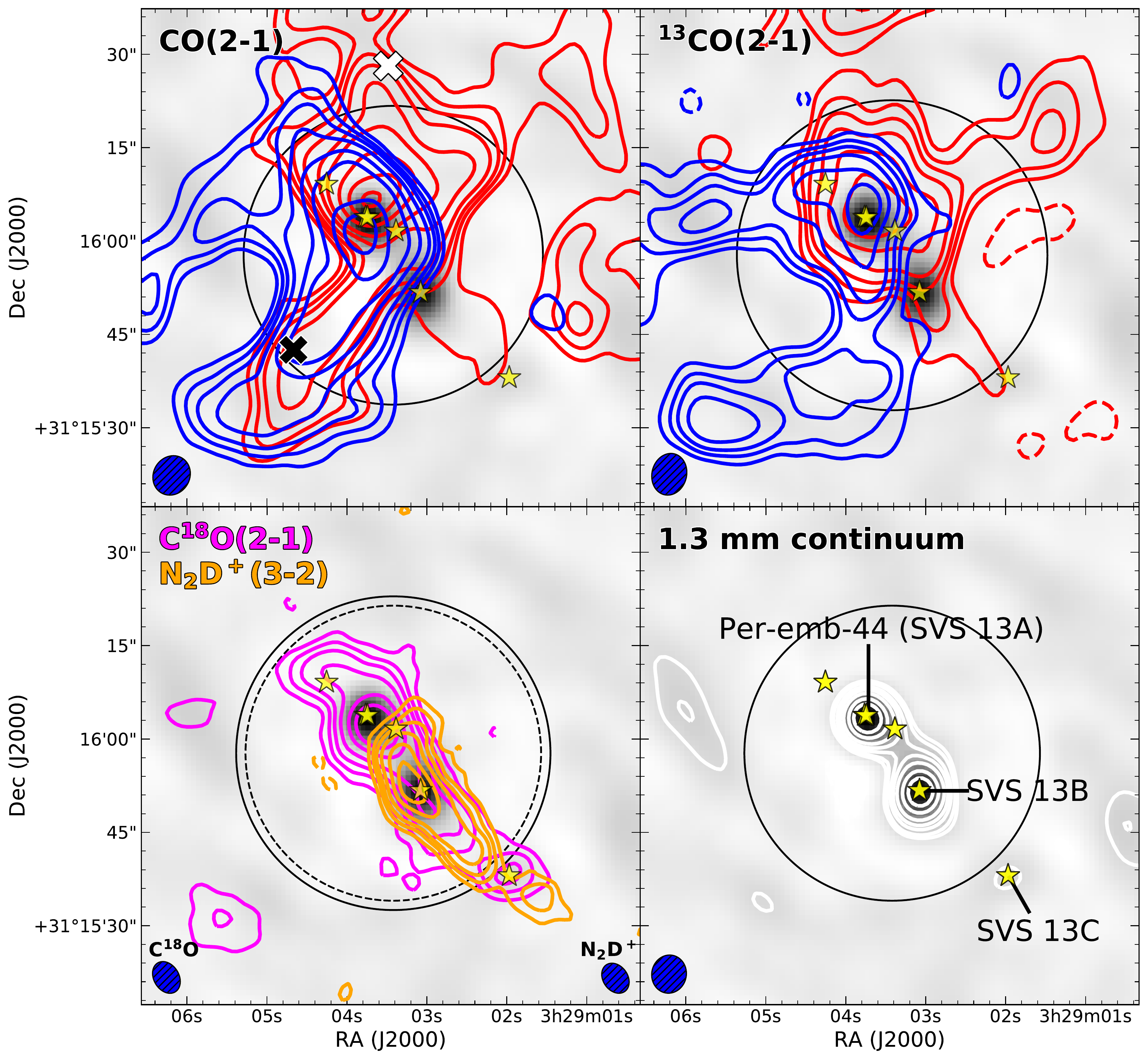}
\end{center}
\caption{MASSES images centered on SVS~13A (Per-emb-44). The images have not been corrected for the primary beam. The gray-scale shows the 1.3\,mm continuum. We show integrated intensity (moment~0) contours for \coto\ (top-left), \ttco\ (top-right), and \ceo\ and \ntdp\ (bottom-left), and we show the 1.3\,mm continuum (bottom-right). For \coto, blue emission is integrated from --153\,\kms\ to 5.5\,\kms\ while red emission is integrated from 9.8\,\kms\ to 164\,\kms. For \ttco, blue emission is integrated from --1.9\,\kms\ to 8\,\kms\ while red emission is integrated from 9.8\,\kms\ to 15.5\,\kms. For \ceo, emission is integrated from 5.4\,\kms\ to 14\,\kms. For \ntdp, emission is integrated from 7\,\kms\ to 9.8\,\kms.
 The black circles show the FWHM primary beam size of the observation; for the bottom left panel, the solid circle shows the primary beam for \ceo\ and the dashed circle shows the primary beam for \ntdp. Yellow stars show the locations of compact sources identified by the VLA in the VANDAM survey \citep{Tobin2016}. The black and white crosses indicate locations in which we show spectra in Figure~\ref{hvc} of high-velocity components of the outflows. For each panel, contours are set to [--5,--3,3,5,7,10,15,20,30,40,50]~$\times$~$f_{\text{x}}$, for $f_{\text{\coto,\text{red}}}$~=~5, $f_{\text{\coto,\text{blue}}}$~=~8, $f_{\text{\ttco,\text{red}}}$~=~0.5, $f_{\text{\ttco,\text{blue}}}$~=~1, $f_{\text{\ceo}}$~=~0.4, and $f_{\text{\ntdp}}$~=~0.16\,Jy\,bm$^{-1}$\,km\,s$^{-1}$ and $f_{\text{1.3~mm}}$~=~16\,mJy\,bm$^{-1}$. Continuum contours also include additional contour levels at $\pm$\,2~$\times$~$f_{\text{1.3~mm}}$. Dashed contours show negative levels.
}
\label{SVS13A} 
\end{figure}

\begin{figure}[ht!]
\begin{center}
\includegraphics[width=0.95\columnwidth]{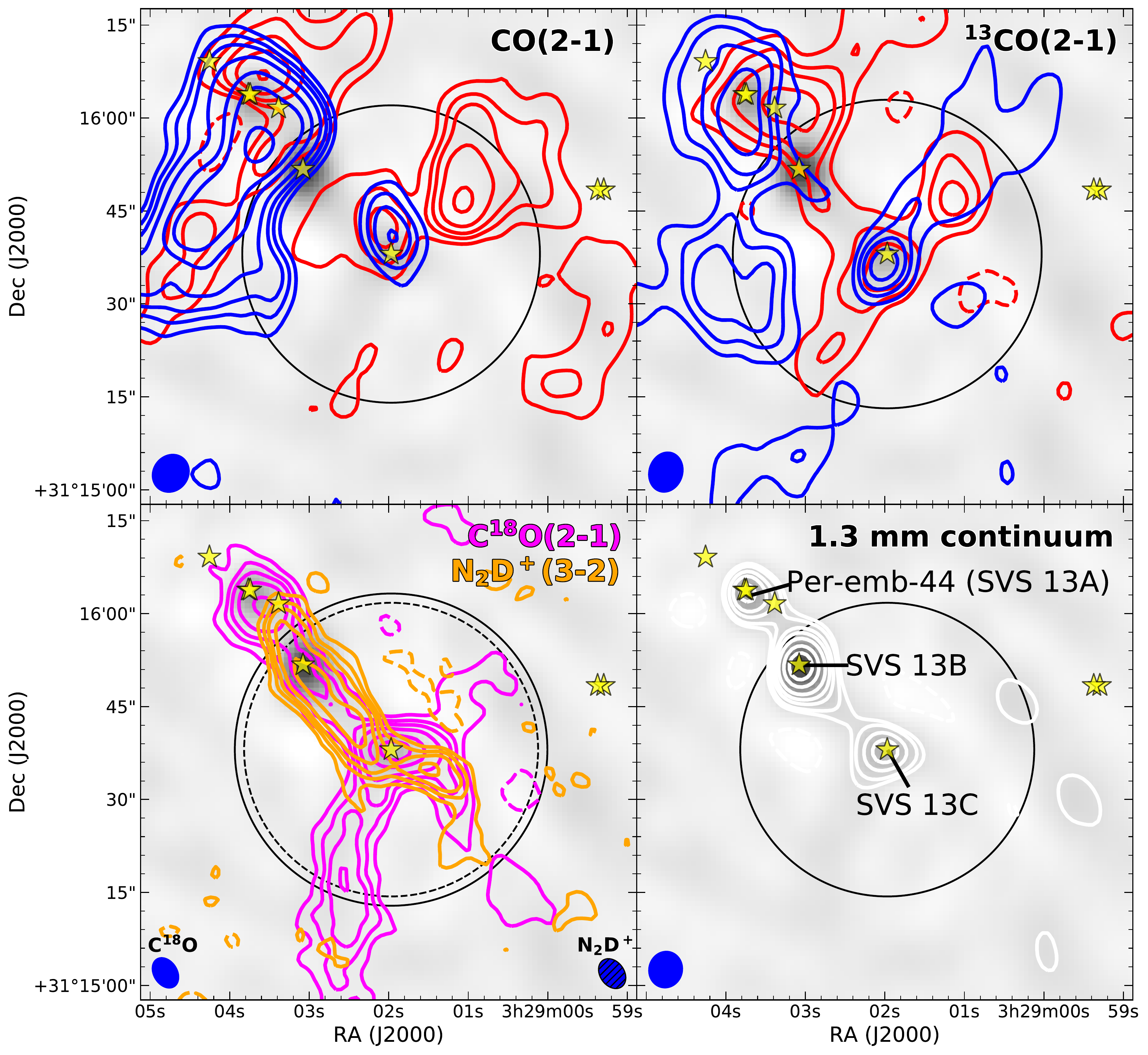}
\end{center}
\caption{Image description is the same as Figure~\ref{SVS13A}, except now the image is centered on SVS~13C, and $f_{\text{\coto,\text{red}}}$ = 5, $f_{\text{\coto,\text{blue}}}$ = 3.7, $f_{\text{\ttco,\text{red}}}$ = 0.5, $f_{\text{\ttco,\text{blue}}}$ = 0.8, $f_{\text{\ceo}}$ = 0.3, and $f_{\text{\ntdp}}$ = 0.14\,Jy\,bm$^{-1}$\,km\,s$^{-1}$ and $f_{\text{1.3~mm}}$ = 12\,mJy\,bm$^{-1}$.
}
\label{SVS13C} 
\end{figure}

\subsection{N$_2$D$^+$ Detections with Evolution}
We do not show the \ntdp\ spectra as we did with \ceo\ because the emission is typically anti-correlated with the continuum emission. Instead, we investigate the detection rates of \ntdp\ near the protostar as a function of evolution.

Figure~\ref{n2dpdetect} shows a histogram of whether or not \ntdp\ is detected toward each protostar, as indicated in Table~\ref{tab:detection}. The approximate age of these protostars are considered to be proportional to the bolometric temperature, $T_{\text{bol}}$ \citep[e.g.,][]{Myers1993}, which is listed for each source in Table~\ref{tab:sources}. Figure~\ref{n2dpdetect} bins the protostars by $T_{\text{bol}}$ in 4 $\sim$equal-sized bins. The detection rates (with marginal detections considered as a non-detection) for the four bins, from left to right, are 84.2~$\pm$~8.4\%, 88.9~$\pm$~7.4\%, 44.4~$\pm$~11.7\%, and 31.6~$\pm$~10.7\%. This indicates that protostars with higher $T_{\text{bol}}$, which are likely to be older protostars, are much more likely to have \ntdp\ undetected. Since \ntdp\ is expected to disappear at $T$ $\gtrsim$ 20\,K \citep[e.g.,][]{Jorgensen2011} or at least in hotter environments \citep{Emprechtinger2009}, such a trend is expected. 

\subsection{MASSES Targets That May Not Be Protostars}\label{fake}
As mentioned in Section~\ref{imaging}, several sources have no spectral lines associated with the target. Some of these non-detected targets also do not have 1.3\,mm continuum associated with them, nor are they detected with the VLA in \citet{Tobin2016}. There are six sources in question: Per-emb-4, Per-emb-39, Per-emb-43, Per-emb-45, Per-emb-59, and Per-emb-60. 

With the exception of Per-emb-4, the derived bolometric luminosities (listed in \citealt{Enoch2009} and \citealt{Tobin2016}) are similar to or smaller than the error. Only Per-emb-39 and Per-emb-60 are marginally detected in the 1.3\,mm continuum (Figure~\ref{outflows}). Based on CSO SHARC-II \citep{Suresh2016} and JCMT SCUBA-2 \citep{MChen2016} single dish observations, Per-emb-39 and Per-emb-60 also have evidence of compact structure, while Per-emb-4 questionably does as well. However, based on these single dish observations and $Herschel$ archival observations, no strong compact emission is evident for Per-emb-43, Per-emb-45, and Per-emb-59.

Envelope masses for these sources are likely to be less than 0.2\,$M_\odot$, and in some cases, much smaller (see Section\,\ref{completeness} and Table\,\ref{tab:fits}). These six sources may have spectral energy distributions that cause them to be undetected at (sub)millimeter wavelengths (e.g., they are not protostellar in nature), are misidentifications in the \citet{Enoch2009} paper, or are simply very low-mass sources that are undetected with the MASSES and VANDAM observations. If these targets are misidentifications, this may indicate that \citet{Enoch2009} sources with poor luminosity fits are less reliably protostars.

\begin{figure*}[ht!]
\begin{center}
\includegraphics[width=2\columnwidth]{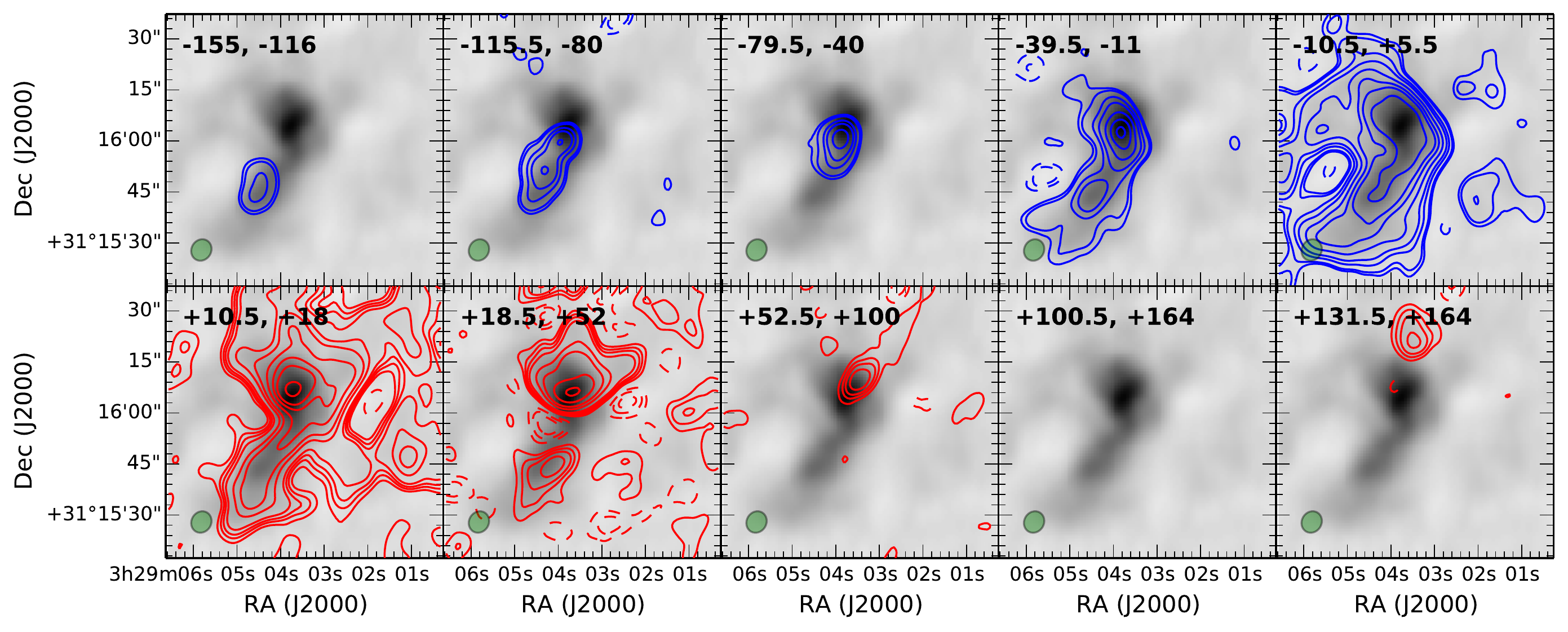}
\end{center}
\caption{Integrated velocity maps for Per-emb-44 (SVS~13A). The grayscale in each panel shows the \coto\ integrated intensity (moment~0) map, which is integrated over the velocity range of --155 to 164\,\kms. Contours show the moment~0 maps integrated over the velocity interval indicated in the top left of each panel. The top panels show the blue-shifted emission from systemic velocities, with contours of \mbox{[--10,--7,7,10,20,30,50,75,100,150]}~$\times$~mJy\,bm$^{-1}$\,\kms, while the bottom panels show red-shifted emission, with contours of \mbox{[--7,--5,--3,3,5,7,10,20,30,50,75,100,150]}~$\times$~mJy\,bm$^{-1}$\,\kms.
}
\label{velmap} 
\end{figure*}

\begin{figure}[ht!]
\begin{center}
\includegraphics[width=0.9\columnwidth]{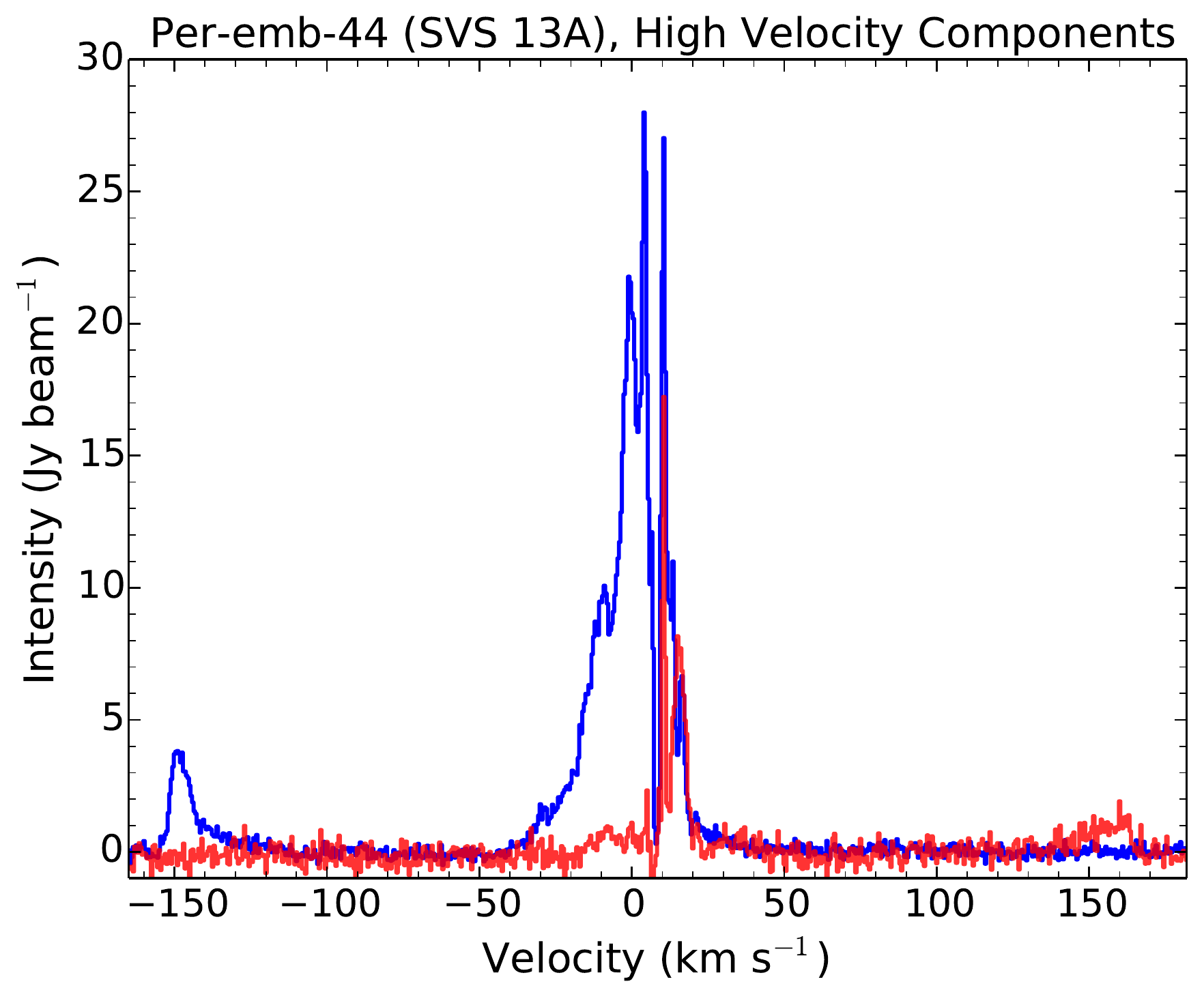}
\end{center}
\caption{Spectra of the high velocity components in the outflow from Per-emb-44 (SVS~13A), which were extracted from the pixels indicated in Figure~\ref{SVS13A}. The blue curve shows a high velocity component near --150\,\kms\ and the red curve shows a component near 160\,\kms. The spectra are from the primary beam-corrected cubes.
}
\label{hvc} 
\end{figure}

\begin{figure*}[ht!]
\begin{center}
\includegraphics[width=1.9\columnwidth]{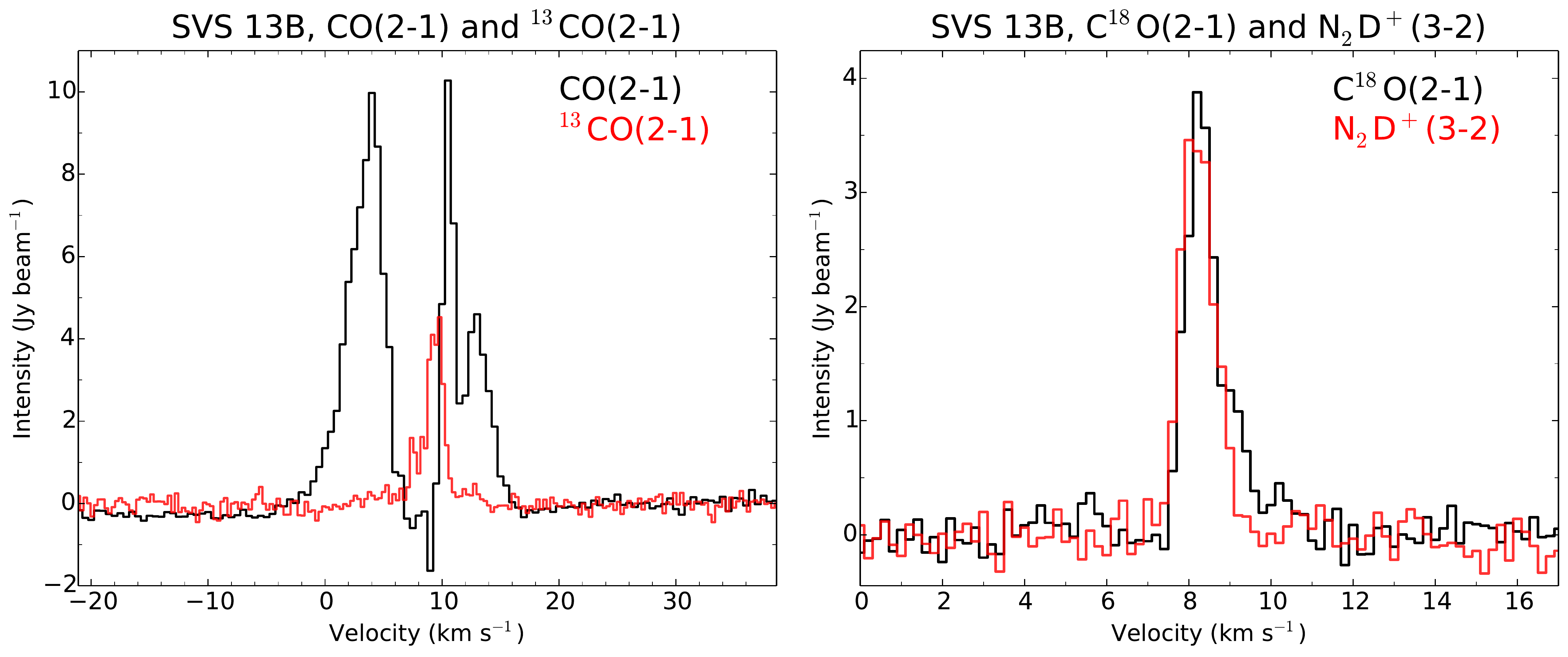}
\end{center}
\caption{Spectra at the 1.3\,mm continuum peak for SVS~13B. \coto\ and \ttco\ are shown on the left, and \ceo\ and \ntdp\ are shown on the right. The spectra are from the primary beam-corrected cubes.
}
\label{spec} 
\end{figure*}

\subsection{The SVS~13 Star-Forming Region}
Many examples of MASSES observations have already been presented in a number of papers \citep{Lee2015,Lee2016,Frimann2017,Stephens2017b,Pokhrel2018}. All of the previous papers showed results from the ASIC correlator. Here, we show an example of one of the brightest regions in the MASSES sample, SVS~13, in which we combine an ASIC-only track with a SWARM-only track. \citet{Looney2000} classified the three envelopes in the SVS~13 region, from northeast to southwest, as SVS~13A, SVS~13B, and SVS~13C. SVS~13A is referred to as Per-emb-44 in the MASSES sample, but we will refer to it as SVS~13A in this section.

In Figures~\ref{SVS13A} and~\ref{SVS13C} we show the MASSES observations of the SVS~13 region. The 1.3~mm continuum, \ceo, and \ntdp\ SWARM observations are combined with ASIC observations, while \coto\ and \ttco\ are SWARM-only (chunks containing \coto\ and \ttco\ were missing from the observations during the ASIC track). The four panels in Figure~\ref{SVS13A} are centered on SVS~13A while four panels in Figure~\ref{SVS13C} are centered on SVS~13C. 


The top panels show integrated intensity (moment~0) maps of \coto\ and \ttco. These maps trace the protostellar outflows of the sources. The bipolar outflow for SVS~13A is quite extended, with an angular extent much larger than the mapped area. Maps of the entire outflows are shown via a CO(1--0) mosaic in \citet{Plunkett2013}. We show \coto\ velocity maps in Figure~\ref{velmap}. For the SVS~13A outflow, we find high velocity components associated with both the blue and red lobes.  We indicate the locations of these high velocity components in Figure~\ref{SVS13A} and show the \coto\ spectra in Figure~\ref{hvc}. Using the MIRIAD task \texttt{maxfit}, we find that the peak emission in the --150\,\kms\ channel is located at RA=03:29:04.6 dec=+31:15:43, and the peak in the 160\,\kms\ channel is located at RA=03:29:03.4 +31:16:29. These high velocity components were originally detected with the Caltech Submillimeter Observatory in \citet{Masson1990}. Given that the systemic velocity of SVS~13 system is about 8\,\kms\ (see Figure~\ref{spec}), both lobes have line-of-sight velocity components of over 150\,\kms\ relative to the protostar's rest frame.




The bottom left panels of Figures~\ref{SVS13A} and~\ref{SVS13C} show moment~0 maps of \ceo\ and \ntdp. Both spectral line moment maps show filamentary structures, with \ceo\ connecting six (SVS~13A is a very close binary) protostellar sources that were identified in the VANDAM survey \citep{Tobin2016}. As seen in Figure~\ref{SVS13C}, the \ceo\ moment~0 map has 4 primary peaks. Three of these peaks are mostly coincident with continuum peaks, while the fourth traces a filamentary structure protruding southeast from SVS~13C. \ntdp\ emission is not detected in this southeast structure. \ntdp\ is absent near SVS~13A, which is the hottest of the three sources ($T_{\text{bol}}$~=~188\,K, c.f., $\sim$30\,K for the other two). \ntdp\ peaks toward SVS~13B, and is reduced toward SVS~13C. The lack of a deuterated species is indeed expected toward hotter sources.

The bottom right panels display the continuum, which show a chain of envelopes. The projected separation for the SVS~13A and SVS~13B 1.3\,mm continuum peaks is 14$\farcs$9 ($\sim$3500~au), while for the SVS~13B and SVS~13C peaks, the projected separation is 19$\farcs$8 ($\sim$4700~au).

Figure~\ref{spec} shows example spectra at the peak of SVS~13B. \coto\ and \ttco\ emission spectra have outflow wings. The \coto\ spectrum has an obvious dip between the two outflow wings, with the emission becoming negative; this emission is not likely absorption, but rather reflects confusion (due to missing zero-spacing in the $uv$-plane) with the large scale emission near the systemic velocity of the local molecular cloud. The \ceo\ and \ntdp\ spectra look very similar to each other. A Gaussian fit to both of these spectra gives linewidths of 1\,\kms, with line centers for \ceo\ and \ntdp\ of 8.3 and 8.2\,\kms, respectively.

In general, MASSES observations of the \ceo\ and \ntdp\ lines are more compact than those seen for the SVS~13 system. More images of MASSES \ceo\ observations can be found in \citet{Frimann2017}.

\section{Summary}\label{summary}
The MASSES survey used the SMA to observe all the known protostars (i.e., younger than Class~II) in the Perseus molecular cloud. This paper summarizes the data release for the subcompact SMA 230\,GHz data (baselines $\sim$\mbox{4 -- 55\,k$\lambda$}) of the MASSES survey, which can be downloaded from \url{https://dataverse.harvard.edu/dataverse/MASSES}. The survey includes observations of the 1.3\,mm continuum, \coto, \ttco, \ceo, and \ntdp. Both $uv$ data and data cube images are provided to the user. 

In this paper, we also present 1.3\,mm continuum and \coto\ subcompact maps for each source. The typical envelope mass is approximately 0.1\,$M_\odot$, and the envelope tends to decrease in mass over time. If \ceo\ is detected toward a protostar, we show the spectrum. We find typical linewidths of 1.45\,\kms, which is higher than the 0.6--1.0\,\kms\ linewidths found in Perseus at core and filament scales \citep{Myers1983, Hatchell2005, Kirk2007}. The larger linewidths seem to be independent of envelope mass, but for some (but certainly not all) protostars, the presence of a close multiple is associated with larger linewidths. We also find that \ntdp\ is significantly more likely to be detected in Perseus toward younger protostars. Moreover, based on these and ancillary observations, we question whether several of these targets (particularly Per-emb-43, Per-emb-45, and Per-emb-59, and perhaps Per-emb-4, Per-emb-39, and Per-emb-60) are actually protostars.

We also present the continuum and the spectral lines for the SVS~13 system. In the SVS~13 system, we show  the location of high velocity components (at --150\,\kms\ and 160\,\kms) of the bipolar outflow. Moment~0 maps of \ceo\ and \ntdp\ show filamentary structures. Moreover, \ceo\ appears to be enhanced toward hotter sources while \ntdp\ disappears.

The subcompact plus extended data, along with 356\,GHz subcompact data, will be presented in a future data release paper.

\acknowledgements
We acknowledge an anonymous referee for a thorough review that significantly helped this paper.
I.W.S. acknowledges support from NASA grant NNX14AG96G. 
E.I.V. acknowledges support form the Russian Ministry of Education and Science grant 3.5602.2017.
The authors thank the SMA staff for executing these observations as part of the queue schedule, Charlie Qi, Glen Petipas, Qizhou Zhang, and Garrett `Karto' Keating for their technical assistance with the SMA data, and Eric Keto for his guidance with SMA large-scale projects.
The Submillimeter Array is a joint project between the Smithsonian Astrophysical Observatory and the Academia Sinica Institute of Astronomy and Astrophysics and is funded by the Smithsonian Institution and the Academia Sinica.
This research has made use of the VizieR catalogue access tool and the SIMBAD database operated at CDS, Strasbourg, France. 
This research made use of APLpy \citep{Robitaille2012} and PySpecKit \citep{Ginsburg2011}, which are open-source plotting packages for Python.

\emph{Facilities:} SMA.

\newpage
\clearpage
\bibliography{stephens_bib}



\end{document}